\documentclass[fleqn,usenatbib,twocolumn]{mnras}

\usepackage{macros}
\usepackage{subfig}
\usepackage{graphicx}
\usepackage{threeparttable}
\usepackage[ruled,vlined]{algorithm2e}
\usepackage{amsmath,amssymb,bm}
\usepackage{tabularx}
\usepackage{orcidlink}
\usepackage{bm}
\usepackage{soul}

\usepackage{mathtools,amssymb,lipsum,widetext}

\usepackage{cuted}
\usepackage{hyperref}
\graphicspath{{./}{./figures/}}

\newcommand{\responsemnras}[1]{{\textcolor{black}{#1}}}
\newcommand{\responsemnrassecond}[1]{{\textcolor{black}{#1}}}

\newcommand{\ve}[1]{\vect{#1}}
\newcommand{\mt}[1]{\bm{\mathsf{#1}}}

\newcommand{\xlrv}[1]{{\color{black}#1}}      

\newcommand{\reGauss}{\texttt{reGauss}}     
\newcommand{\galsim}{\texttt{GalSim}}       
\newcommand{\res}{\mathcal{R}}              
\newcommand{\dNNz}{\texttt{dNNz}}           

\newcommand{\mizuki}{\texttt{mizuki}}

\DeclareRobustCommand{\VAN}[3]{#2}
\let\VANthebibliography\thebibliography
\def\thebibliography{\DeclareRobustCommand{\VAN}[3]{##3}\VANthebibliography}

\title[PSF bias in shear analysis]{A General Framework for Removing Point Spread Function Additive Systematics in Cosmological Weak Lensing Analysis}

\author[]{Tianqing Zhang$^1$\thanks{\tt tianqinz@andrew.cmu.edu}\orcidlink{0000-0002-5596-198X}, Xiangchong Li$^1$\orcidlink{0000-0003-2880-5102}, Roohi Dalal$^2$\orcidlink{0000-0002-7998-9899}, Rachel Mandelbaum$^1$\orcidlink{0000-0003-2271-1527}, Michael A. Strauss$^2$\orcidlink{0000-0002-0106-7755} ,\newauthor Arun Kannawadi$^2$\orcidlink{0000-0001-8783-6529}, Hironao Miyatake$^{3,4,5}$\orcidlink{0000-0001-7964-9766},  
Andrina Nicola$^{2,6}$\orcidlink{0000-0003-2792-6252}, Andrés A. Plazas Malagón$^{7,8}$\orcidlink{0000-0002-2598-0514},\newauthor Masato Shirasaki$^{9,10}$\orcidlink{0000-0002-1706-5797}, Sunao Sugiyama$^{3}$\orcidlink{0000-0003-1153-6735}, Masahiro Takada$^3$\orcidlink{0000-0002-5578-6472} 
Surhud More$^{3,11}$\orcidlink{0000-0002-2986-2371}
 \\
$^1$McWilliams Center for Cosmology, Department of Physics, Carnegie Mellon University, 5000 Forbes Ave, Pittsburgh, PA 15213, USA. \\
$^2$Department of Astrophysical Sciences, Princeton University, Peyton Hall, Princeton, NJ 08544, USA. \\
$^3$Kavli Institute for the Physics and Mathematics of the Universe (WPI), The University of Tokyo Institutes for Advanced Study (UTIAS), \\
The University of Tokyo, Chiba 277-8583, Japan\\
$^4$Kobayashi-Maskawa Institute for the Origin of Particles and the Universe (KMI), Nagoya University, Nagoya, 464-8602, Japan\\
$^5$Institute for Advanced Research, Nagoya University, Nagoya 464-8601, Japan\\
$^6$Argelander Institut f\"ur Astronomie, Universit\"at Bonn, Auf dem H\"ugel 71, 53121 Bonn, Germany\\
$^7$Kavli Institute for Particle Astrophysics and Cosmology, P.O. Box 20450, MS29, Stanford, CA 94309, USA\\
$^8$SLAC National Accelerator Laboratory, 2575 Sand Hill Road, MS29, Menlo Park, CA  94025, USA\\
$^9$National Astronomical Observatory of Japan (NAOJ), National Institutes of Natural Sciences, Osawa, Mitaka, Tokyo 181-8588, Japan\\
$^{10}$The Institute of Statistical Mathematics, Tachikawa, Tokyo 190-8562, Japan\\
$^{11}$The Inter-University Centre for Astronomy and Astrophysics, Post bag 4, Ganeshkhind, Pune 411007, India
}

\date{\today}

\begin{document}

\label{firstpage}
\pagerange{\pageref{firstpage}--\pageref{LastPage}}
\maketitle

\begin{abstract}
Cosmological weak lensing measurements rely on a precise measurement of the shear two-point correlation function (2PCF) along with a deep understanding of systematics that affect it.  In this work, we demonstrate a general framework for \responsemnras{detecting and modeling} the impact of PSF systematics on the cosmic shear 2PCF, and mitigating its impact on cosmological analysis. Our framework can detect PSF leakage and modeling error from all spin-2 quantities contributed by the PSF second and higher moments, rather than just the second moments\responsemnras{, using the cross-correlations between galaxy shapes and PSF moments}. We interpret null tests using the HSC Year 3 (Y3) catalogs with this formalism, and find that leakage from the spin-2 combination of PSF fourth moments is the leading contributor to additive shear systematics, with total contamination that is an order of magnitude higher than that contributed by PSF second moments alone. We conducted a mock cosmic shear analysis for HSC Y3, and find that, if uncorrected, PSF systematics can bias the cosmological parameters $\Omega_m$ and $S_8$ by $\sim$0.3$\sigma$. The traditional second moment-based model can only correct for a 0.1$\sigma$ bias, leaving the contamination largely uncorrected. We conclude it is necessary to model both PSF second and fourth moment contamination for HSC Y3 cosmic shear analysis. We also reanalyze the HSC Y1 cosmic shear analysis with our updated systematics model, and identify a 0.07$\sigma$ bias on $\Omega_m$ when using the more restricted second moment model from the original analysis. We demonstrate how to self-consistently use the method in both real space and Fourier space, assess shear systematics in tomographic bins, and test for PSF model overfitting.
\end{abstract}


\section{Introduction}
\label{sec:int:0}

In the past two decades, weak gravitational lensing, the slight distortions of the shape and size of the background (source) galaxies due to deflection of light rays by the foreground matter distribution, has  become one of the most powerful probes to study the distribution of dark matter in the Universe due to its sensitivity to the matter density field along the line of sight \citep{Hu:2001fb, 2010GReGr..42.2177H, 2013PhR...530...87W}. Measurements of cosmic shear, the coherent shape distortions of the source galaxies quantified via two-point correlation functions of galaxy shear estimates, are one of the most effective ways to measure the Large Scale Structure (LSS) and constrain the cosmological model. Stage-III imaging surveys \citep{Albrecht:2006um} such as the Hyper Suprime-Cam survey \citep[HSC;][]{2018PASJ...70S...4A}, the Dark Energy Survey \citep[DES;][]{des_review}, and the Kilo-Degree Survey \citep[KiDS;][]{deJong:2017bkf} all conduct cosmic shear analysis \citep[e.g.,][]{2020arXiv200715633A,2019PASJ...71...43H, 2020PASJ...72...16H,2022PhRvD.105b3514A,2022PhRvD.105b3515S}.  
Future galaxy surveys such as the \textit{Vera C.\ Rubin} Observatory Legacy Survey of Space and Time (LSST)  \citep[LSST;][]{Ivezic:2008fe, 2009arXiv0912.0201L}, the \textit{Nancy Grace Roman} Space Telescope High Latitude Imaging Survey \citep{2015arXiv150303757S, 2019arXiv190205569A} and \textit{Euclid} \citep{Euclid_overview} will measure cosmic shear with smaller statistical uncertainties by increasing the survey area, and in some cases by increasing the depth, and thus the number of galaxies, therefore putting more stringent requirements on controlling and modeling the systematic biases and uncertainties that affect cosmic shear measurements \citep{Mandelbaum:2017jpr}.
Another major motivating factor for improving our ability to control systematic uncertainties in weak lensing is the potential tension in the lensing amplitude, $S_8$  \citep{2021APh...13102604D}, an important parameter of the $\Lambda$CDM cosmological model, between the weak lensing cosmology and the Cosmic Microwave Background (CMB) cosmology \citep{2020A&A...641A...6P}.

The Point Spread Function (PSF) describes the image response to the light of a point source, after passing through atmospheric turbulence and the telescope optics \citep{2000PASP..112.1360A,2013A&A...551A.119P}. The PSF effectively acts as a convolution on the images of all observed objects, including galaxies. Therefore, the PSF is a major source of systematic biases and uncertainties in the measured galaxy shape, from which the weak lensing shear information is extracted. Biases in the estimated PSF size can give rise to multiplicative biases in the weak lensing shear signal as well, because the biases in PSF size result in an incorrect estimate of how much the PSF has rounded the observed galaxy shape (an effect for which we implicitly or explicitly correct).  The PSF shape (ellipticity) can contaminate cosmic shear in two different ways: First, ``PSF leakage'' arises when the shape of the PSF coherently contaminates the inferred shear even when the PSF model is perfect. This effect originates from an imperfect shear estimation method. Second, when the PSF model inaccurately describes the actual PSF shape (``PSF modeling error''), the inferred shear can get an additive systematics term  \citep[e.g.,][]{2008A&A...484...67P}. This second effect arises even for principled shear inference methods that should be unbiased with a perfect PSF model \citep[e.g.,][]{2016MNRAS.459.4467B,metaDet-Sheldon2020,FPFS_Li2022b}.  In many previous cosmic shear analyses, coherent biases in the PSF second moments (i.e., the shape and size) were monitored through the $\rho$ statistics  \citep{2010MNRAS.404..350R,2016MNRAS.460.2245J}. Null tests designed to identify potential additive shear systematics are typically conducted by cross-correlating the galaxy shapes, the PSF shape, and its modeling error \citep[e.g.,][]{HSC3-catalog, 2021MNRAS.501.1282J}, so that corresponding corrections can be made to the cosmic shear two-point correlation function (2PCF) through forward modeling. However, these PSF systematics formalisms have been limited to PSF second moments only.

\cite{2022MNRAS.510.1978Z, 2022arXiv220507892Z} showed that modeling error in PSF higher moments causes additive and multiplicative shear bias. \cite{2022MNRAS.510.1978Z} found sub-percent level multiplicative shear bias due to biases in a single PSF higher moment (radial kurtosis), while \cite{2022arXiv220507892Z} provided a formalism for measuring PSF higher moments more generally, and studied the shear additive bias and its impact on cosmology analysis based on the HSC Public Data Release 1  \citep{2018PASJ...70S...8A}. \cite{2022arXiv220507892Z} suggested that the higher moments of the PSF can cause additive shear biases on a comparable level to the second moments, thereby motivating null tests involving PSF higher moments, and the development of a PSF systematics forward modeling formalism that considers PSF higher moments for current Stage-III surveys.

In this study, we develop a more rigorous and self-consistent framework for testing and modeling PSF systematics in the cosmic shear analysis. We generalize how PSF moments contaminate weak lensing shears by introducing the concept of a ``spin-2 PSF quantity''. Specifically, we derive the spin-$2$ quantities associated with PSF higher moments, which affect the inferred galaxy shears. We make a star catalog including PSF higher moment measurement, and inspect the overfitting issue of the PSF model. We articulate and carry out a more comprehensive set of null tests by correlating galaxy shapes in the HSC three-year (later referred to as HSC Y3, or Y3) shape catalog \citep{HSC3-catalog} with the PSF second and higher moment spin-2 quantities that impact the galaxy shapes. We compare different models for modeling the additive shear biases associated with the PSF second and higher moments, provide a method to select models based on its complexity and level of impact on the cosmological probe, and propose the best-suited model for the HSC Y3 cosmic shear analysis. More importantly, we provide general guidelines for inspecting and modeling PSF systematics in future cosmic shear analyses. We demonstrate the impact of the new PSF systematics model on cosmological weak lensing analysis by re-analysing the HSC first-year (later referred to as HSC Y1) cosmic shear data and conducting a mock analysis of Y3, comparing models with or without the inclusion of the PSF higher moments. Finally, we investigate several aspects that complicate the PSF systematics model, including the redshift dependency of how PSF systematics affect galaxy shape measurements, a constant systematic shape, impact on $\xi_-$, and second order spin-2 terms. While this paper focuses on the real space analysis of weak lensing, we also provide a PSF formalism for the Fourier space analysis using cosmic shear power spectra, and study the internal consistency between the real and Fourier space formalisms.

The layout of this paper is as follows: we review the background of shear estimation and the associated PSF systematics in Section~\ref{sec:bgd:0}. We describe the HSC Y3 galaxy shape and mock galaxy catalogs, which we use to demonstrate the methods introduced in this work, in Section \ref{sec:shape:0}. We describe the HSC Y3 star catalog, the moment measurements we conducted, and the spin-2 quantities associated with the PSF, an important concept throughout the paper, in Section~\ref{sec:star:0}.  We describe the methodology and results of modeling the PSF higher moments in shear, conducting cross-correlation null tests, and tests for potential redshift dependency of the model in Section~\ref{sec:model:0}. We demonstrate the impact on cosmological parameter analysis due to these PSF systematics by conducting a reanalysis of the  HSC Y1 cosmic shear data vectors and an HSC Y3 mock analysis in Section~\ref{sec:cosmo:0}. Our method is summarized with a concise list of steps in Section~\ref{sec:recipe:0}. In Section~\ref{sec:conclu}, we draw conclusions from the results of this paper and discuss its future implications and applicability to other weak lensing shear surveys.

\section{Background}
\label{sec:bgd:0}

In this section, we briefly review the background to this study. In Section \ref{sec:bgd:cosmic_shear}, we introduce cosmic shear: how it is estimated from the galaxy shapes, and how the likelihood analysis is carried out to extract cosmological information from the data. In Section~\ref{sec:bgd:psf_sys}, we introduce PSF-related systematic effects on weak lensing shear estimation.

\subsection{Cosmic Shear}
\label{sec:bgd:cosmic_shear}

Cosmic shear is a way of measuring cosmological weak lensing, the coherent distortions of large ensembles of background galaxies by the foreground Large Scale Structure (LSS) of the Universe \citep[For a review, see][]{Kilbinger:2014cea}. Since these distortions are induced by all matter along the line of sight, cosmic shear is a powerful probe of the dark matter distribution, which is otherwise challenging to observe. Cosmic shear was first measured in the early 2000s \citep[e.g.][]{2000astro.ph..3338K,2000MNRAS.318..625B,2000Natur.405..143W,2001A&A...374..757V,2004ApJ...605...29R} and consolidated in the late 2000s to early 2010s \citep[e.g.][]{2007ApJS..172..219L,2012MNRAS.427..146H,2014MNRAS.440.1322H}
with larger volumes of survey data, improved redshift estimation \citep[e.g.,][]{2007ApJS..172..219L,2007ApJS..172..239M}
and statistical analysis \citep[e.g.][]{2010A&A...516A..63S}. Multiple ongoing and recently completed surveys have conducted successful cosmic shear analyses \citep[e.g.,][]{2020arXiv200715633A,2019PASJ...71...43H, 2020PASJ...72...16H,2022PhRvD.105b3514A,2022PhRvD.105b3515S}.
Future imaging surveys such as LSST \citep{Ivezic:2008fe}, \textit{Euclid} \citep{Euclid_overview}, and \textit{Nancy Grace Roman} Space Telescope \citep{2019arXiv190205569A}
will provide unprecedented statistical constraining power for cosmic shear observation, making requirements for controlling systematic biases and uncertainties more stringent. The decrease in statistical uncertainties and improvement in control of systematics may provide insights into the apparent $S_8$ tension between the cosmic microwave background and weak lensing \citep{2021APh...13102604D}.

In this section, we briefly describe how weak lensing shear is measured in imaging surveys (Section~\ref{sec:bgd:shear_est}) and is used to constrain cosmological parameters (Section~\ref{sec:bgd:analysis}).

\subsubsection{Shear Estimation}
\label{sec:bgd:shear_est}
%

Galaxy ellipticity is widely used to quantify the spin-$2$ aspect of galaxy
shape and infer the weak lensing shear distortion.  We adopt the `distortion'
definition of ellipticity, \begin{equation}\label{eq:e}
    (e_1,e_2)=\frac{1-(b/a)^2}{1+(b/a)^2} (\cos 2\phi,\sin 2\phi),
\end{equation} where $a$ and $b$ are the major and minor axes and $\phi$ is the position angle
of the galaxy major axis with respect to the $x$-axis of the sky coordinates
taking the flat-sky approximation (with North being $+y$ and East being $+x$).
Here we use this ellipticity definition as an example, but we note that other
spin-$2$ observables (with two components) can also be used for shear
inference, e.g., moments or derivatives of a galaxy's light profile
\citep{Z08,BFD-Bernstein2014}, projections of a galaxy's light profile onto
basis functions \citep{shapeletsII-Refregier2003,FPFS_Li2018} or parameters
used to fit a galaxy's light profile \citep{im3shape,lensfit2}.

For an isotropically-oriented galaxy ensemble distorted by a constant shear,
the shear can be estimated as a weighted average of the distortion of all
galaxies:
\begin{equation}\label{eq:g}
    \hat{g}_\alpha = \frac{1}{2\res} \left\langle e_\alpha \right\rangle,
\end{equation}
where the shear responsivity ($\res$) is the linear response of the average
galaxy ellipticity to a small shear distortion
\citep{KSB-Kaiser1995,Shapes-Bernstein2002}, and $\alpha=1,2$ are the indices
for the two components of the ellipticity. Note that shear
$\hat{g}_\alpha$ in this work is sometimes referred to as the ``reduced
shear'', corresponding to the part of the shear that only changes the galaxy shape rather than the size. Since the galaxy detection and selection are
dependent on the underlying shear distortion, an accurate shear responsivity
should  include the shear response of galaxy detection
\citep{metaDet-Sheldon2020} and galaxy sample selection \citep{KaiserFlow2000}.
In addition, since galaxy images are noisy, noise bias from the nonlinearity in
the ellipticity and responsivity should be estimated and corrected for an
accurate shear estimation \citep{noiseBiasRefregier2012}. These biases can be
corrected empirically by shearing each observed galaxy and adding artificially sheared noise to galaxy images
\citep{metacal-Huff2017,metacal-Sheldon2017,metaDet-Sheldon2020}; analytically
by correcting for the perturbations from shear and noise on the galaxy number
distribution in the space of galaxy properties
\citep{FPFS_Li2018,FPFS_Li2022,FPFS_Li2022b}; or by calibrating the shear
estimates with artificially sheared galaxy image simulations that are
representative of the observed galaxy sample
\citep{HSC1-GREAT3Sim,HSC3-catalog,2022MNRAS.509.3371M}.

Moreover, in order to eliminate shear estimation bias due to PSF smearing,
one can deconvolve the PSF from the galaxy image in Fourier space
\citep{Z11,BFD-Bernstein2014,metacal-Huff2017,FPFS_Li2018}; construct the PSF
correction term based on analytic formalisms that connect this term to second moments of the galaxy and PSF
\citep{KSB-Kaiser1995,Regaussianization,shapeletsII-Refregier2003}; or convolve
models fitted to each galaxy with the PSF
\citep{polar_shapelets,im3shape,lensfit2}. This paper focuses on the PSF-related systematics (including PSF
leakage and PSF modelling error, see Section~\ref{sec:bgd:psf_sys} for more detail) after the PSF correction step in the shear
estimation and the shear calibration with image simulations.

Throughout this work, we will use the terms ``additive bias'' and ``multiplicative bias'' to quantify shear systematics. The observed shear $\hat{g}$ can be generally expressed by
\begin{equation}
\label{eq:m_and_c}
\hat{g} = (1+m)g + c.
\end{equation}
Here $g$ represents the true shear, $m$ is called the multiplicative bias, and $c$ is called the additive bias. Generally, any source of systematics that correlates with the shear or the galaxy shapes would contribute to the multiplicative bias, and systematics that are independent of the galaxy shape would enter as an additive bias.

\subsubsection{Cosmic Shear Analysis}
\label{sec:bgd:analysis}

In this section, we describe how cosmic shear analyses allow one to constrain cosmological parameters starting from a galaxy catalog. The steps include measuring summary statistics of the shear catalog, forward modeling the summary statistics based on cosmology and systematics, and conducting likelihood analysis. The PSF systematics model described in this work is an integrated part of the forward model in the likelihood analysis, therefore impacting the overall results of the cosmological analysis.

A common method to extract summary statistics from a galaxy catalog is to measure the shear-shear two-point correlation function (2PCF) $\xi^{ij}_+$ and $\xi^{ij}_-$ of the galaxy shape \citep[for reference, see][]{Kilbinger:2014cea}, where $i$ and $j$ are the indices of the tomographic redshift bins used for the analysis \citep{1999ApJ...522L..21H}. The shear-shear 2PCF is used in real-space cosmic shear analyses \citep[e.g.,][]{2020PASJ...72...16H, Joudaki:2019pmv, 2022PhRvD.105b3514A}. Other works use Fourier space and measure angular power spectra $C_\ell^{ij}$ as the summary statistics \citep{2019PASJ...71...43H,2022MNRAS.515.1942D,2022A&A...665A..56L}; we discuss the formalism relevant to power spectra in Appendix~\ref{sec:ap:fourier_formalism}.
To increase statistical constraining power, the $\xi^{ij}_\pm (\theta)$ measurements are averaged within angular bins with a range of separations $\theta$ for the galaxy pairs. The angular bins and tomographic bin-pairs form a cosmic shear data vector, which we denote $\ve{D}_{gg}$.

The next stage of the cosmic shear analysis is the likelihood analysis \citep[see, e.g.,][]{2017arXiv170609359K}, where the cosmic shear data vector $\ve{D}_{gg}$ is compared with a theoretical data vector $\ve{T}_{gg}$. $\ve{T}_{gg}$ is computed using a forward model that predicts the data vector based on cosmological parameters and any needed nuisance parameters, including PSF parameters, which will be discussed in Section~\ref{sec:model:formalism}.
The log-likelihood is defined as
\begin{equation}
\label{eq:likelihood}
\log(\mathcal{L}(\ve{\Omega} | \ve{D}_{gg})) =  - \frac{1}{2}(\ve{D}_{gg}-\ve{T}_{gg}(\ve{\Omega}))^T \mt{\Sigma}_{gg}^{-1} (\ve{D}_{gg}-\ve{T}_{gg}(\ve{\Omega})),
\end{equation}
where $\ve{\Omega}$ is a set of parameters, and $\mt{\Sigma}_{gg}^{-1}$ is the inverse of the covariance matrix of $\ve{D}_{gg}$. A sampling algorithm, e.g., \textsc{emcee} \citep{2013PASP..125..306F} or \textsc{MultiNest} \citep{Feroz_2008, 2009MNRAS.398.1601F, Feroz_2019}, traverses the parameter space to approximate the likelihood function across the parameter space, and eventually provides parameter constraints from the data vector. Models for additive shear systematics induced due to the PSF enter in the likelihood analysis; typically these are informed by null tests that reveal what types of systematics may be present, with priors on model parameters determined using those tests.

In Section~\ref{sec:cosmo:0}, we describe the forward cosmological model we used to reanalyze the HSC Y1 cosmic shear data vector and to conduct a HSC Y3 mock analysis. Our new model for additive PSF systematics and their impact on weak lensing shear, which is part of the forward model, is described in Section~\ref{sec:model:formalism}.

\subsection{PSF-related systematics}
\label{sec:bgd:psf_sys}

For a ground-based imaging survey telescope, e.g., the HSC and LSST, the PSF describes the smearing of the image  by the turbulent atmosphere and the telescope optics. In this paper, the PSF also involves the pixelization effect in the CCD, described as a convolution by a \responsemnras{unit-square} function; the PSF including this effect is referred to as the ``effective'' PSF.
The single exposures within the survey footprint are combined to produce a ``coadded'' image, as are the PSF models \citep{2018PASJ...70S...5B}. For the HSC survey, shear estimation is carried out on the coadded image, making the coaddition procedure and production of the coadded PSF a crucial step that can affect shear estimation \citep{2022arXiv220909253M}.

After convolution with the PSF, the observed size and shape of the galaxy differ from the true values, in a way that depends on the galaxy and PSF properties \citep{2008A&A...484...67P}. Since almost all weak lensing science heavily relies on the precise measurement of the galaxy shape or some other spin-2 quantity based on galaxy second moments \citep{Mandelbaum:2017jpr}, it is crucial to precisely model the PSF and correct for its impacts on the galaxy shear estimate during the shape measurement phase of the image processing. Imperfections in PSF modeling cause a PSF modeling bias in shear \citep[see Appendix~A in][]{2008A&A...484...67P}, even for principled shear estimation methods that should be unbiased.
If the shear estimation method is imperfect, it causes PSF leakage bias in shear \citep{2018PASJ...70S..25M}. 
Typically, it is prudent to test for both leakage and modeling error, even when using a shear estimation method that should not have any leakage.

Weak lensing shear systematics related to the PSF second moments are well-studied in previous work. The PSF leakage of the second moment-based \reGauss{} method is characterized in  \citet{Regaussianization,SDSS-shape-Mandelbaum2005}. In  \cite{2008A&A...484...67P}, expressions for the additive shear systematics due to PSF modeling errors  are derived under the assumption that both the galaxy and PSF profiles are Gaussian. In \citet{2010MNRAS.404..350R,2016MNRAS.460.2245J}, the propagation of the previously mentioned leakage and modeling error to the cosmic shear 2PCF is quantified using the ``$\rho$ statistics ''.

Recent cosmic shear analyses made different choices for how to model these additive shear systematics due to the PSF. In \cite{2020PASJ...72...16H}, the additive bias on $\xi_+$ in HSC Y1 due to PSF second moment leakage and modeling error were modeled and marginalized with two parameters $\alpha_{\rm PSF}$ and $\beta_{\rm PSF}$ (see Eq.~\ref{eq:psf-sys-second}), which are taken to be the same across all tomographic bins. The additive bias on $\xi_-$, and any mean shear, are neglected after confirming that the mean shear in each survey region was consistent with zero within the uncertainty due to cosmic variance. In \cite{2022PhRvD.105b3514A}, the PSF second moment leakage, shape modeling error, and size modeling error were investigated.  However, the additive bias on shear was not included in the fiducial analysis because the $\rho$ statistics  were within the survey requirement \citep{2021MNRAS.504.4312G}.
In Section~3 of \cite{2021A&A...645A.105G}, the PSF contamination on $\xi_+$ was modeled by a leakage term, a modeling error term, and a constant term.

In \cite{2022MNRAS.510.1978Z}, multiplicative bias induced by the modeling error of the PSF fourth radial moments, or kurtosis, were found using image simulations and the HSC Y1 dataset. The multiplicative bias predicted for the cosmic shear due to this effect is on the order of $10^{-3}$, which is sub-dominant to other sources of multiplicative bias (e.g., the PSF size modeling error). For this reason, directly modeling it in the HSC Y3 analysis is not urgent. In \cite{2022arXiv220507892Z}, additive biases from the PSF higher moments modeling error were found to be of a similar magnitude to the second moment additive biases in the HSC Y1 data. Therefore, testing for and modeling additive biases due to PSF higher moment modeling errors in cosmic shear analyses is necessary. These two previous works are the main motivation for developing a framework to self-consistently identify and model PSF additive systematics to higher order than those caused by second moments.  Crucially, our method includes a step to limit the model to only those terms that turn out to be present at a significant level in a given dataset, enabling different data-motivated choices of what terms to model in different datasets.

\begin{figure*}
    \centering
    \includegraphics[width=1.9\columnwidth]{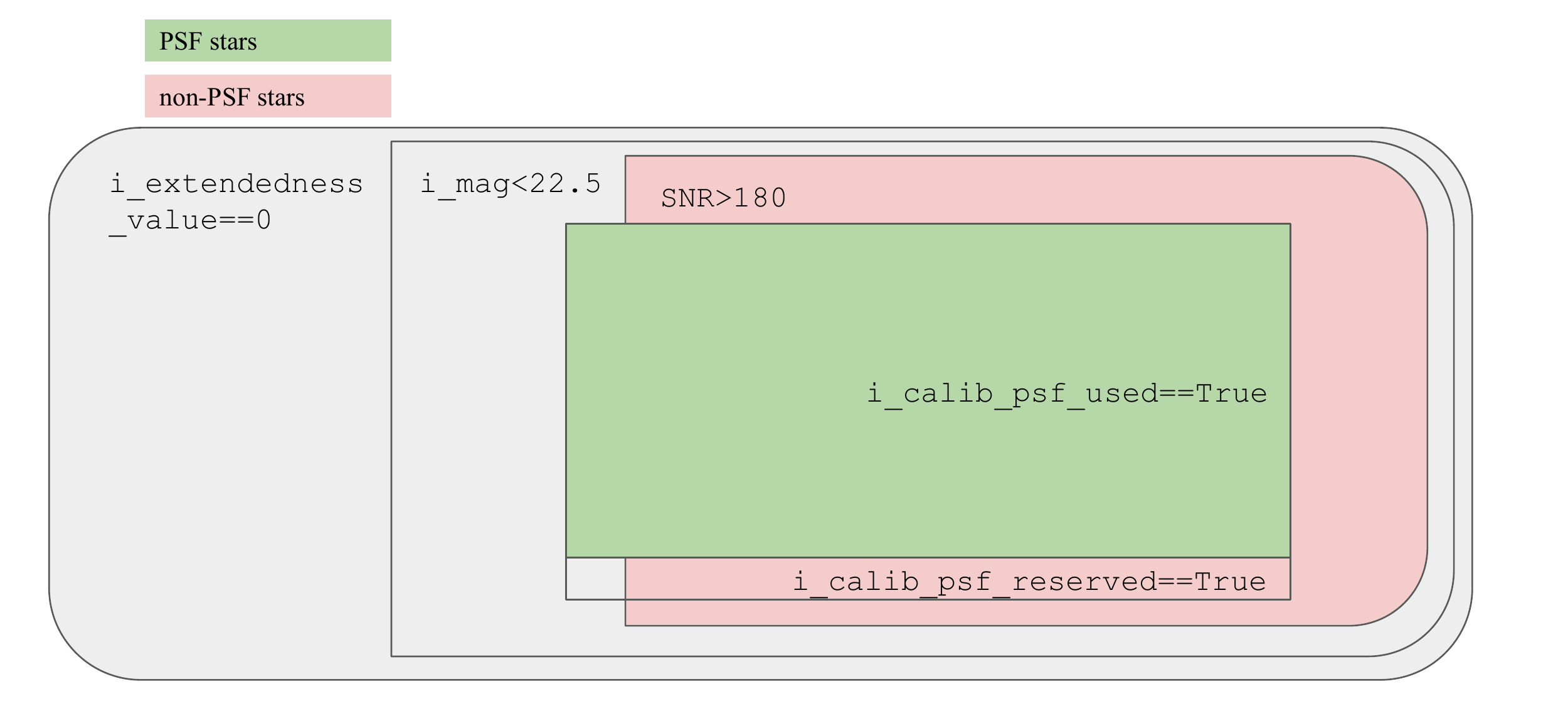}
    \caption{\responsemnras{Schematic diagram describing the selection of the PSF and non-PSF star catalogs in this work. The selection on the $i$-band extendedness, magnitude and signal-to-noise ratio are done at the coadd level, while the selections of PSF used stars and reserved stars are done on the single visit level. The green region is the PSF star catalog, while the pink region is the non-PSF star catalog. We can see that the PSF and non-PSF stars are selected using different criteria, resulting in different results for the PSF systematics parameters (see Section~\ref{sec:model:psf_vs_nonpsf}). However, the impacts on cosmology are similar, as shown in Section~\ref{sec:model:cosmology}. Note that a box within another box does not imply that one is a subset of the other; instead, it indicates a sequence of selections we imposed on our samples.}  }
    \label{fig:star_selection}
\end{figure*}

\section{HSC Shape Catalog}
\label{sec:shape:0}

In this section, we describe the galaxy catalog we used to explore PSF systematics modeling in HSC Y3 cosmic shear analysis. In Section~\ref{sec:shape:shape}, we describe the HSC Y3 shape catalog \citep{HSC3-catalog}. In Section~\ref{sec:shape:mock}, we describe the mock catalogs we used for uncertainty estimation.

\subsection{HSC Y3 Shape Catalog}
\label{sec:shape:shape}

In this section, we summarize the HSC three-year (Y3) \citep{HSC3-catalog} shear
catalog for weak lensing science. In the HSC shear catalog, galaxy
ellipticities are estimated from $i$-band coadded images with the
re-Gaussianization ($\reGauss$) shear estimator and PSF correction method
\citep{Regaussianization}, which is implemented in $\galsim$ \citep{Galsim}, an
open-source package for image simulation and image processing. \reGauss{} has
been developed and used extensively on data from the Sloan Digital Sky Survey
\citep[SDSS;][]{SDSS-shape-Mandelbaum2005,SDSS-gglensDR7-Mandelbaum2013} and
the first HSC shape catalog \citep{2018PASJ...70S..25M}. 
The $\reGauss$ estimator measures the two ellipticity
components for each galaxy using its spin-$2$ elements in the second-order
moment matrix.

\reGauss{} also computes the resolution factor, $R_2$\, which is used to
quantify the extent to which the galaxy is resolved compared to the PSF. The
resolution factor is defined for each galaxy using the trace of the second
moments of the PSF ($T_{{\rm PSF}}$) and those of the observed galaxy image
($T_{{\rm gal}}$):
\begin{equation}
    R_2=1-\frac{T_{{\rm PSF}}}{T_{{\rm gal}}}\,.
\end{equation}

The inverse variance weights to be used while performing the ensemble average
are the galaxy shape weights ($w_i$) defined as
\begin{equation}\label{eq:optimalW}
    w_i=\frac{1}{\sigma_{e;i}^2+e_{{\rm RMS};i}^2},
\end{equation}
where $i$ is an index over galaxies, $\sigma_e$ is the per-component $1\sigma$
uncertainty of the shape estimation error due to image noise, and
$e_{\rm{RMS}}$ denotes the per-component root-mean-square ($\texttt{RMS}$) of
the galaxy intrinsic ellipticity\footnote{While the RMS ellipticity is
ostensibly associated with the entire sample, it does depend on the particular
subpopulation within the catalog.  To enable division of the catalog into
subsamples (e.g., for tomographic analysis), information is provided on this
variation to enable a correct estimate of the RMS ellipticity for the selected
subsample.} (often referred to as `shape noise'). The parameters $e_{\rm{RMS}}$
and $\sigma_e$ are modeled and estimated for each galaxy using image
simulations \citep{HSC1-GREAT3Sim,HSC3-catalog}. The shear responsivity for the
source galaxy population is estimated as
\begin{equation}\label{eq:response}
    \res=1-\frac{\sum_i w_i e^2_{{\rm RMS};i}}{\sum_i w_i}\,.
\end{equation}

The measured shears are calibrated with realistic image simulations downgrading
the galaxy images from COSMOS Hubble Space Telescope
\citep{2007ApJS..172..219L} to the HSC observing conditions
\citep{HSC1-GREAT3Sim}.
The calibration removes the galaxy property-dependent (galaxy resolution,
galaxy SNR, and galaxy redshift) estimation bias and the detection and
selection bias due to the correlation between detection/selection and the
underlying shear distortion.
The image simulation used for calibration includes the blending of light from
neighboring galaxies; therefore, the calibration removes biases related to
blending.
The resulting systematic uncertainties in the shear estimation are below $1\%$
after the calibration \citep{HSC3-catalog}.

With conservative selection cuts on each galaxy's $i$-band magnitude (brighter
than $24.5$) and resolution (greater than $0.3$), the full galaxy shear catalog
has a raw (effective\footnote{See \citet{WLsurvey_neffective_Chang2013} for the
definition of effective number density.}) number density of
$23~\mathrm{arcmin}^{-2}$ ($20~\mathrm{arcmin}^{-2}$) covering
$417~\mathrm{deg}^2$\,, after removing a $20~\mathrm{deg}^2$ region that failed
the cosmic shear B-mode test (more information found in
Appendix~\ref{sec:ap:cutaway}).
The full galaxy catalog is divided into $4$ tomographic bins by selecting
galaxies within redshift intervals of $(0.3, 0.6]$, $(0.6, 0.9]$, $(0.9, 1.2]$
and $(1.2, 1.5]$ using the best point estimate \citep{2018PASJ...70S...9T} of
the Deep Neural Net Photometric Redshift (\dNNz{}; Nishizawa et. al {\it in
prep.}) conditional density estimates of individual galaxy redshift posteriors,
where \dNNz{} is a template based inference method.
We found that some \mizuki{} \citep{2018PASJ...70S...9T} and \dNNz{} photometric redshift posteriors have
a secondary peak at $z\gtrsim3.0$\,. These photometric redshift posteriors are
difficult to calibrate using spatial cross-correlations, since the secondary
peak lies outside the redshift coverage of the CAMIRA sample (Cluster finding
algorithm based on Multi-band Identification of Red-sequence gAlaxies;
\citealt{CAMIRA_HSC_Oguri2018}), which we use as a reference sample and which
is limited to $z<1.2$\,. In order to prevent the secondary solution from
biasing the sample redshift distribution inference, we remove galaxies with
double solutions in the estimated photo-$z$ posteriors (see Rau et. al {\it in
prep.} for details).
The cuts that are used to remove the galaxies with secondary peaks reduce the
number of galaxies in the first (second) bin by $30\%$ ($8\%$). After the
region cut and double solution cut, we have 5,889,826, 8,445,233, 7,023,314,
and 3,902,504 galaxies in the corresponding four redshift bins, respectively.
The corresponding raw (effective) galaxy number densities are $3.92~(3.77)$,
$5.63~(5.07)$, $4.68~(4.00)$ and $2.60~(2.12)~\mathrm{arcmin}^{-2}$,
respectively.


\subsection{HSC Mock Catalogs}
\label{sec:shape:mock}
We use the  HSC three-year mock shear catalog to accurately quantify the
uncertainties of our measured 2PCFs  due to cosmic variance, galaxy shape
noise, measurement errors due to photon noise, and photometric redshift
uncertainties. The mock catalogs are created following
\citet{HSC1-mock-Shirasaki2019}, but with updates to incorporate the
survey footprint, galaxy shape noise and shape measurement error of the
three-year HSC shear catalog.

The mock shear catalog uses full-sky lensing simulations generated by
\citet{raytracingTakahashi2017} with $108$ full-sky simulations. To
increase the number of total realizations of the mock catalogs, we extract $13$
separate regions from each  full-sky simulation,  obtaining $108
\times 13 = 1404$ mock catalogs in total.

These realizations of the lensing simulations are combined with the observed
photometric redshifts, angular positions, and shapes  of real galaxies
\citep{HSC3-catalog} to generate mock shear catalogs. To be more specific,
source galaxies are populated on the light-cone of the lensing simulations
using the original angular positions and ``best-fit" redshifts of the galaxies
(estimated with \dNNz{}) in the HSC three-year shear catalog. Each galaxy is
assigned a source redshift estimate in the mock following the posterior
distribution of photometric redshift estimated by the \dNNz{} algorithm. The
shape noise on each galaxy is generated with a random rotation of the galaxy's
intrinsic shape according to the intrinsic shape dispersion estimated in the
HSC shear catalog, and the measurement error is generated as a zero-mean
Gaussian random number with the standard deviation measured in the HSC shear
catalog \citep[see Section~4.2 in][]{HSC1-mock-Shirasaki2019}.

\begin{figure}
    \centering
    \includegraphics[width=1.0\columnwidth]{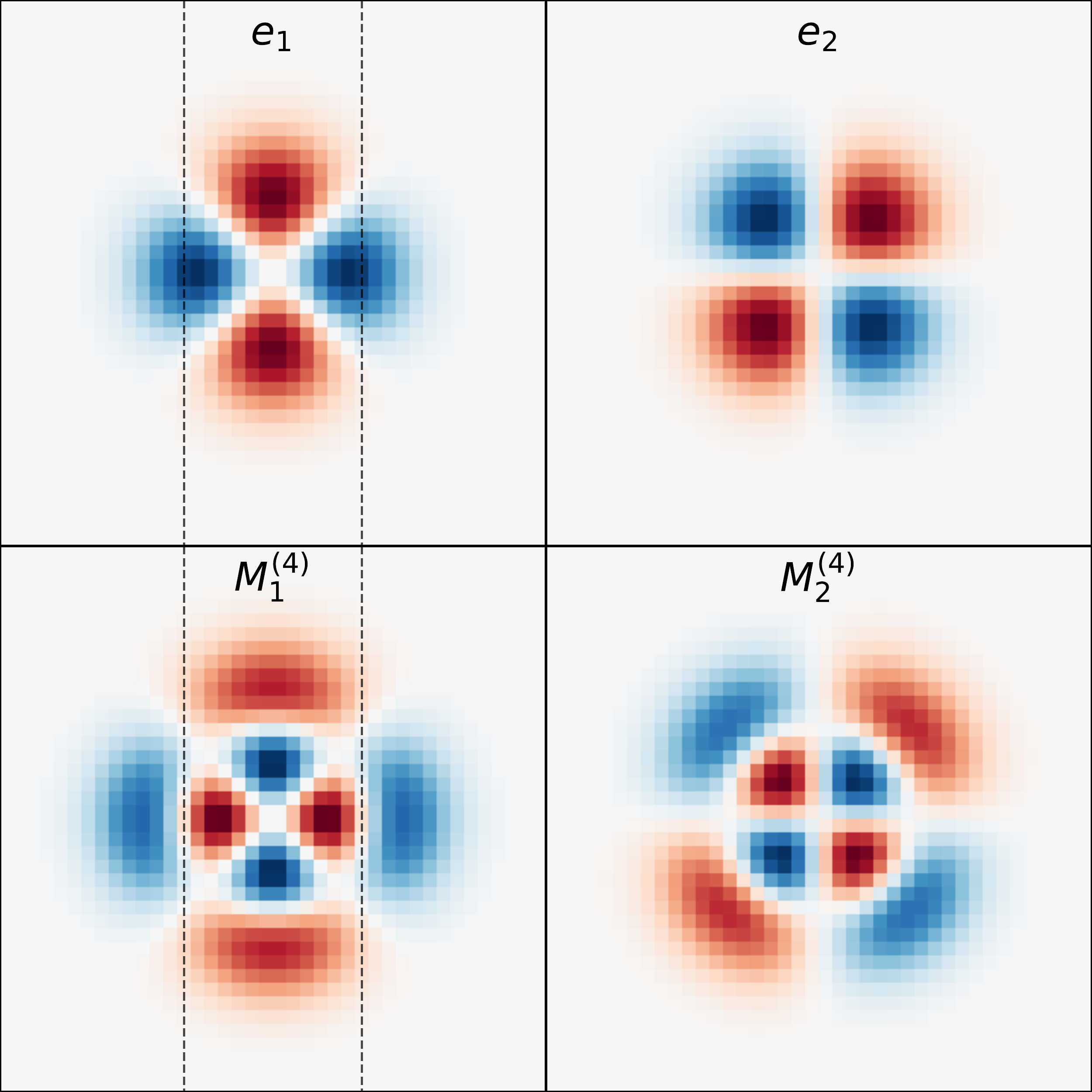}
    \caption{The image response to the spin-$2$ quantities of the second moments $e_1$ and $e_2$, and fourth moments $M^{\rm (4)}_1$ and $M^{\rm (4)}_2$. The fourth moment spin-$2$ quantities are sensitive to scales larger and smaller than those to which the second moment spin-2 quantities are sensitive, as the dashed reference lines show. The color scale for each base covers $[-A,A]$, where $A$ is the maximum of the absolute value of the basis function.
    }
    \label{fig:image_response}
\end{figure}

\begin{figure*}
    \centering
    \includegraphics[width=1.3\columnwidth]{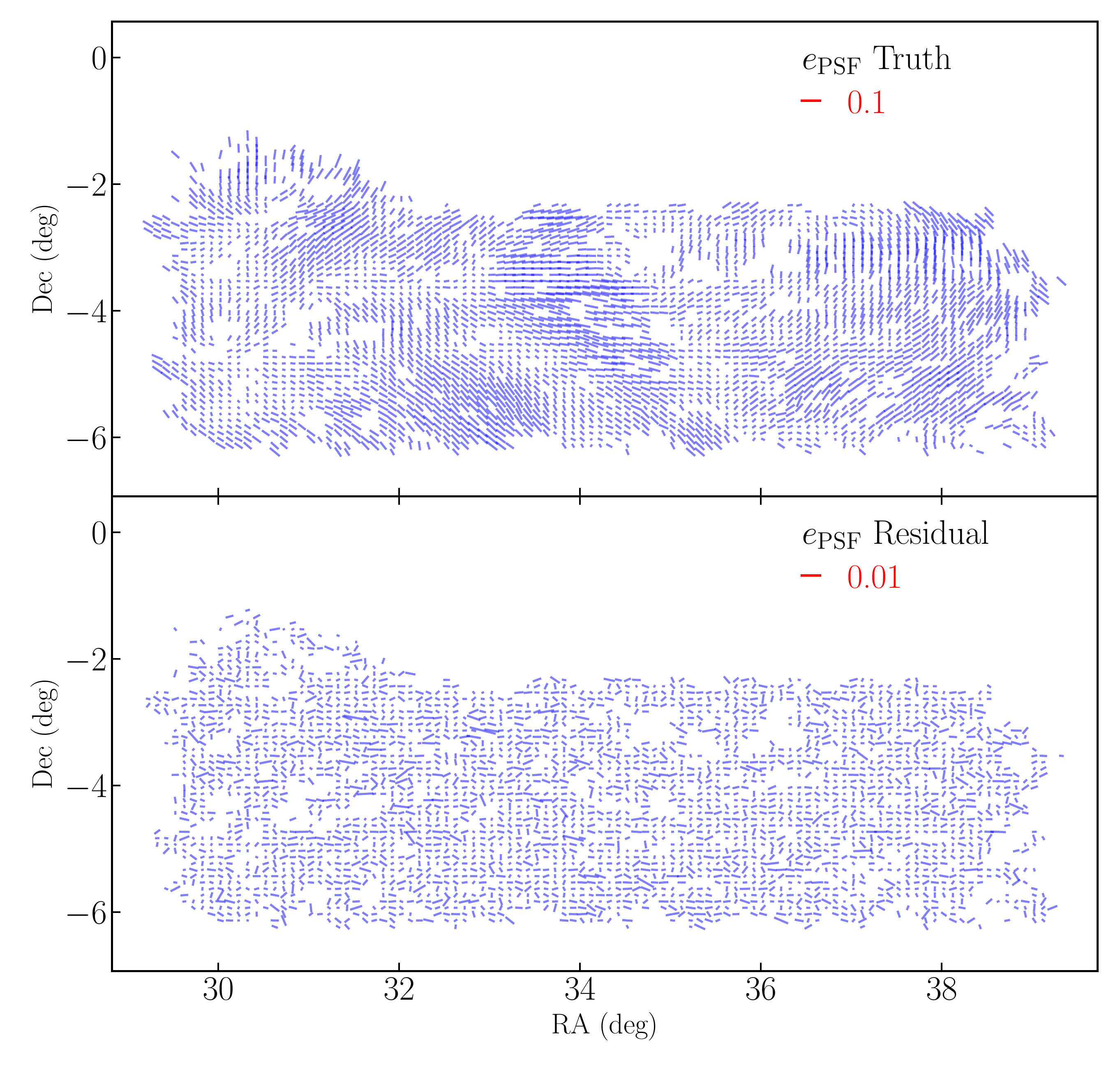}
    \includegraphics[width=1.3\columnwidth]{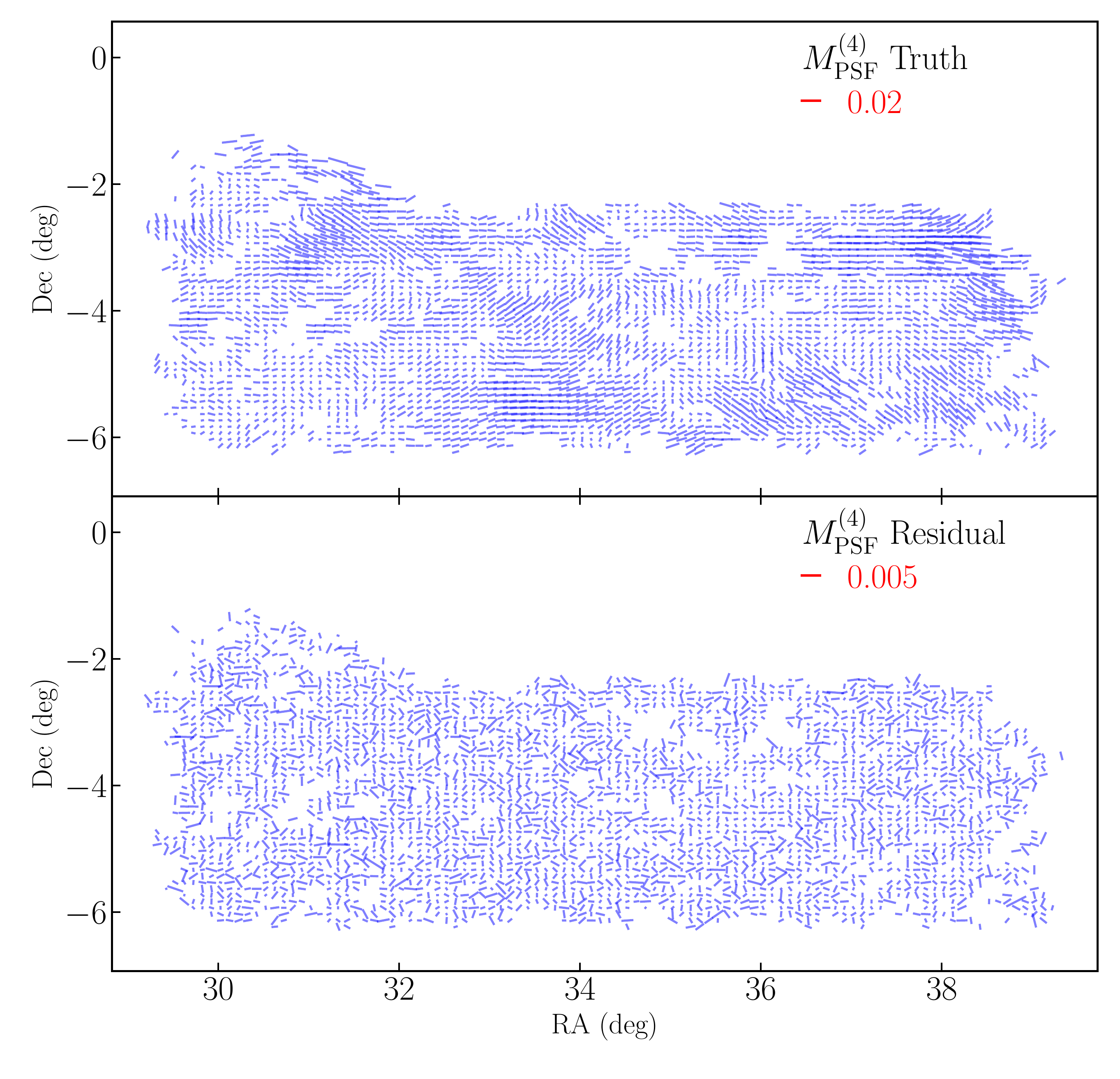}
    \caption{The whisker plots of the true and residual spin-2 components of the PSF second (top) and fourth moments (bottom) in the \code{XMM} field. There is an obviously coherent pattern in the  whisker plots for the true moments, while the pattern is less visible in the moment residuals (later, we will see that the correlation length of the residual field is smaller, which makes the coherence less visible in the whisker plots). The spin-2 pattern of the true fourth moments is clearly different from that of the second moments, which indicates that contamination in the PSF higher moments must be separately modeled in cosmic shear, as we explore in Section~\ref{sec:model:0}.
    }
    \label{fig:whisker_plots}
\end{figure*}

\section{Star Catalogs and Moments Measurements}
\label{sec:star:0}

In this section, we introduce the HSC Y3 star catalogs with the measurements of higher moments. In Section~\ref{sec:star:selection}, we describe how the star sample is selected. In Section~\ref{sec:star:moments}, we summarize the measurement of the PSF second and higher moments.
In Section~\ref{sec:star:spin2}, we introduce a key concept in the paper: how to identify the PSF higher moments that form ``spin-$2$'' quantities, which will be included in the PSF systematics formalism in Section~\ref{sec:model:formalism}.

\subsection{Sample Selections}
\label{sec:star:selection}

\responsemnras{In this section, we describe the star catalogs used in this work. The overall selection processes are shown in Figure~\ref{fig:star_selection}, in which the PSF and non-PSF star catalogs are marked in green and pink, respectively. We describe each of these selections in this subsection.}

The HSC Y3 star catalog used in this work is selected from the sample of point sources based on the star samples in Section~5.1 of \cite{HSC3-catalog} covering the same footprint as the galaxy shape catalog described in Section~\ref{sec:shape:shape}. Unlike \citet{2022arXiv220507892Z}, we measure the star moments on postage stamp images after deblending, as described in \cite{2018PASJ...70S...5B}. Therefore, we did not apply any selection criteria to omit stars based on their blendedness.

The HSC data processing and PSF modeling is carried out independently in multiple exposures. Each exposure has a different random subset of the stars used for PSF model estimation. The exposures are then stacked to make the coadded image \citep[see][]{2018PASJ...70S...5B}.  \responsemnrassecond{In each single exposure, a set of candidate PSF stars is selected using the $k$-means clustering algorithm in the magnitude-size plane. A randomly-selected $80\%$ of the candidate stars is used for PSF modeling, with the random selection carried out independently for each exposure. On the coadd level, the stars with \code{i\_calib\_psf\_used==True} are those that were used as PSF stars in more than $20\%$ of the contributing $i$-band exposures, while those labelled as \code{i\_calib\_psf\_reserved==True} were used as PSF stars in fewer than $20\%$ of the contributing exposures. Because the random selection of stars for PSF modeling in single exposures is carried out independently, the \code{i\_calib\_psf\_reserved==True} stars in the coadded image are very rare.} 

To mitigate the scarcity of the PSF reserved stars, we use a more lenient \code{i\_extendedness==False} $\&$ \code{i\_mag<22.5} cut on the coadd catalog to pre-select a star catalog. The \code{i\_extendedness} flag is a star-galaxy selection procedure only based on the model magnitude and PSF magnitude. Within that catalog, those with \code{i\_calib\_psf\_used=True} are defined to be the ``PSF star'' catalog. Stars with \code{i\_calib\_psf\_used=False} are candidates for the ``non-PSF stars''.    In addition, we find that the low SNR non-PSF stars have larger sizes and lower ellipticities to a statistically significant degree, making them unrepresentative samples of the true PSF. This could potentially be caused by the increasing fraction of galaxy contamination at low SNR, which only affects non-PSF stars because the PSF stars have a preliminary SNR cut in the image processing pipeline.
Therefore, we applied an empirical SNR cut to the non-PSF sample, requiring flux SNR$>180$, so that the non-PSF star sample has a nearly identical size distribution as the PSF star sample. \responsemnras{Although the PSF star catalog also has low SNR samples, those samples have a similar size and ellipticity distribution to the rest, due to the strict star-galaxy separation done on the single visits for the PSF stars.} 
This eliminated $23\%$ of the potential non-PSF star sample. After these selections, of the coadd star samples, about $6\%$ of the stars are ``non-PSF'' stars. 

We also removed an area of $20$~deg$^2$ (at ${\rm RA} \in [132.5,140]$, ${\rm Dec} \in [1.6,5]$) in the \code{GAMA09H} field that generated a strong B-mode shear signal, as will be described in Li et al. {\it in prep}. We explored this region of the sky and found a significant PSF fourth moment modeling error, described in Appendix~\ref{sec:ap:cutaway}.

After the previously mentioned cuts, there are 2,118,183 PSF stars and 132,687 non-PSF stars in our sample, where the former has an average density of $1.42$ arcmin$^{-2}$ and the latter, $0.09$ arcmin$^{-2}$.

\subsection{Second and Higher Moments}
\label{sec:star:moments}

In this section, we briefly review the measurement of the PSF second and higher moments from a pixelized i-band postage-stamp image. This formalism follows the one in \cite{2022arXiv220507892Z}.
We define the adaptive second moment matrix $\mt{M}$ of a light profile $I(x,y)$ in the image coordinate system with origin at the centroid of $I(x,y)$ as
\begin{equation}
\label{eq:second_moment measurement}
    M_{pq} = \frac{\int \mathrm{d}x \mathrm{d}y \, x^p y^q \, \omega(x,y)
    I(x,y)}{\int \mathrm{d}x \mathrm{d}y \,  \omega(x,y) I(x,y)}.
\end{equation}
Here ($p$, $q$) take the values of $(2,0), (1,1), \text{and} (0, 2)$, and $\omega(x,y)$
is the adaptive Gaussian weight \citep{Regaussianization}, defined as
\begin{equation}
\label{eq:weight_function}
    \omega(x,y) = \exp \left( - \frac{1}{2} \begin{bmatrix} x & y\end{bmatrix}
    \begin{bmatrix} M_{20} & M_{11} \\  M_{11} & M_{02}\end{bmatrix}^{-1}
    \begin{bmatrix} x \\ y\end{bmatrix}
    \right).
\end{equation}
The second moment trace $T_{\rm PSF}$ and shape $e_{\rm PSF} = e_{\rm PSF,1} + i e_{\rm PSF,2}$ are defined based on
$\mt{M}$\,
\begin{align}
\label{eq:def_trace}
    T_{\rm PSF} &= M_{20}+M_{02}\\\label{eq:def_e1}
    e_{\rm PSF,1} &= \frac{M_{20} - M_{02}}{M_{20}+ M_{02}}\\\label{eq:def_e2}
    e_{\rm PSF,2} &= \frac{2 M_{11}}{M_{20} + M_{02}}.
\end{align}
Notice that there is a different definition for the PSF second moment size $\sigma_{\rm PSF}$, which approximates the standard deviation of the Gaussian that best fits the PSF profile. It is defined by
\begin{equation}
\label{eq:def_sigma}
\sigma_{\rm PSF} = \left[\text{det}(\mt{M})\right]^{\frac{1}{4}}.
\end{equation}

{A natural way to define the higher moments is to integrate over $x^py^q$, as in Eq.~\eqref{eq:second_moment measurement}. However, the resulting higher moments will depend on the size and shape of the PSF. There are two approaches to disentangling the higher moments from the second moments: one is through a combination of the higher and second raw moments as defined above; the other is to define the higher moments in a transformed coordinate that normalizes second moments. In this work, we discuss both approaches, although the second approach is used in most parts of this work.  To connect the two approaches, we describe the formalism of raw and standardized higher moments in Appendix \ref{sec:ap:alt_def:0}, including the \responsemnrassecond{second moments and higher moments parts} of the raw moments. We also demonstrate empirically for the HSC survey data that you can use the raw moments to track PSF additive bias in shear-shear 2PCF, and get consistent results from the results using standardized moments, in Section~\ref{sec:ap:alt_def:cosmology}.}

\responsemnras{In the ``normalizing approach'', the higher moments are defined by} integrating over the image using $u^p\, v^q$, where $(u,v)$ is a standardized coordinate defined by
\begin{equation}
    \begin{bmatrix} u \\ v\end{bmatrix} =
    \mt{M}^{-\frac{1}{2}}
    \begin{bmatrix} x \\ y\end{bmatrix}.
\end{equation}
In the $(u,v)$ coordinate system, the second moment shapes of $I(x,y)$ are $e'_1 = e'_2 = 0$, and the second moment size $\sigma' = 1$.
The higher moments, defined using the standardized coordinates, are
\begin{equation}
\label{eq:moment_define}
    M_{pq} = \frac{\int \mathrm{d}x \, \mathrm{d}y \, u^p \, v^q \, \omega(x,y)
    \, I(x,y)}{\int \mathrm{d}x \, \mathrm{d}y \, \omega(x,y) \, I(x,y) }.
\end{equation}
Note that here integrating in $\mathrm{d}x \, \mathrm{d}y$ is the same as in $\mathrm{d}u \, \mathrm{d}v$ since the $\mt{M}^{-1}$ factor cancels out between the denominator and the numerator. \responsemnras{The connection between the standardized and raw higher moments is described in Appendix~\ref{sec:ap:alt_def:connection}.}

For the $n$th moments, $p$ takes integer values from 0 to $n$, while $q = n - p$. Therefore, there are $n+1$ $n$th moments. The standardized higher moments are not sensitive to any of the lower moments from $n=0$ to 2 (the flux, centroid, size, or shape). They describe the non-Gaussian morphology of the PSF profile.

The HSC Y3 star catalog in this work contains second to sixth moments, 25 in total, measured using the $i$-band deblended coadded images of the stars and PSF model.
The PSF model is a modified version of $\textsc{PSFEx}$ \citep{2011ASPC..442..435B}, initially described in \cite{2018PASJ...70S...5B} and later updated in \cite{2022PASJ...74..247A}.
We measure the moments of the PSF model images evaluated at the star positions as the model moments.

The residual for the moment $M_{pq}$ is defined as
\begin{equation}
    \label{eq:moment_residual}
    \Delta M_{pq} = M_{pq,\text{model}} - M_{pq,*},
\end{equation}
where $M_{pq,\text{model}}$ is the moment of the PSF model, and $M_{pq,*}$ is the true moment measured on the star image.

\begin{figure*}
    \centering
    \includegraphics[width=1.8\columnwidth]{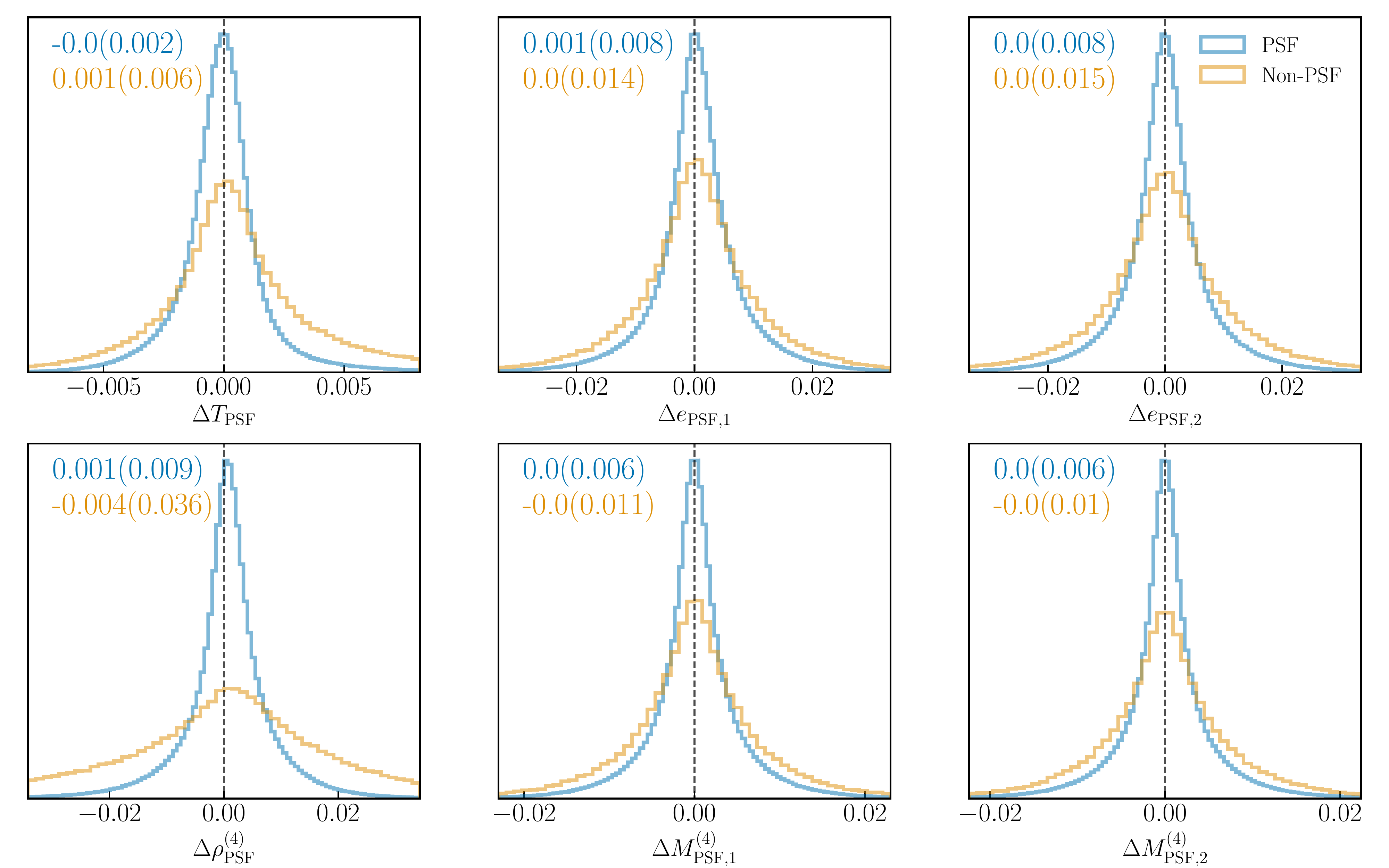}
    \caption{The modeling errors, defined in Eq.~\eqref{eq:moment_residual}, in the spin-2 and spin-0 components of the PSF second and fourth moments, as defined in Section~\ref{sec:star:spin2}, are shown in the top and bottom rows, respectively. The results using the PSF (non-PSF) stars are in  blue (orange).
    Text labels on each panel show the mean and standard deviation of the distributions with matching color.
    The PSF stars have a narrower residual distribution than the non-PSF stars, especially in the spin-0 components. We concluded that this is caused by PSF model overfitting, as described in Section~\ref{sec:star:spin2}.
    }
    \label{fig:moment_dist}
\end{figure*}

\subsection{Spin-2 and Spin-0 PSF Moment Combinations}
\label{sec:star:spin2}

The image simulations in \cite{2022arXiv220507892Z} provide early evidence that (a) PSF even moments are far more important for weak lensing shear than PSF odd moments, and (b) the shear response to PSF higher moments exhibits symmetries in moment indices. In this section, we show that one linear combination of all higher moments at a given even order  
represents the ``spin-2'' contribution of those moments to the first order; and we prove that only the even moments can combine to form spin-$2$ moments in Appendix~\ref{sec:ap:spin2}.  This demonstration is important, because spin-2 PSF quantities are assumed to be the main contributor to the additive shear systematics (given that shear is spin-2). Additionally, the product of a spin-$0$ and spin-$2$ quantity is also a spin-2 property \citep{2016MNRAS.460.2245J}. Therefore, it is also relevant for us to define the spin-0 combinations of the second and higher moments of the PSF.

A spin-$2$ complex quantity, e.g., weak lensing shear, negates when coordinates are rotated by $\pi/2$ (see, e.g., Appendix~A of \citealt{FPFS_Li2022b}). We are interested in the spin-$2$ components of the PSF's fourth moments. Therefore, we find the spin-$2$ component of the $4$th order complex polynomials in polar coordinates $(r,\phi)$:
\begin{align}
\label{eq:spin-2-polynomial}
\nonumber r^4 e^{2i \phi} =& \, r^4 \left[ \cos(2\phi) + i \sin(2\phi) \right]\\\nonumber
    =& \,r^4 \left[ \cos^4(\phi) - \sin^4(\phi) \right] \\\nonumber
    &+i r^4 \left[ 2\sin(\phi)\cos^3(\phi) + 2\sin^2(\phi)\cos(\phi) \right] \\
    =& \, (x^4-y^4) + i (2x^3y + 2xy^3).
\end{align}
The first parenthetical  polynomial leads to the  combination of two moments,
$M_{40} - M_{04}$, while the second parenthetical polynomial leads to an imaginary combination of two moments, $2M_{13} + 2M_{31}$\,. We therefore define the spin-2 combination
of the PSF $4$th moments as
\begin{align}
\label{eq:spin-2-fourth}
M^{\rm (4)}_{\rm PSF} = (M_{40} - M_{04}) + i (2M_{13} + 2M_{31}).
\end{align}
In support of this definition, Figure~6 of \cite{2022arXiv220507892Z} provides numerical evidence that
$M_{40}$ and $M_{04}$ are the only fourth moments that impact $g_1$, and
$M_{31}$ and $M_{13}$ are the only ones that impact $g_2$\,. In Fig.~\ref{fig:image_response}, we show the image responses to the spin-2 quantities of second and fourth moments. The image responses show the variation of a Gaussian PSF when only a specific spin-2 quantity is changed, while other moments remain constant. It is computed by \textsc{PSFHOME}\footnote{\url{https://github.com/LSSTDESC/PSFHOME}} \citep{2022arXiv220507892Z}. Fig.~\ref{fig:image_response} shows that the $M^{\rm (4)}_{\rm PSF}$ values are sensitive to pixels with radius both larger and smaller than the pixels to which the $e_{\rm PSF}$ values are sensitive. The sensitivities of $M^{\rm (4)}_{\rm PSF}$ to smaller and larger radii with the same polar angle have opposite signs, which means $M^{\rm (4)}_{\rm PSF}$ is sensitive to the difference in spin-2 between pixels with small and large radii.

As shown in Appendix~\ref{sec:ap:spin2}, there will in general be a spin-2 combination of even moments at any order. For example, for the $6$th moments, we can expand $r^6 e^{2i\phi}$ as in  Eq.~\eqref{eq:spin-2-polynomial} to define the spin-$2$ combination of the PSF $6$th moments as $M^{\rm (6)} = (M_{60} + M_{42} -M_{24} -  M_{06}) + i (2M_{51} + 4M_{33} + 2M_{15})$\,.  In Appendix~\ref{sec:ap:sot}, we demonstrate that sixth moments do not need to be modeled in the PSF systematics for the HSC analysis in practice, since they are noise dominated and highly correlated with fourth moments.

In Fig.~\ref{fig:whisker_plots}, we visualize  the spin-$2$ combination of PSF second moments, i.e., the shape (upper panel), and of the PSF fourth moments, i.e., $M^{\rm (4)}$ (lower panel), in one of the six HSC fields. In both cases, we show the true moments measured using star images and their residuals defined in Eq.~\eqref{eq:moment_residual}. We can see distinctive patterns in the true second and fourth moment distributions, which suggest that they must both be modeled in the weak lensing shear analysis. The pattern in the residuals is less visible, mainly because they are coherent on a smaller angular scale than the resolution of these whisker plots, as we will see later in Section~\ref{sec:model:null} through the two-point correlation functions.

As stated previously, the product of a spin-0 and spin-2 quantity is also spin-2. Therefore, it is relevant for us to define the spin-0 quantities of the PSF moments. We can find the spin-0 components of the second and fourth moments by doing a similar exercise for $r^2$ and $r^4$ instead of $r^4 e^{2i\phi}$ as in Eq.~\eqref{eq:spin-2-polynomial}. For the second moments, that process yields the trace of the second moment matrix $\mt{M}$,
\begin{equation}
\label{eq:define_trace}
T_{\rm PSF} = M_{20} + M_{02}.
\end{equation}
For the fourth moments, it yields the radial kurtosis,
\begin{equation}
\label{eq:define_kurtosis}
\rho^{\rm (4)}_{\rm PSF} = M_{40}+2M_{22}+M_{04}.
\end{equation}
In Fig.~\ref{fig:moment_dist}, we show the residual distributions of the spin-2 and spin-0 moment combinations (for the second and fourth moments) for the PSF and non-PSF stars in the Y3 star catalog. We see that the non-PSF stars have a wider spread in all moments\responsemnras{, which can be caused by either overfitting of the PSF model, or different SNR distributions of the PSF and non-PSF stars}. To \responsemnras{rule out the SNR explanation}, we inspected the moment residuals for SNR in the range $[300, 500]$, where the PSF and non-PSF stars have very similar SNR distributions, and also found a similarly wider spread for the non-PSF stars \responsemnras{compared to the results shown in Fig.~\ref{fig:moment_dist}}. Therefore, we conclude that the PSF model is overfitting the PSF, which means that the PSF stars have underestimated PSF model residuals compared to other locations (such as those where we expect to find galaxies).

\responsemnras{The noise in the image can cause noise bias in the PSF higher moments for low signal-to-noise samples. We conducted a simple numerical test with an HSC-like PSF profile to ensure that the multiplicative noise bias to our higher moments measurement is on or below the order of $10^{-3}$ within the SNR range of the star samples used here. Due to this finding, we do not expect noise to cause significant bias in the higher moments measurement.}

\begin{figure*}
\centering
\includegraphics[width=2.0\columnwidth]{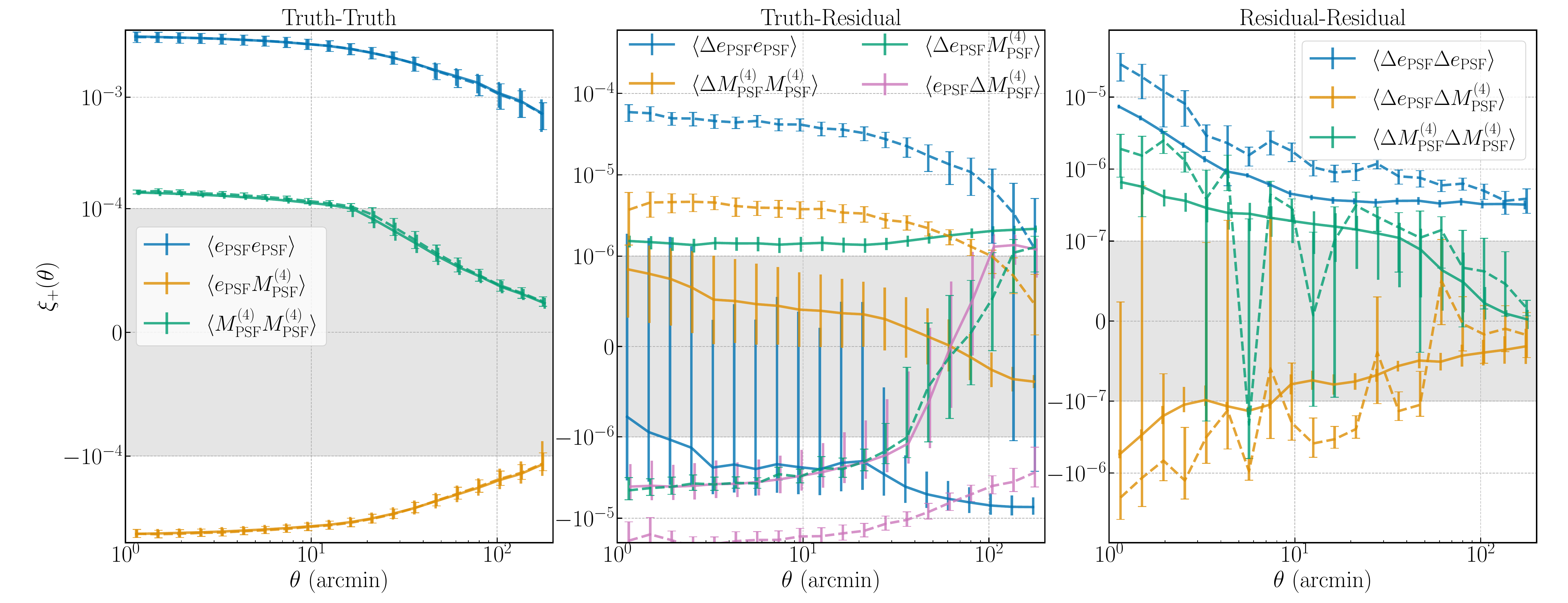}
\caption{
    The PSF-PSF correlation functions between $[e_{\rm PSF}, \Delta e_{\rm
    PSF}, M^{\rm (4)}_{\rm PSF}, \Delta M^{\rm (4)}_{\rm PSF}]$ of the PSF
    stars (solid lines) and the non-PSF stars (dashed lines) in all six HSC Y3 fields. The 10
    correlation functions are divided into truth-truth (left panel),
    truth-residual (middle panel), and residual-residual (right panel). The PSF
    and non-PSF stars have identical truth-truth correlations, as expected
    since they trace the same survey area in consistent ways.  The
    truth-residual and residual-residual correlations of the non-PSF stars are
    all larger than those for PSF stars. This result is consistent with the
    evidence for PSF model overfitting that was identified in
    Figure~\ref{fig:moment_dist}. \responsemnras{We use the}
    symmetrical-logarithmic scale on the y-axes, with the linear region shaded
    in grey. All errorbars are obtained by the jackknife in
    \textsc{TreeCorr} \citep{2004MNRAS.352..338J} after dividing the entire HSC Y3 fields
    into 20 patches using $k$-means. Note that the truth are orders of magnitude larger than the residual, therefore the three panels have very different scale in y-axis.
    The errorbars on the correlation functions for PSF (non-PSF) stars have (do not have) caps.
    }
    \label{fig:pp_corr}
\end{figure*}

The identification of spin-2 combinations of PSF higher moments is a powerful tool to reduce the dimensionality of the data vector of PSF higher moments that must be considered as potential contaminants to weak lensing shear, which can greatly simplify the cosmic shear analysis while still allowing for effective mitigation of all relevant PSF systematics. In Section~\ref{sec:model:formalism}, we will build the PSF systematics model including spin-2 higher moment combinations.

\section{PSF Systematics in Cosmic Shear}
\label{sec:model:0}

In Section~\ref{sec:model:formalism}, we present our formalism for describing PSF systematics in cosmic shear.  
In Section~\ref{sec:model:null}, we describe the process for model selection (demonstrating it by determining our fiducial model for HSC) and for determining the priors on the corresponding parameters. 
In Section~\ref{sec:model:zbin}, we describe the process for accounting for how PSF systematics may affect tomographic bins in different ways due to evolution in galaxy properties and shear with redshift. 

We have also confirmed that a number of factors are subdominant and need not be included in our model.  These aspects include the PSF systematics impact on $\xi_-$, PSF sixth order spin-2 quantity, and impact of  second-order systematics terms. These are discussed in Appendix~\ref{sec:ap:consistency}. These outcomes are specific to the HSC Y3 dataset, and we recommend that other surveys  carry out these tests when determining their PSF systematics model as well.  
This section derives the PSF systematics models in the real space cosmic shear analysis. We provide the equivalent formalism in Fourier space and discuss the consistency between the real and Fourier space analyses in Appendix~\ref{sec:ap:fourier}.

\subsection{Formalism}
\label{sec:model:formalism}

The observed galaxy ellipticity can be expressed as
\begin{equation}
\label{eq:obs_gal}
\hat{g}_{\rm gal} = g_{\rm gal} + g + g_{\rm sys}.
\end{equation}
Here $g_{\rm gal} = e_{\rm gal}/(2\mathcal{R})$ is the shear of the intrinsic shape of the galaxy, $g$ is the cosmic shear, introduced in Section~\ref{sec:bgd:cosmic_shear}, $\mathcal{R}$ is the responsivity of the shape to shear, and $g_{\rm sys}$ is the additive systematic shear due to the PSF. In this formalism, the multiplicative bias, which normally is a pre-factor of the shear, is absorbed in the responsivity matrix. The spin-2 quantities related to the PSF, described in Section~\ref{sec:star:spin2}, contribute to the additive bias $g_{\rm sys}$.
Note that in some literature, the additive shear bias $g_{\rm sys}$ is referred to as $e_{\rm sys}$ \citep{2019PASJ...71...43H,2018PhRvD..98d3528T}.

The past treatment of PSF systematics due to second moments in $g_\text{sys}$ has included two terms: PSF leakage and PSF shape modeling error.
The PSF leakage refers to the imperfect correction for the shear estimation method, which correlates the galaxy shape $\hat{g}_{\rm gal}$ with the PSF shape $e_{\rm PSF}$. For example, it is found in previous studies that \reGauss{} is susceptible to PSF leakage \citep{2018PASJ...70S..25M, 2020PASJ...72...16H}. The PSF modeling error term originates from the residual in PSF shape modeling and therefore the unavoidable bias in the galaxy shape estimation \citep{2008A&A...484...67P}, which correlates the galaxy shape $\hat{g}_{\rm gal}$ with the PSF shape residual $\Delta e_{\rm PSF} = e_{\rm PSF,model} - e_{\rm PSF,*}$. Previous work often used $\alpha$ and $\beta$ as prefactors for the leakage and modeling error terms \citep{2020PASJ...72...16H,2022PhRvD.105b3514A,2021A&A...645A.105G}. When only considering the PSF second moments,
\begin{equation}
\label{eq:psf-sys-second}
g_{\rm sys} = \alpha e_{\rm PSF} + \beta \Delta e_{\rm PSF}.
\end{equation}
Note that in \cite{2020PASJ...72...16H}, instead of $(\Delta) e_{\rm PSF}$, $(\Delta) g_{\rm PSF} = (\Delta) e_{\rm PSF}/2$ is used. We decided to use the distortion ($e_{\rm PSF}$) 
directly throughout the paper so that second and higher moments would be treated consistently. This choice only results in a factor of $2$ difference in the second moment PSF parameters, and do not impact the cosmological prediction.

We found a spin-2 quantity consisting of PSF fourth moments in Section~\ref{sec:star:spin2}. Therefore, a logical generalization of the PSF systematics formalism is to add and test for fourth moment leakage and modeling error terms as part of $g_{\rm sys}$. We also want to check the necessity of including a constant ellipticity parameter $e_{c} = e_{c,1} + i e_{c,2}$ in the formalism, to model the systematics from other sources that generate a non-zero mean shape in the catalog, other than that from the cosmic variance.
Therefore, the full model for $g_{\rm sys}$ is
\begin{equation}
\label{eq:psf-sys-full}
g_{\rm sys} = \alpha^{\rm (2)} e_{\rm PSF} + \beta^{\rm (2)} \Delta e_{\rm PSF}
+ \alpha^{\rm (4)} M^{\rm (4)}_{\rm PSF} + \beta^{\rm (4)} \Delta M^{\rm (4)}_{\rm PSF}
+ e_{c}.
\end{equation}
Here $\alpha^{\rm (2)}$ and $\beta^{\rm
(2)}$ are leakage and modeling error coefficients for second moments, and $\alpha^{\rm (4)}$ and $\beta^{\rm (4)}$ are comparable quantities for fourth moments. This formalism could in principle extend to all spin-$2$ quantities, including PSF sixth moments, and product of spin-$0$ and spin-$2$ quantities, etc. However, higher moments and higher order terms are increasingly noise dominated. In Appendix~\ref{sec:ap:six}, we show that extending to sixth moments does not increase the overall estimated additive bias significantly, and therefore is not needed for HSC Y3. Similarly, we show in Appendix~\ref{sec:ap:sot} that second order terms do  not significantly contribute to additive shear biases for HSC Y3. However, we recommend that other surveys with more stringent requirement on systematics also test for the impact of these quantities when defining their PSF systematics model.

Since  $g_{\rm sys}$ and $ g_{\rm gal} + g$ are uncorrelated, the 2PCF of the observed galaxy
shape is
\begin{equation}
\langle \hat{g}_{\rm gal} \hat{g}_{\rm gal} \rangle = \langle (g_{\rm gal} + g)(g_{\rm gal} + g) \rangle + \langle g_{\rm sys} g_{\rm sys} \rangle\,.
\end{equation}
We focus on the last term, which is the additive shear contamination due to the PSF in  the shear-shear 2PCF. To efficiently express $\langle g_{\rm sys} g_{\rm sys} \rangle$, we define the parameter vector $\ve{p} = [\alpha^{\rm (2)},
\beta^{\rm (2)},\alpha^{\rm (4)}, \beta^{\rm (4)}, e_{c}]$, and
define the PSF moments vectors $\ve{S} = [e_{\rm PSF}, \Delta e_{\rm PSF},
M^{\rm (4)}_{\rm PSF}, \Delta M^{\rm (4)}_{\rm PSF}\responsemnras{, \ve{1}}]$. Here $\ve{p}$ is a parameter set defined for the galaxy ensemble, while $\ve{S}$ is a set of PSF quantities that varies across the position on the sky. We include $e_c$ in the PSF parameter vector to simplify the formalism for likelihood analysis.
The expansion of
$\langle g_{\rm sys} g_{\rm sys} \rangle$ from Eq.~\eqref{eq:psf-sys-full} becomes
\begin{equation}
\label{eq:expand-esys_esys}
\langle g_{\rm sys} g_{\rm sys} \rangle =
\sum_{k=1}^5 \sum_{q=1}^5 \ve{p}_k \ve{p}_q \langle \ve{S}_k \ve{S}_q \rangle
\end{equation}
\responsemnras{Here the double summation includes the impact of (a) the 10 unique PSF-PSF correlation functions (p-p correlations), (b) the product of the mean shape systematic term $e_c$ and mean PSF moments, and (c) the mean shape systematic term $e_c$ itself. When two complex numbers are multiplied together, the complex conjugate must be used for one of them. }

\begin{figure*}
    \centering
    \includegraphics[width=2.0\columnwidth]{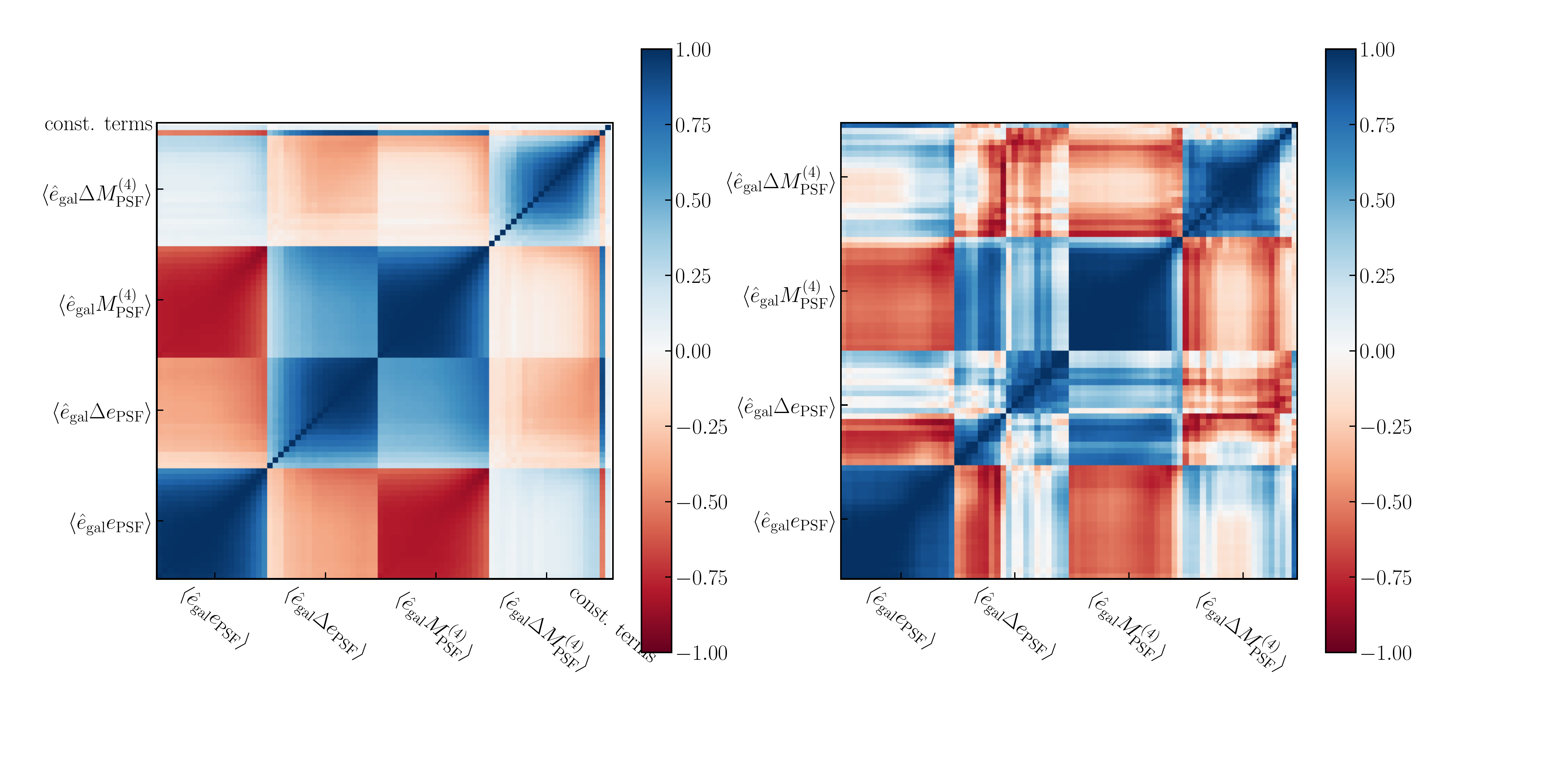}
\caption{
We show the correlation matrix of $\ve{D}_{gp}$ as defined in Eq.~\eqref{eq:correlation_matrix} in the left panel, and the correlation matrix of $\mt{K}(\ve{p})\ve{D}_{pp}$ in the right panel.  We see that $\ve{D}_{gp}$ values at different angular scales are highly correlated, and there are significant anti-correlations between $\langle \hat{g}_{\rm gal} e_{\rm PSF} \rangle$ and $\langle \hat{g}_{\rm gal} M^{\rm (4)}_{\rm PSF} \rangle$. $\mt{K}\ve{D}_{pp}$ across angular bins are also highly correlated for the correlation with the PSF truth. These significantly affect the outcome of the maximum-likelihood fitting process by penalizing cases where the theory data vector is such that the sign of $\ve{D}_{gp} - \ve{T}_{gp}$ differs across angular bins, or where the sign of $\ve{D}_{gp} - \ve{T}_{gp}$ is the same for $\langle \hat{g}_{\rm gal} e_{\rm PSF} \rangle$ and $\langle \hat{g}_{\rm gal} M^{\rm (4)}_{\rm PSF} \rangle$. Notice that the correlation matrix of the p-p correlation is more noisy than that of the g-p data vector, because the former is calculated using the jackknife method, while the latter is calculated using a large number of the mock catalogs. \responsemnras{We use the best-fitting parameters of the ``4+c'' model (listed in Table~\ref{tab:p_value_nobin}) to construct the correlation matrix of $\mt{K}(\ve{p})\ve{D}_{pp}$.}
On average, the covariance matrix from the p-p correlation contributes about $20\%$ of $\mt{\Sigma}_{gp}$ to the total covariance matrix $\tilde{\mt{\Sigma}}_{gp}(\ve{p})$ at the best-fitting parameters of the fiducial model, introduced in Section~\ref{sec:model:define}.
}
    \label{fig:correlation_nobin}
\end{figure*}

In Fig.~\ref{fig:pp_corr}, we show the p-p correlation functions between
all PSF moment pairs in $\ve{S}$, for the PSF stars (solid lines) and non-PSF
stars (dashed lines). We denote the moments of the PSF as ``truth'', and the difference expressed in Eq.~\eqref{eq:moment_residual} as ``residual''. The PSF and non-PSF samples are similar in truth-truth
correlations. However, because the moment residuals are much larger for the non-PSF
samples, all of the truth-residual and residual-residual correlations are significantly larger for the non-PSF stars.

\subsection{Building a Data-Driven PSF Systematics Model}
\label{sec:model:null}

\begin{figure*}
\centering
\includegraphics[width=1.9\columnwidth]{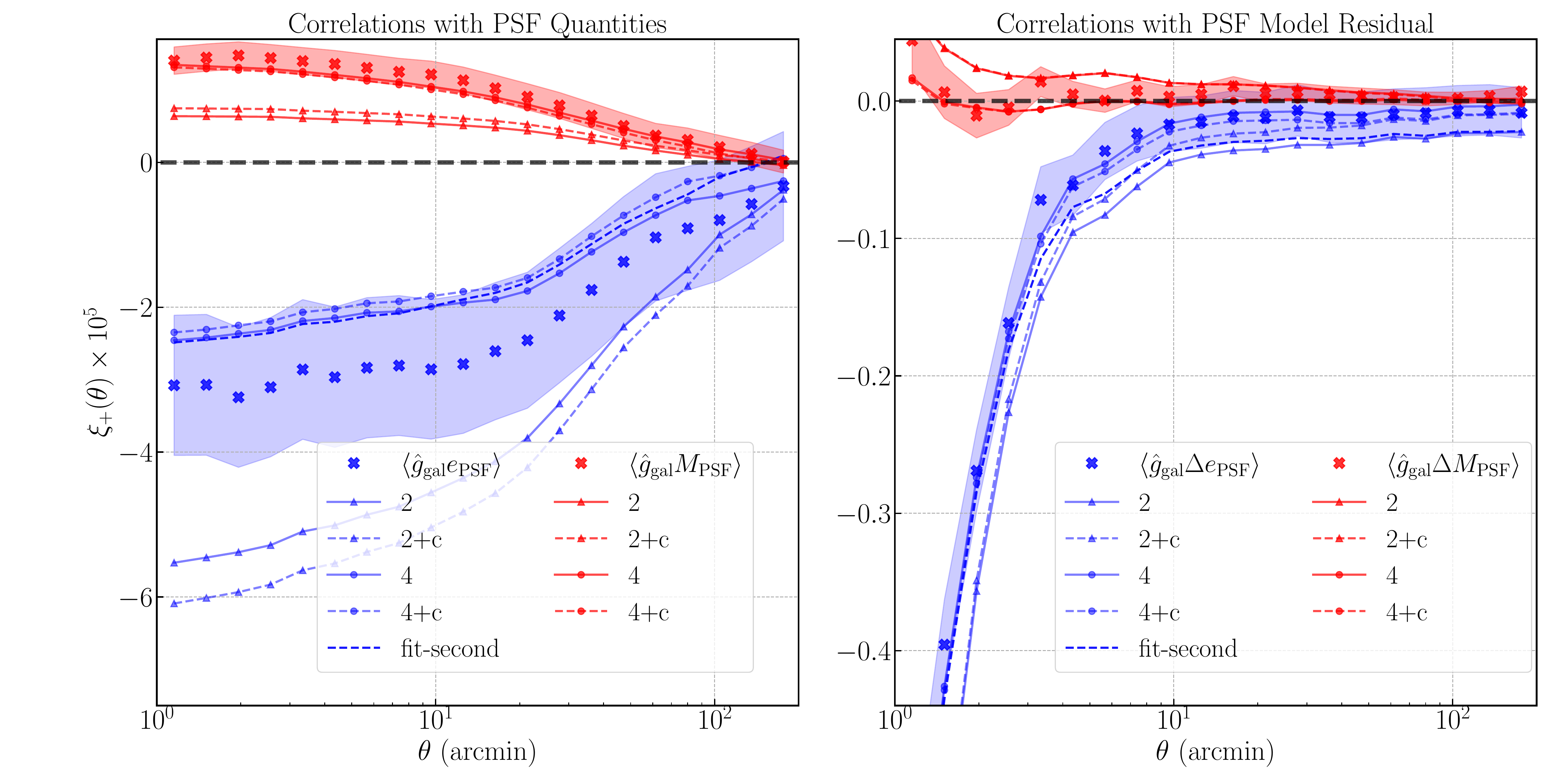}
    \caption{
    The correlations functions  of galaxy shapes with PSF quantities (left panel, Eqs.~\ref{eq:null1} and \ref{eq:null3}) and with
    PSF modeling residuals (right panel, Eqs.~\ref{eq:null2} and \ref{eq:null4})
    and the best-fitting PSF systematics
    models for the PSF stars. The correlations between the Y3 star
    catalog and shape catalog are shown as ``x'', with the shaded region
    representing the 1$\sigma$ uncertainty. The best-fitting correlations from the
    models are shown in the solid and dashed lines, where the quantity being modelled is reflected by the color.
    ``2'' means that the
    model only includes second moments leakage and modeling error terms, ``+c'' means
    that the model includes the constant galaxy shape term, and ``4'' stands for the
    fiducial model, which includes both the PSF second and fourth moments.  All
    models are fitted to all four galaxy-PSF correlation functions and to the
    average galaxy shape, except for ``fit-second'', which only fits to $\langle  \hat{g}_{\rm gal} e_{\rm PSF} \rangle$ and $\langle  \hat{g}_{\rm gal} \Delta e_{\rm PSF} \rangle$.
    }
    \label{fig:gp_corr}
\end{figure*}

\begin{figure}
    \centering
    \includegraphics[width=0.9\columnwidth]{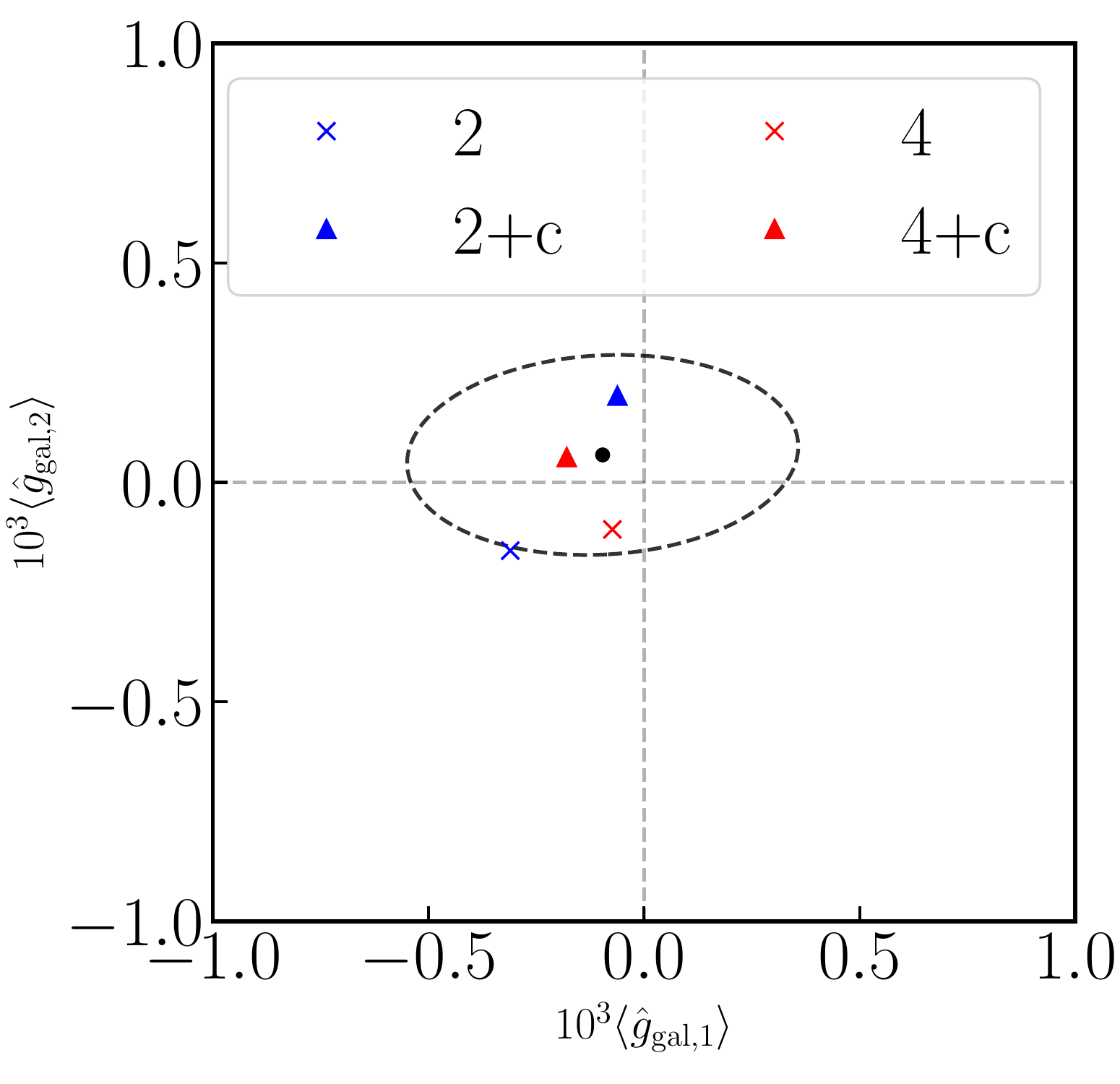}
\caption{The dot shows the average galaxy
    shape $\langle \hat{g}_{\rm gal,1} \rangle$ and $\langle \hat{g}_{\rm gal,2} \rangle$
    , and its 1$\sigma$ contour estimated using the Y3
    mock catalog. The crosses and triangles show the best-fitting $\langle \hat{g}_{\rm gal,1} \rangle$ and $\langle \hat{g}_{\rm gal,2} \rangle$ predicted by Eq.~\eqref{eq:null_ave}, for the models labeled using the same notation as in  Fig.~\ref{fig:gp_corr}.
    }
    \label{fig:constant_shape}
\end{figure}

In this section, we use data-driven approaches to define the PSF systematics model.
In Section~\ref{sec:model:gp_corr}, we use the galaxy-PSF correlation as an observable to infer the posterior of the PSF parameter, by building models to predict galaxy-PSF correlations with the formalism in Section~\ref{sec:model:formalism}. In Section~\ref{sec:model:define}, we define the full models and their submodels, as well as the traditional second-moment-only model for comparison.  In Section~\ref{sec:model:psf_vs_nonpsf}, we discuss the difference between PSF and non-PSF results. In Section~\ref{sec:model:cosmology}, we show the impact on cosmological observable by the PSF systematics.In Section~\ref{sec:model:model_selection}, we describe our model selection criteria.

\subsubsection{Galaxy-PSF Cross Correlation and Model Fitting}
\label{sec:model:gp_corr}

Cross correlating galaxy and PSF spin-2 components is a common approach to identifying and quantify additive PSF systematics. When used for identifying systematics, these calculations are referred to as ``null tests'' \citep[e.g.,][]{2016MNRAS.460.2245J, 2018PASJ...70S..25M}. The PSF-PSF correlations are referred to as the $\rho$ statistics.
Here we employ galaxy-PSF correlation (g-p correlation) for two purposes: (a) identifying the contamination from PSF leakage and modeling error on weak lensing shears, and (b) quantifying the prior on any PSF systematics parameters that need to be included in the cosmological likelihood analysis.

We cross-correlate galaxy shapes with the PSF spin-2 quantities and their residuals, both from second and fourth moments, and the constant systematics term.
Since the true correlation between the galaxy's intrinsic shape and shear with any PSF moment $M_{\rm PSF}$ should be zero, we can assume that $\langle \hat{g}_{\rm gal} M_{\rm PSF} \rangle = \langle g_{\rm sys} M_{\rm PSF} \rangle$. Therefore, the cross-correlation between $\hat{g}_{\rm gal}$ in  Eq.~\eqref{eq:obs_gal} and PSF moments in Eq.~\eqref{eq:psf-sys-full} becomes

\begin{widetext}
\begin{align}
    \label{eq:null1}\langle \hat{g}_{\rm gal} e_{\text{PSF}}\rangle &= \alpha^{\rm (2)} \langle e_{\text{PSF}} e_{\text{PSF}} \rangle  + \beta^{\rm (2)}\langle \Delta e_{\text{PSF}}  e_{\text{PSF}} \rangle  + \alpha^{\rm (4)} \langle M^{\rm (4)}_{\text{PSF}}  e_{\text{PSF}} \rangle  + \beta^{\rm (4)} \langle \Delta M^{\rm (4)}_{\text{PSF}}  e_{\text{PSF}}\rangle + e_c \langle  e_{\text{PSF}} \rangle \\
    \label{eq:null2} \langle \hat{g}_{\rm gal} \Delta e_{\text{PSF}} \rangle &= \alpha^{\rm (2)} \langle  e_{\text{PSF}} \Delta e_{\text{PSF}}\rangle  + \beta^{\rm (2)} \langle \Delta e_{\text{PSF}} \Delta e_{\text{PSF}}\rangle + \alpha^{\rm (4)} \langle M^{\rm (4)}_{\text{PSF}} \Delta e_{\text{PSF}} \rangle  + \beta^{\rm (4)} \langle \Delta M^{\rm (4)}_{\text{PSF}} \Delta e_{\text{PSF}} \rangle + e_c \langle  \Delta e_{\text{PSF}} \rangle\\
    \label{eq:null3} \langle \hat{g}_{\rm gal}  M^{\rm (4)}_{\text{PSF}}\rangle &= \alpha^{\rm (2)} \langle  e_{\text{PSF}}  M^{\rm (4)}_{\text{PSF}} \rangle  + \beta^{\rm (2)} \langle \Delta e_{\text{PSF}} M^{\rm (4)}_{\text{PSF}} \rangle + \alpha^{\rm (4)} \langle  M^{\rm (4)}_{\text{PSF}}  M^{\rm (4)}_{\text{PSF}} \rangle  + \beta^{\rm (4)} \langle \Delta M^{\rm (4)}_{\text{PSF}}  M^{\rm (4)}_{\text{PSF}} \rangle + e_c \langle  M^{\rm (4)}_{\text{PSF}}\rangle\\
    \label{eq:null4}\langle \hat{g}_{\rm gal} \Delta M^{\rm (4)}_{\text{PSF}}\rangle &= \alpha^{\rm (2)} \langle  e_{\text{PSF}} \Delta M^{\rm (4)}_{\text{PSF}} \rangle  + \beta^{\rm (2)} \langle \Delta e_{\text{PSF}} \Delta M^{\rm (4)}_{\text{PSF}} \rangle  + \alpha^{\rm (4)} \langle  M^{\rm (4)}_{\text{PSF}} \Delta M^{\rm (4)}_{\text{PSF}} \rangle  + \beta^{\rm (4)} \langle \Delta M^{\rm (4)}_{\text{PSF}} \Delta M^{\rm (4)}_{\text{PSF}} \rangle + e_c \langle   \Delta M^{\rm (4)}_{\text{PSF}}\rangle.
\end{align}
\end{widetext}
Here the correlation functions on the left-hand-side (LHS) of the equations are what
we call ``galaxy-PSF correlations'' (g-p correlations), and the correlation functions on the right
are ``PSF-PSF correlations'' (p-p correlations).  Additionally, we check if the average galaxy shape in the catalog follows the model
\begin{align}
\label{eq:null_ave}
\nonumber\langle \hat{g}_{\rm gal} \rangle &= \alpha^{\rm (2)} \langle e_{\rm PSF}\rangle + \beta^{\rm (2)} \langle\Delta e_{\rm PSF}\rangle + \alpha^{\rm (4)} \langle M^{\rm (4)}_{\rm PSF}\rangle \\&\, + \beta^{\rm (4)}\langle \Delta M^{\rm (4)}_{\rm PSF}\rangle + e_c.
\end{align}

In this work, we measure the p-p and g-p correlations in 20 angular bins from 1-200 arcmin. The range of angular bins was defined so that it covers the scales used from the HSC Y1 cosmic shear analyses, while also ensuring the small scales are not affected by blending. The upper scale cuts are extended to 200 arcmin to provide more constraining power on the PSF parameters.
The data vector $\ve{D}_{gp} = [\langle \hat{g}_{\rm gal} e_{\rm PSF} \rangle, \langle \hat{g}_{\rm gal} \Delta e_{\rm PSF} \rangle, \langle \hat{g}_{\rm gal} M^{\rm (4)}_{\rm PSF} \rangle, \langle \hat{g}_{\rm gal} \Delta M^{\rm (4)}_{\rm PSF} \rangle, \langle\hat{g}_{\rm gal}\rangle]$, with 82 data points in total, is fitted by the theory data vector $\ve{T}_{gp}(\ve{p})$ predicted from Eqs.~\eqref{eq:null1}--\eqref{eq:null_ave}, by maximizing the log-likelihood function
\begin{equation}
    \label{eq:likelihood_define}
    \log(\mathcal{L}(\ve{p} | \vect{D}_{gp})) \propto  - \frac{1}{2}\chi^2  - \frac{1}{2}\log(\det( \mt{\tilde{\Sigma}}_{gp})),
\end{equation}
where
\begin{equation}
    \chi^2 = (\vect{D}_{gp}-\vect{T}_{gp}(\ve{p}))^T \mt{\tilde{\Sigma}}_{gp}^{-1}(\ve{p})(\vect{D}_{gp}-\vect{T}_{gp}(\ve{p})).
\end{equation}
Here $\mt{\tilde{\Sigma}}^{-1}_{gp}(\ve{p})$ is the parameterized inverse covariance matrix that includes the Gaussian covariance matrix of the p-p correlation functions
\begin{equation}
    \label{eq:parameterized_covariance}
    \mt{\tilde{\Sigma}}_{gp}(\ve{p}) = \mt{\Sigma}_{gp} + \mt{K}(\ve{p}) \mt{\Sigma}_{pp} \mt{K}(\ve{p})^T.
\end{equation}
$\mt{\Sigma}_{gp}$ is the covariance matrix of $\ve{D}_{gp}$ computed using the HSC Y3 mock catalog described in Section~\ref{sec:shape:mock}, $\mt{\Sigma}^{-1}_{pp}$ is the covariance matrix of the p-p correlation vector $\ve{D}_{pp}$, which consists the p-p correlation functions ordered in the reading order of the RHS of Eq.~\eqref{eq:null1}--\eqref{eq:null4}. $\mt{K}(\ve{p})$ is the linearized transformation matrix of the RHS of Eq.~\eqref{eq:null1}--\eqref{eq:null4}.  By having a parameterized covariance matrix in the likelihood, we effectively marginalize over the uncertainty of the p-p correlation function \citep{2020MNRAS.491.5498M}. In Eq.~\eqref{eq:likelihood_define}, the second term comes from the normalizing factor in the Gaussian likelihood, which changes during the fitting because of the parameterized covariance matrix. 

In Fig.~\ref{fig:correlation_nobin}, we show the correlation matrix ${\rm Cor}(\ve{D}_{gp})$ of $\ve{D}_{gp}$, where
\begin{equation}
\label{eq:correlation_matrix}
{\rm Cor}(\ve{D}_{gp})[i][j] = \frac{\mt{\Sigma}_{gp}[i][j]  }{\sqrt{\mt{\Sigma}_{gp}[i][i] \mt{\Sigma}_{gp}[j][j]}}
\end{equation}
in the left panel, and the correlation matrix of $\mt{K}(\ve{p}) \ve{D}_{pp}$ at the right panel. The elements in the covariance matrix contributed by the p-p correlation $\mt{K}(\ve{p}) \mt{\Sigma}_{pp} \mt{K}(\ve{p})^T$ are typically $20\%$ of $\mt{\Sigma}_{gp}$ at the best-fitting parameters of fiducial model, introduced in Section~\ref{sec:model:define}. Therefore these are not negligible in the model fitting.

By maximizing Eq.~\eqref{eq:likelihood_define}, we get the best-fitting value of the parameters $\ve{p}$. We also used  Markov Chain Monte Carlo (MCMC), implemented in \textsc{emcee} \citep{2013PASP..125..306F}, to measure the posterior of the PSF parameters $P(\vect{p} | \ve{D}_{gp})$. The priors on all PSF systematics parameters are flat from $-\infty$ to $+\infty$.

\subsubsection{Model Definition}
\label{sec:model:define}

Now we define the models we included in the model fitting and selection, assuming the PSF parameters are independent of the tomographic bins.
The full model (``4+c'') includes all 6 parameters in $\ve{p}$. We define sub-models by setting some parameters in $\ve{p}$ to zero while still fitting the entire data vector $\vect{D}_{gp}$. The fiducial model (``4'') is a sub-model that only includes the first four parameters in $\ve{p}$; later in this section, we explain the statistical criteria used to identify this model as the fiducial one. The second-moments-only model, denoted as ``2'', only has the first two parameters in $\ve{p}$. The ``2+c'' model adds the $e_c$ parameters to the  second-moments-only model. The ``4'', ``2'' and ``2+c'' are all sub-models of the full model ``4+c'', defined in Eq.~\eqref{eq:null1}--\eqref{eq:null_ave}.

The ``fit-second'' model is not a sub-model of the full model: in particular, this corresponds to taking the ``2'' model and only fitting it to the second moment g-p correlations. The ``fit-second'' model is a logical choice if the fourth moment g-p correlation is ignored in the null testing.
We introduced ``fit-second'' because it mirrors what was done in past shear analyses that did not consider fourth moments.

We validated our statistical inference on the PSF parameters by adding PSF shear bias to the mock catalog with known PSF parameters, and attempted to recover the PSF parameters through our inference. This process is described in Appendix~\ref{sec:ap:mock_test}.

\subsubsection{PSF and non-PSF stars}
\label{sec:model:psf_vs_nonpsf}

We carried out the same analysis, now using the non-PSF stars. Since the PSF star sample shows that the models with only second moment leakage and modeling error cannot predict the galaxy shape correlations with PSF fourth moments, and the constant terms are later deemed unnecessary, we only used the fiducial model and the fit-second model. The $p$-values of these models as applied to the non-PSF star correlation functions are also included in Table~\ref{tab:p_value_nobin}. The fiducial model still performs well for the non-PSF star sample, as well as the fit-second model fitted to the g-p correlations with the second moments of the non-PSF stars.

\begin{figure}
    \centering
    \includegraphics[width=0.9\columnwidth]{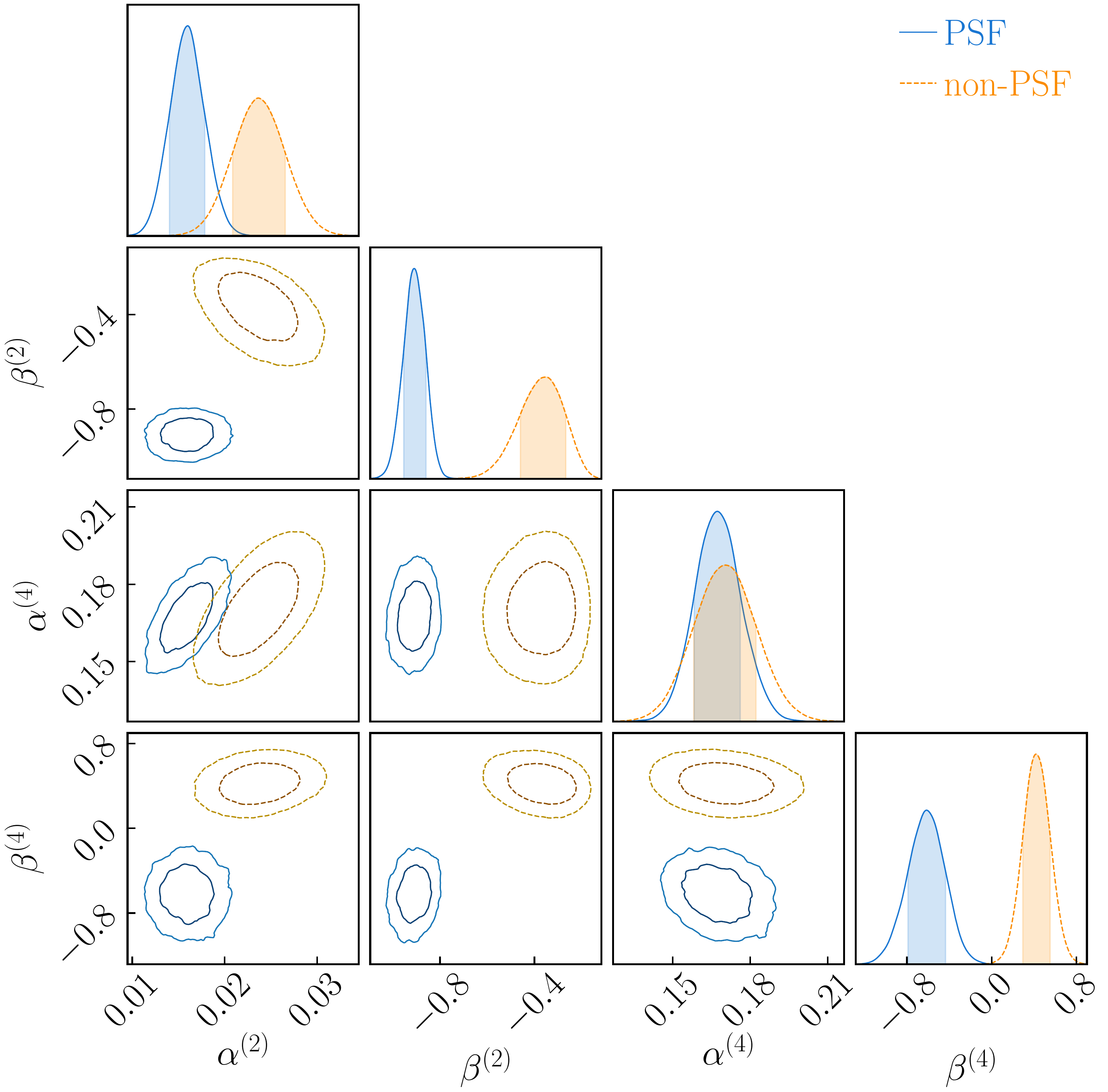}
\caption{The posterior of the PSF systematics model parameters for the fiducial model, using the PSF and non-PSF stars as indicated in the legend. The PSF stars provide significantly  larger estimates for both $\beta^{\rm (2)}$ and $\beta^{\rm (4)}$, which could be explained by the overfitting of the PSF model.
}
    \label{fig:prior_nobin}
\end{figure}

In Fig.~\ref{fig:prior_nobin}, we show the PSF parameter posteriors for our fiducial model, for the PSF and non-PSF stars. The best-fitting parameters and their errorbars are shown in Table~\ref{tab:p_value_nobin}. We see that the results from the two datasets provide statistically consistent values for $\alpha^{\rm (4)}$, but inconsistent ones for $\alpha^{\rm (2)}$, $\beta^{\rm (2)}$ and $\beta^{\rm (4)}$. \responsemnras{Overall, the inconsistency between the PSF and non-PSF results is 8.2$\sigma$, ignoring the correlation between the two results.}
The mismatch of the $\beta$ values can be explained by the overfitting of the PSF model:
Fig.~\ref{fig:moment_dist} shows that the second moment residual is overfitted by a factor of $\sim 2$, which means the $\Delta e_{\rm PSF}$ of the PSF stars are underestimated by a factor of $2$. To compensate for this in the model fitting, the underestimation of $\Delta e_{\rm PSF}$ gives rise to a $\beta^{\rm (2)}$ for PSF stars $\sim 2$ times larger than the one for non-PSF stars, as $g_{\rm sys}$ is ultimately the source of the inferred $\beta$ values.  In other words, we are fitting to correlation functions that carry information about the true systematic uncertainties in the galaxy shears, and hence using a star sample that underestimates the magnitude of $\Delta e_\text{PSF}$ leads to a correspondingly higher value for $\beta$ but effectively the same actual $\Delta \xi_+$ (which is the product of the two factors).\responsemnras{ The inconsistency in the $\alpha^{\rm (2)}$ values is roughly $1.7\sigma$, without considering the potential correlation between the two results.}
Next we will directly demonstrate that the two samples nonetheless predict a consistent impact on cosmic shear.

\subsubsection{Impact on Cosmic Shear}
\label{sec:model:cosmology}

\begin{figure}
    \centering
    \includegraphics[width=0.9\columnwidth]{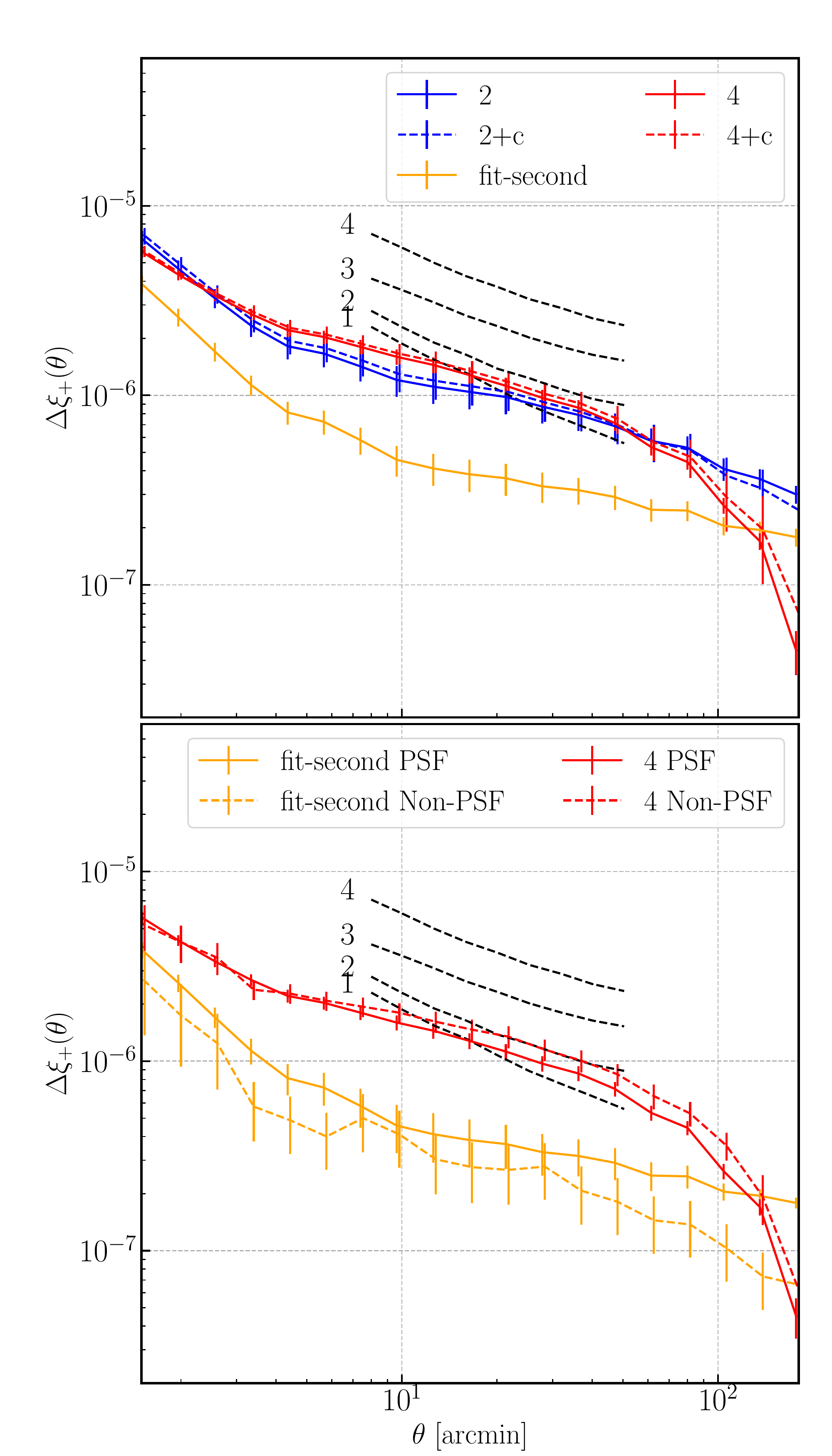}
\caption{The additive bias on the cosmic shear 2PCF $\xi_+$ for the redshift-independent models. 
\responsemnras{The statistical uncertainties on the shear-shear auto correlation are shown as the black dashed lines, with an index for the tomographic bin. } 
The upper panel shows the $\Delta \xi_+$ of the different nested models of ``4+c'' using their best-fitting parameters. The model naming convention follows Fig.~\ref{fig:gp_corr}. We can see that including the galaxy shape correlations with the PSF fourth moments increases the estimated $\Delta \xi_+$ on some angular scales by almost an order of magnitude compared to when we fit to second moments only. And the $\Delta \xi_+$ predicted by the fiducial model \responsemnras{on par with the statistical uncertainty of the first and second bin's auto correlation function}, which is a significant contamination level. The bottom panel shows the comparison of the estimated $\Delta\xi_+$ for just two of the models using the PSF stars (solid lines) and non-PSF stars (dashed lines).
}
\label{fig:delta_xip_nobin}
\end{figure}

With the best-fitting value \responsemnras{and uncertainty contour} for $\ve{p}$ for each model, we can also predict the impact on cosmic shear by the PSF systematics as a whole, expressed in Eq.~\eqref{eq:expand-esys_esys}. In Fig.~\ref{fig:delta_xip_nobin}, we show the additive bias on cosmic shear $\Delta \xi_+$, defined in Eq.~\eqref{eq:expand-esys_esys}, predicted by these models. In the upper panel, we compare the $\Delta \xi_+$ predicted by different models fitted to the galaxy shape correlations with PSF star moments. The traditional ``fit-second'' model omits the g-p correlation functions with PSF fourth moments and residuals, and  its $\Delta \xi_+$ is therefore underestimated by up to an order of magnitude, which is significant. Although ``2'' and ``2+c'' predict a similar magnitude for $\Delta \xi_+$ as ``4'' and ``4+c'', they fail to fit the g-p correlations, according to Fig.~\ref{fig:gp_corr} and Table~\ref{tab:p_value_nobin}, and thereby are suboptimal.
The difference between the full model and the fiducial model is insignificant in terms of $\Delta\xi_+$, \responsemnras{ compared to the statistical uncertainty of the shear-shear 2PCF}. Therefore, we can drop the $e_c$ parameters in the HSC Y3 cosmic shear analysis. In the lower panel, we compare the $\Delta \xi_+$ predicted using PSF versus non-PSF stars, for the fiducial and fit-second model. We notice that the $\Delta \xi_+$ predicted by the fiducial model is very similar for both star samples, while there is a larger discrepancy between $\Delta\xi_+$ predicted using the fit-second model fits to the PSF and non-PSF stars. This is because, when including fourth moments, the PSF and non-PSF stars' predicted $\Delta \xi_+$ is dominated by the fourth moment leakage, which (unlike second order modeling error terms) is less affected by the difference between PSF and non-PSF stars. 

For reference, we also plotted the \responsemnras{statistical uncertainty of the} shear-shear auto-correlation function \responsemnras{predicted for the HSC Y3 cosmic shear analysis, and computed the statistical significance of the PSF systematics bias by}
\begin{equation}
\label{eq:stat_significance}
b^i = \sqrt{\Delta \xi_+^T \Sigma^{-1}_{ii} \Delta \xi_+}.
\end{equation}
\responsemnras{Here $\Sigma^{-1}_{ii}$ is the estimated covariance matrix of the shear-shear auto correlation function of bin $i$. We find the statistical significance of the additive shear systematics for our fiducial model from bins $1-4$ to be $1.74, 1.10, 0.66, 0.42$ for the PSF stars, and $2.03, 1.27, 0.75, 0.49$ for the non-PSF stars. Overall, the statistical significance for all 10 $xi^{ij}_+$ is 2.0$\sigma$ for the PSF stars, and 2.3$\sigma$ for the non-PSF stars. Note that the statistical significance here might not directly correspond to the bias on the cosmological parameters. Rather, serves as an approximate indicator for the significance of the PSF systematics. }

\subsubsection{Model Comparison}
\label{sec:model:model_selection}

Here we show the model fitting results and describe the methodology to select among the models.

In Fig.~\ref{fig:gp_corr}, we show the galaxy-PSF correlation functions (LHS of Eq.~\eqref{eq:null1}--\eqref{eq:null4}) with their 1-$\sigma$ uncertainties. The correlations with PSF quantities  are shown in the left panel (second and fourth moments in blue and red, respectively), and the correlations with PSF model residuals are shown. Fig.~\ref{fig:constant_shape} shows the average galaxy shape in the right panel.
The 1-$\sigma$ uncertainty of $\ve{D}_{gp}$ as assessed using mock catalogs (including cosmic variance) is shown with shaded regions for the correlation functions and an ellipse for $\langle\hat{g}_{\rm gal}\rangle$.

The best-fitting theory vectors $\hat{\ve{T}}_{gp}$ in Fig.~\ref{fig:gp_corr} show that both the full model (``4+c'') and the fiducial model (``4'') can fit the data vectors within 1-$\sigma$ in the full angular range from $1$-$200$ arcmin.
The models involving only second moments (``2'' and ``2+c'') cannot fit the g-p correlations with PSF second moments and residuals nearly as well.
The ``fit-second'' model also fits the second moment correlations well. All the models fit the average shape $\langle \hat{g}_{\rm gal} \rangle$ within 1-$\sigma$, as shown in Fig.~\ref{fig:constant_shape}.

We measure the goodness of fit using $p$-values, assuming the number of degrees of freedom is $82-k$, where $k$ is the number of model parameters, and $82$ is the length of $\ve{D}_{\rm gp}$. The $p$-values are shown in Table~\ref{tab:p_value_nobin}. A $p$-value over 0.05 is considered a good fit to the data, and our results show that we need to include the fourth moments explicitly (``4'' or ``4+c'') to fit all g-p correlations.

\responsemnras{Ultimately, we use the impact on the cosmic shear data vector to select which model we should use. The most efficient model should include the minimum number of parameters needed to capture most of the contamination to $\xi_+$. In our case, the statistical significance of the estimated contamination changed from 0.7$\sigma$ for the second-moment model to 2.0$\sigma$ for the fiducial model, with only two additional parameters. Therefore, the fiducial model is our preferred choice, so as to avoid underestimating the additive systematics by more than a factor of two. In Appendix~\ref{sec:ap:six} and Appendix~\ref{sec:ap:sot}, we will see that none of the sixth-order moments or second-order terms can contribute enough additive bias to be worth using.  }

\begin{table*}
\centering
\begin{tabular}{llllllllll}
\hline
Sample  &  Model   & $\alpha^{\rm (2)}$  &  $\beta^{\rm (2)}$ & $\alpha^{\rm (4)}$ & $\beta^{\rm (4)}$  & $e_{\rm c,1}\times10^4$  & $e_{\rm c,2}\times10^4$  & $p$-value     \\ \hline
 & ``2'&   $-0.022\pm0.002$  &   $-1.08\pm0.06$    &   $0$    &   $0$    &   $0$    &   $0$    &  $0.0$    \\
PSF & ``2+c''& $-0.023\pm0.002$  &   $-1.09\pm0.06$    &   $0$    &   $0$    &   $2\pm2$    &   $3\pm1$    &  $0.0$    \\
 & ``4'' & $0.016\pm0.002$  &   $-0.88\pm0.05$    &   $0.17\pm0.01$    &   $-0.6\pm0.2$    &   $0$    &   $0$    &  $0.92$  \\
 & ``4+c''& $0.016\pm0.002$  &   $-0.88\pm0.05$    &   $0.17\pm0.01$    &   $-0.6\pm0.2$    &   $-1\pm2$    &   $2\pm1$    &  $0.54$    \\
 & ``fit-second'' & $-0.007\pm0.002$  &   $-0.83\pm0.05$'    &   $0$    &   $0$    &   $0$    &   $0$    &  $0.72$   
\\\hline
non-PSF & ``4'' & $0.024\pm0.003$  &   $-0.4\pm0.1$    &   $0.17\pm0.01$    &   $0.4\pm0.1$    &   $0$    &   $0$    &  $0.84$    \\
 & ``fit-second'' & $-0.004\pm0.002$  &   $-0.4\pm0.1$    &   $0$    &   $0$    &   $0$    &   $0$    &  $0.78$    \\
\end{tabular}
\caption{The best-fitting parameters, $p$-value of the models fitted to galaxy-PSF correlation functions in a single redshift bin (no tomography).
A $p$-value indicates the probability that the data may be a random realization of the model given the uncertainties, and a threshold of $0.05$ is commonly adopted. 
The models are defined in Section~\ref{sec:model:define}. The ``4+c'' model is the parent model among the first four models.
The second-moment models (``2'' and ``2+c'') failed when fitted to all g-p correlations, but provide an acceptable fit to the second moments' g-p correlation functions on their own (``fit-second'').
}
\label{tab:p_value_nobin}
\end{table*}

\begin{figure}
    \centering
    \includegraphics[width=0.98\columnwidth]{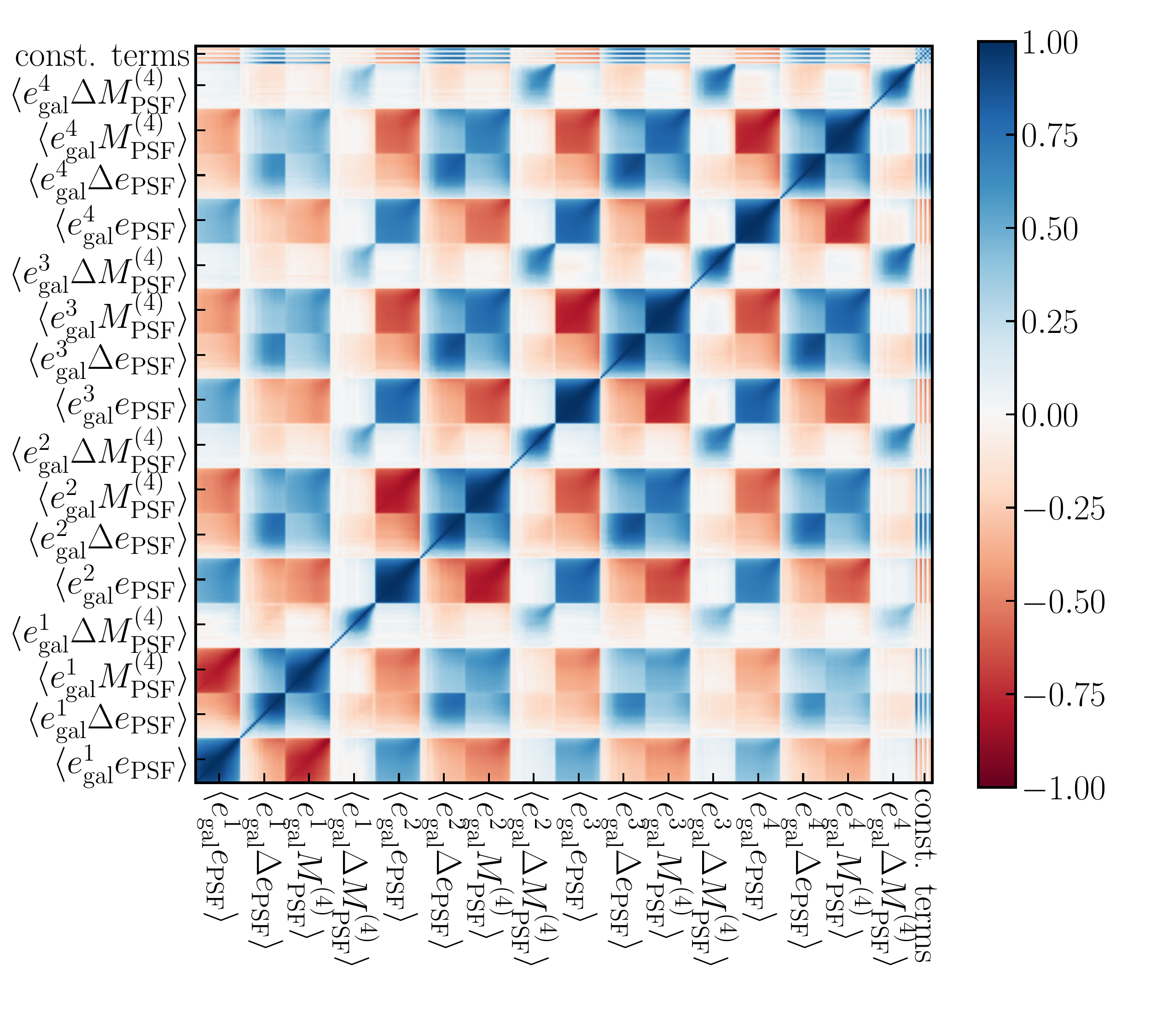}
\caption{ The correlation matrix of the data vector in the redshift-dependent model fitting. The quantity of the section in the data vector are shown in the x- and y-axis. Due to the correlation between the shear in different tomographic bins, g-p correlation functions across tomographic bins are also highly correlated. This is the primary reason that the PSF parameters for the 4 tomographic bins need to be jointly fitted, rather than individually fitted.
}
    \label{fig:correlation}
\end{figure}

\begin{figure*}
    \centering
    \includegraphics[width=2.0\columnwidth]{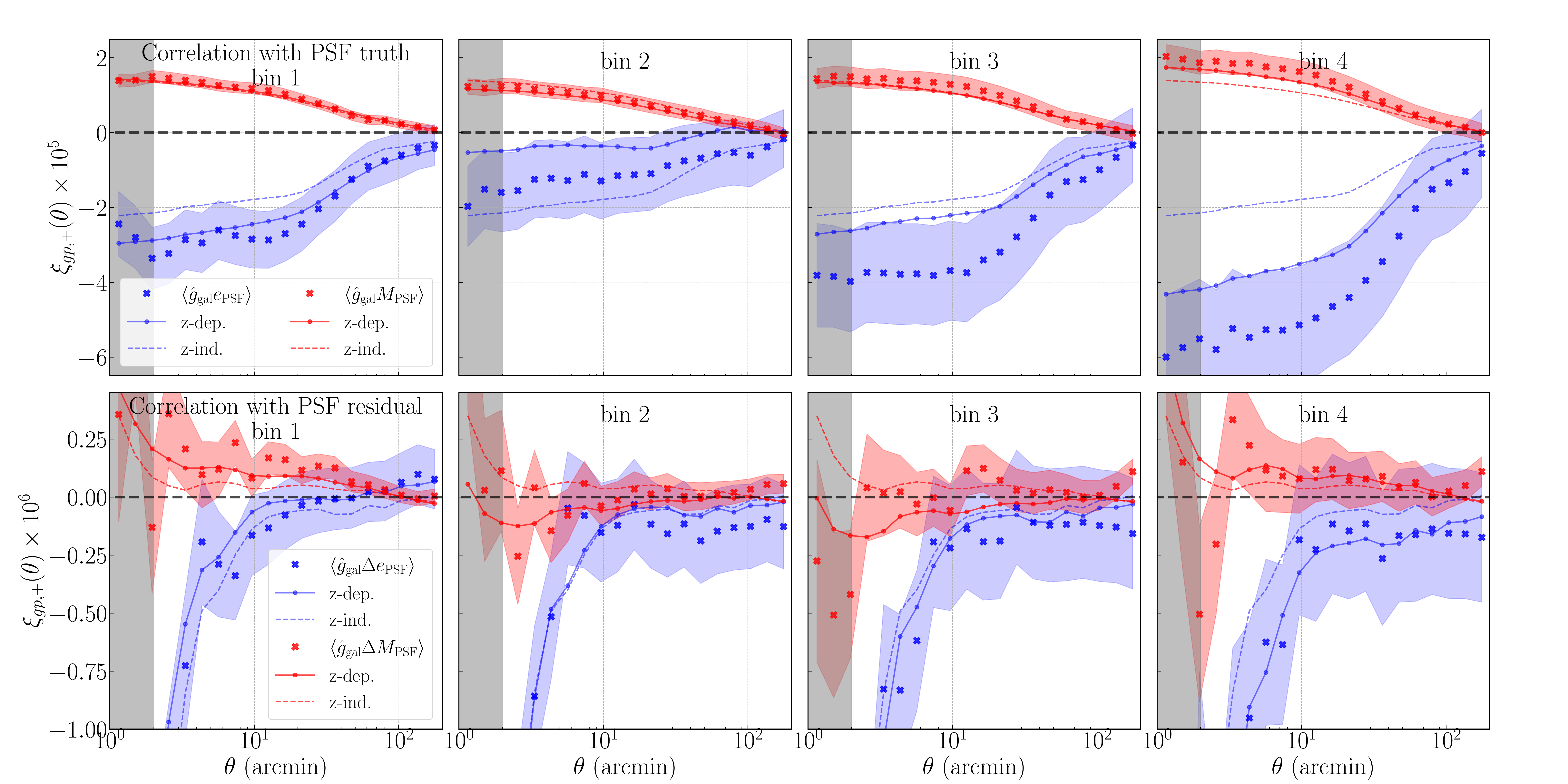}
\caption{
Galaxy-PSF correlation functions with galaxy samples subdivided into four tomographic bins as defined for the HSC Y3 cosmic shear analysis. The first row shows the correlations with the PSF truth terms, and second row with the PSF residual terms. The four columns correspond to the four tomographic bins.
The stars are the best-fitting values for the redshift-dependent model, the dashed lines are the best-fitting values for the redshift-independent model. The shaded regions are excluded from the fits because the model is not able to fit the data there, as assessed using $p$-values. 
}
    \label{fig:gp_redshift}
\end{figure*}

\begin{figure}
    \centering
    \includegraphics[width=1.0\columnwidth]{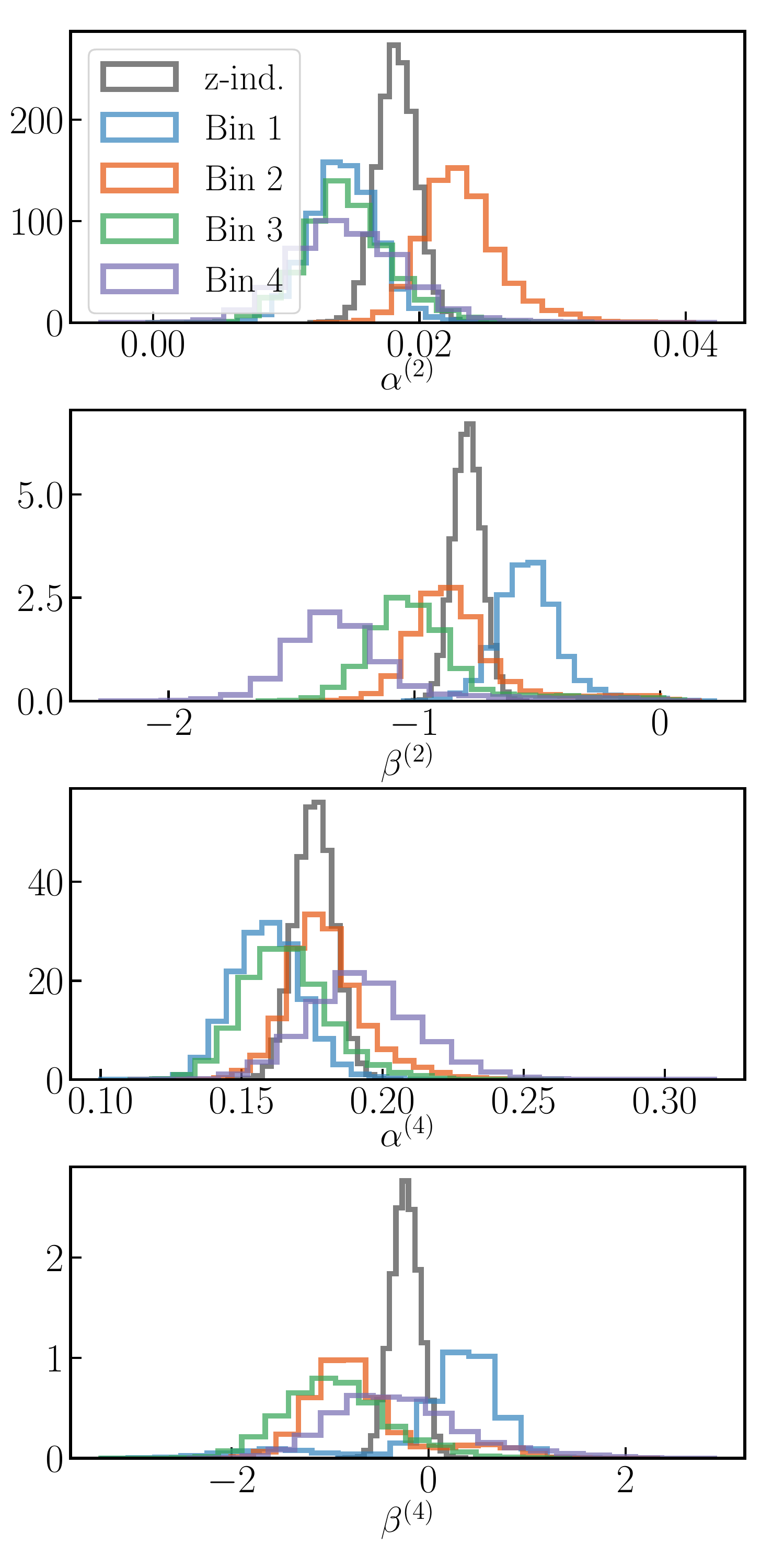}
\caption{ Marginalized 1D posterior distributions of the PSF systematics parameters for the redshift-dependent analysis. The parameters corresponding to different tomographic bins are color coded.  The differences in the distributions for different tomographic bins may be caused by the differences in galaxy property distributions and the resulting difference in sensitivity to PSF systematics. The posterior of the redshift-independent model, shown in grey, corresponds roughly to the average of the distributions of the redshift-dependent model.
}
    \label{fig:prior_redshift}
\end{figure}

\begin{figure}
    \centering
    \includegraphics[width=0.98\columnwidth]{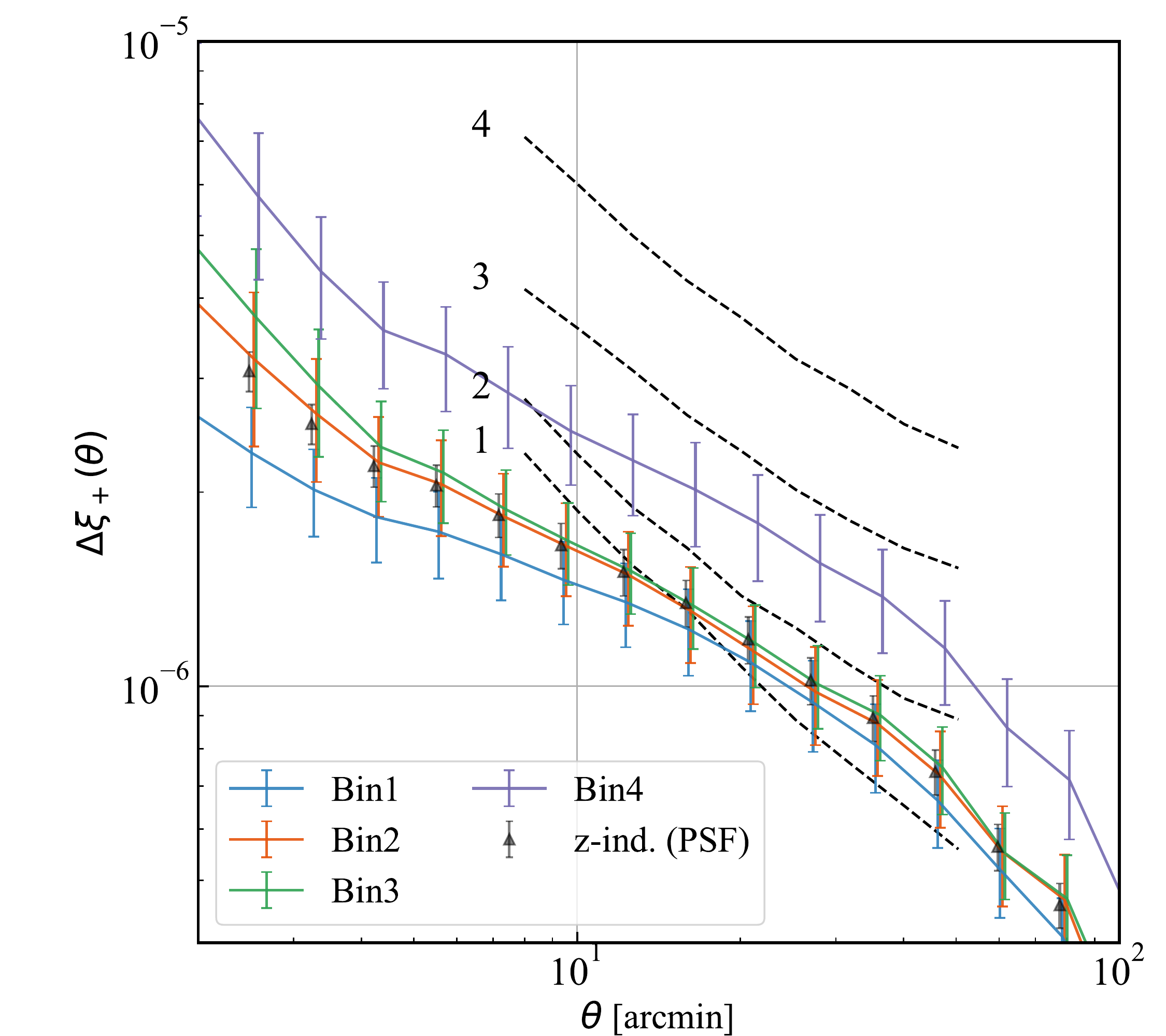}
\caption{ The additive bias on the auto-correlations of the cosmic shear 2PCF $\xi_+$ for the redshift-dependent models and redshift-independent model. We compare the $\Delta \xi_+(\theta)$ with the statistical uncertainty of $\xi_+(\theta)$. To avoid overcrowding, we only show the model fitted to the PSF stars. The black triangle line shows the $\Delta \xi_+(\theta)$ redshift-independent model fitted to PSF stars. We discuss the redshift dependency of the model in Section~\ref{sec:model:zbin}.
}
    \label{fig:delta_xip_redshift}
\end{figure}

\subsection{Redshift Dependency}
\label{sec:model:zbin}

In this section, we discuss the necessity of including redshift dependence in our PSF systematics model. In principle, a dependence on redshift could arise because the PSF leakage and modeling error parameters $[\alpha^{\rm (2)},\beta^{\rm (2)},\alpha^{\rm (4)}, \beta^{\rm (4)}]$ may depend on the ensemble galaxy properties, e.g., galaxy size, S\'ersic index distribution \citep{2022MNRAS.510.1978Z}, which vary across tomographic bins. In past work in DES \citep{2022PhRvD.105b3514A}, the redshift dependence of the PSF systematics model parameters was investigated for the second moments model. Although the overall level of PSF systematics in that work is small, the redshift dependence of the parameters was found to be statistically significant.

We investigated the redshift dependency of our model by joint fitting all the g-p correlations and average galaxy shape per bin by defining one set of parameters $\ve{p}^i = [\alpha^{\rm (2),i},\beta^{\rm (2),i},\alpha^{\rm (4),i}, \beta^{\rm (4),i}]$ for each tomographic bins, where $i$ stands for the tomographic bin index from 1-4.  The redshift-dependent data vector $\ve{D}_{gp}^z = [\langle \hat{g}^1_{\rm gal} e_{\rm PSF} \rangle, \dots, \langle \hat{g}^1_{\rm gal} \Delta M^{\rm (4)}_{\rm PSF} \rangle, \langle \hat{g}^2_{\rm gal} e_{\rm PSF} \rangle, \dots,\\ \langle \hat{g}^2_{\rm gal} \Delta M^{\rm (4)}_{\rm PSF} \rangle,\dots,  \langle \hat{g}^4_{\rm gal} \Delta M^{\rm (4)}_{\rm PSF} \rangle, \langle \hat{g}^1_{\rm gal} \rangle, \dots, \langle \hat{g}^4_{\rm gal} \rangle  ]$, which has a total length of $4\times4\times 20 + 4\times 2 = 328$. The parameter set $\ve{p}^z = [\alpha^{\rm (2), 1}, \dots, \beta^{\rm (4), 1},\alpha^{\rm (2), 2}, \dots, \beta^{\rm (4), 2}, \dots, \beta^{\rm (4), 4}  ]$, which has a total length of $16$. We call this the ``redshift-dependent fiducial model''.
We conducted the joint fitting rather than fitting the data separately in each tomographic bin to account for the covariance between the tomographic bins. In Fig.~\ref{fig:correlation}, we showed the correlation matrix of $\ve{D}_{gp}^z$ and found the correlation between the tomographic bins are significant. For the redshift-dependent model, we use angular scales from 2-200 arcmin, because including the smaller angular scales will result in the model fits to non-PSF stars failing the $p$-value test.
In comparison, we also fit a ``redshift-independent'' model to the same data vector $\ve{D}_{gp}^z$, by enforcing the PSF parameters to be the same across the 4 tomographic bins.

\begin{table*}
\centering
\begin{tabular}{llllllll}
\hline
Sample& Bin & $\alpha^{\rm (2),i}$ & $\beta^{\rm (2),i}$ & $\alpha^{\rm (4),i}$ & $\beta^{\rm (4),i}$ & p-value  \\ \hline
          & 1 & $0.014\pm0.002$ & $-0.5\pm0.1$ & $0.16\pm0.01$ & $0.1\pm0.7$\\
PSF stars & 2 & $0.023\pm0.002$ & $-0.8\pm0.2$ & $0.18\pm0.01$ & $-0.7\pm0.6$ & 0.91\\
          & 3 & $0.014\pm0.003$ & $-1.0\pm0.2$ & $0.17\pm0.02$ & $-0.9\pm0.5$\\
          & 4 & $0.014\pm0.004$ & $-1.3\pm0.3$ & $0.20\pm0.02$ & $-0.3\pm0.75$\\
          & all & $0.018\pm0.002$ & $-0.86\pm0.06$ & $0.176\pm0.007$ & $-0.2\pm0.1$ & 0.28 
\\\hline
              & 1 & $0.023\pm0.004$ & $-0.4\pm0.1$ & $0.16\pm0.013$ & $0.2\pm0.3$\\
non-PSF stars & 2 & $0.028\pm0.004$ & $-0.3\pm0.1$ & $0.16\pm0.013$ & $0.5\pm0.2$ & 0.63  \\
              & 3 & $0.018\pm0.005$ & $-0.0\pm0.1$ & $0.14\pm0.016$ & $0.9\pm0.3$\\
              & 4 & $0.020\pm0.007$ & $-0.1\pm0.2$ & $0.17\pm0.021$ & $1.4\pm0.3$\\
              & all & $0.022\pm0.002$ & $-0.17\pm0.06$ & $0.156\pm0.008$ & $0.57\pm0.1$ & 0.12 
\\\hline
\end{tabular}
\caption{The best-fitting parameters, $p$-values of the models fitted to the set of g-p correlation function across all tomographic bins. The first section shows the best-fitting parameters and the $p$-values using the PSF stars, while the second section shows the results for the non-PSF stars. The last line of each section shows the results for a  redshift-independent model that was fitted to the tomographic data vector. The results show a mild preference for the redshift-dependent model, but the redshift-independent model cannot be ruled out.  }
\label{tab:params_redshift}
\end{table*}

In Fig.~\ref{fig:gp_redshift}, we show the g-p correlations of the four tomographic bins of the HSC Y3 shape catalog \citep{HSC3-catalog}, and their best-fitting values according to the redshift-dependent fiducial model, using the PSF stars. The 1d marginal posteriors for the PSF parameters are shown in Fig.~\ref{fig:prior_redshift}.   The best-fitting parameters, $p$-values fitted using both PSF and non-PSF stars are listed in Table~\ref{tab:params_redshift}. We see a slight statistical significance in the redshift-dependency in the PSF parameters, especially with the decreasing trend of $\beta^{\rm (4)}$ with redshift. However, the redshift-independent model also has an acceptable $p$-value, while significantly decreasing the number of parameters needed to model PSF systematics, which is a practical issue of some importance.   For this reason, we will want to use mock cosmic shear analyses to quantitatively assess the model performance for the simpler model and determine whether it is acceptable, even if not statistically preferred.

The impact on the cosmic shear 2PCF in bin-$i$ and bin-$j$ predicted by the redshift-dependent model is
\begin{equation}
    \Delta \xi_+^{ij} = \sum^{4}_{k=1} \sum^{4}_{q=1} \ve{p}^i_k \ve{p}^j_q \langle \ve{S}_k \ve{S}_q \rangle
\end{equation}
In Fig.~\ref{fig:delta_xip_redshift}, we show the impact on the cosmic shear auto-correlation functions in tomographic bins due to the PSF systematics, fitted by PSF stars, comparing the redshift-dependent model (colored lines) versus the redshift-independent model (black \responsemnrassecond{circle}). We also show the statistical uncertainty of the HSC Y3 cosmic shear $\xi_+$, which is predicted by the covariance matrix used in Section~\ref{sec:cosmo:y3_ana}.
We also show the redshift-independent model fitted to the tomographic g-p correlations in black circles for PSF stars and black triangles for the non-PSF stars.
The $\Delta \xi_+^{ii}$ from bin 1 to 3 are statistically consistent with each other, but the bin 4 correlation is significantly higher than the others (in absolute value, not in its ratio to the cosmic shear signal).
For the non-PSF stars, the predictions for $\Delta \xi_+^{ii}$ increase gradually with redshift from bin 1 to 3, and likewise increase quite sharply for bin 4, probably due to the fact that bin 4 has the largest $\alpha^{\rm (4)}$ and $\beta^{\rm (2)}$. To avoid overcrowding the plot, we do not show the lines for non-PSF stars. 
The redshift-independent models predict equal $\Delta \xi_+$ for all tomographic bin-pairs.  Evaluating the model at its best-fitting parameters yields to a prediction for $\Delta \xi_+$ comparable to the amplitude of the redshift-dependent prediction from bin 1 to bin 3, while underestimating the $\Delta \xi_+$ for bin 4 by a factor of $\sim 2$.

Overall, the prediction of $\Delta \xi_+^{ii}$s by the redshift-dependent model are not statistically consistent with each other across tomographic bins. 
However, modeling the redshift dependence by assigning a separate set of parameters to each tomographic bin will significantly increase the number of PSF parameters from 4 to 16. While the redshift-independent model remains competitive in terms of $p$-value (see Table~\ref{tab:params_redshift}), we think the redshift-independent is still a potentially acceptable model of choice for the HSC Y3 analysis.
Whether modeling the redshift dependency is worth increasing the number of nuisance parameters by 12 should be determined based on the impact on the cosmological results, which is inferred in Section~\ref{sec:cosmo:y3_ana}.

An option to model the redshift dependence of the PSF systematics in shear without a drastic increase in the number of model parameters is to introduce parametrized models for the redshift dependence of selected  PSF parameters. For example, based on our results for HSC Y3 analysis, a reasonable choice might be to  model $\beta^{\rm (2)}(z^i)$ as $\beta^{\rm (2)}_0 f_1(z)$, where $f_1(z)$ is a simple single-parameter function of redshift. Another option is to subtract the mean redshift-dependent $\Delta \xi_+^{ij}$ the cosmic shear data vector, and model a few principal components of the uncertainty of the parameters. However, this approach relies on the assumption that the uncertainties of the PSF parameters are highly correlated with 
each other, so that a principal component analysis can be effective.

\section{Cosmological Impact}
\label{sec:cosmo:0}

In this section, we test the impact of the new PSF systematics model described in Section~\ref{sec:model:0} in cosmological analyses. In Section~\ref{sec:cosmo:y1_ana}, we present a re-analysis of  HSC Y1 cosmic shear, using the Y1 cosmic shear data vector, covariance matrix, and redshift distribution from \cite{2020PASJ...72...16H}. In Section~\ref{sec:cosmo:y3_ana}, we present a mock cosmological analysis for HSC Y3 cosmic shear using a noiseless mock data vector and covariance, and the galaxy-PSF correlations from the real HSC Y3 star and shape catalogs described in Section~\ref{sec:star:selection} and Section~\ref{sec:shape:0}.

\subsection{HSC Y1 Re-analysis}
\label{sec:cosmo:y1_ana}

\begin{figure}
    \centering
    \includegraphics[width=0.98\columnwidth]{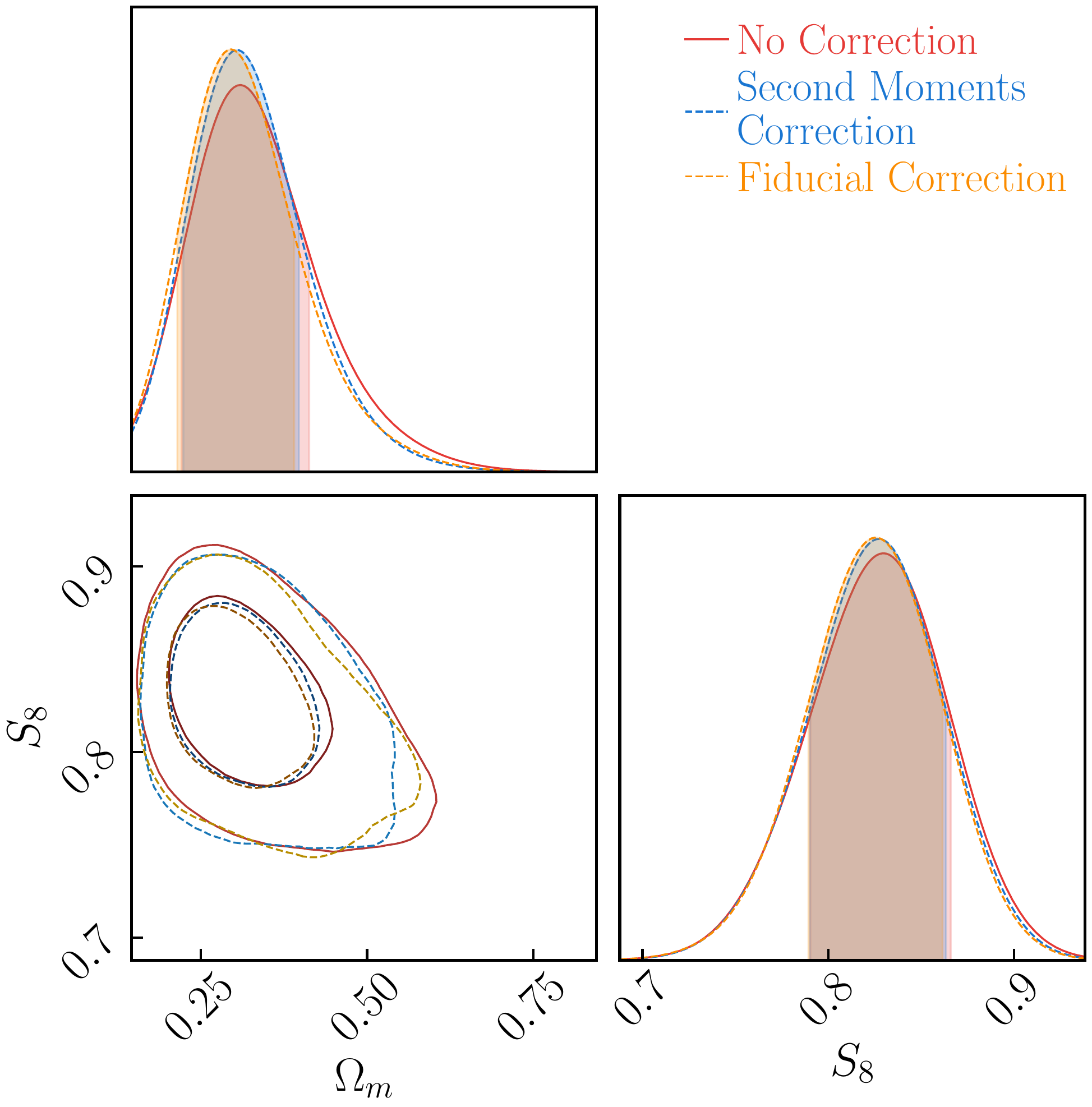}
    \includegraphics[width=0.98\columnwidth]{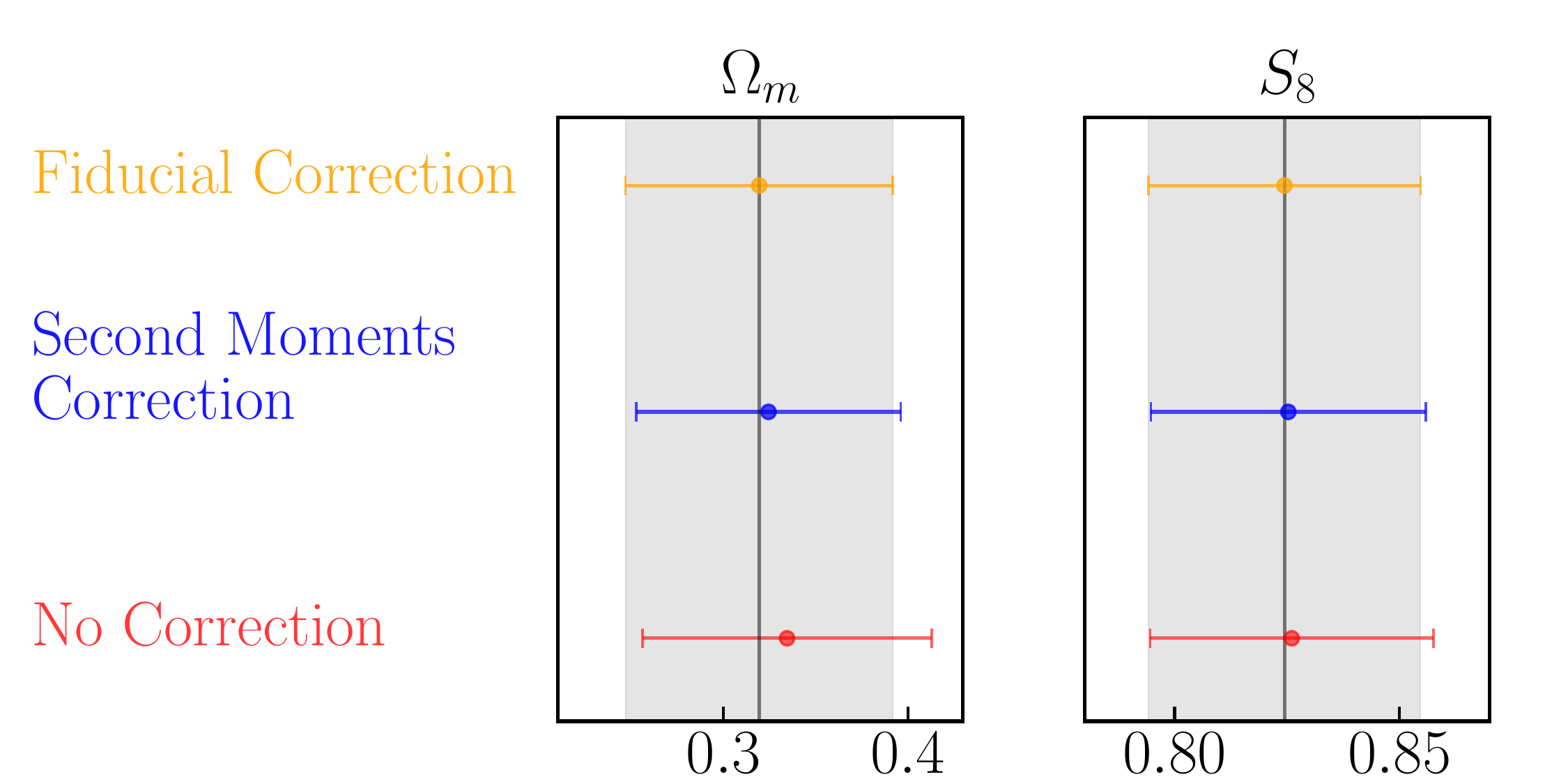}
\caption{ The $\Omega_m$-$S_8$ constraints of the HSC Y1 cosmic shear re-analysis. The upper panel shows the 2d contours of the $68\%$ and $95\%$ confidence interval 
and the 1d marginal posterior distributions, while the lower panel shows the 1d marginalized $1\sigma$ errorbars. The vertical lines are the mean values of the posterior of the fiducial correction method, while, the shaded areas indicate the marginalized $1\sigma$ errorbars of the fiducial correction. Compared to the case of no correction for PSF systematics, the fiducial model correction shift the mean $S_8$ by 0.2$\sigma$. However, the correction based on only PSF second moments shifts $\Omega_m$ by $0.05\sigma$, resulting a $0.15\sigma$ bias on $\Omega_m$ compared to our fiducial model.  The impact on $S_8$ is more modest.
}
    \label{fig:y1_contour}
\end{figure}

\begin{table}\centering
\begin{tabular}{lllll}
\hline
Parameter            & Fiducial & Prior (Y1)  & Prior (Y3)          \\ \hline
$\log(A_s\times 10^9)$                & $0.322$    & $U[-1.5,2.0]$   & $U[-1.5,2.0]$            \\
$\Omega_b$           & $0.0489$ & $U[0.038, 0.053]$   & $U[0.038, 0.053]$     \\
$n_s$                & $0.967$  & $U[0.87,1.07]$     & $U[0.87,1.07]$    \\
$h_0$                & $0.677$  & $U[0.64,0.82]$    & $U[0.64,0.82]$     \\
$\Omega_m$           & $0.311$  & $U[0.039, 0.953]$  & $U[0.039, 0.953]$    \\
$\tau$               & $0.0561$   & const.     & const.       \\
$\Omega_\nu$         & $0.06$    & const.     & const.          \\
$w$                  & $-1.0$   & const.    & const.      \\
$w_a$                & $0.0$    & const.  & const.     \\\hline
$A_{\text{IA}}$      & $1.0$  & $U[-5,5]$  & $U[-5,5]$      \\
$\eta$             & $0.0$  & $U[-5,5]$   & $U[-5,5]$       \\
$z_0$                & $0.62$   & const.    & const.       \\\hline
$m^1$                & $0.0$   & $\mathcal{N}(0.0086,0.01)$  & const. \\
$m^2$                & $0.0$   & $\mathcal{N}(0.0099,0.01)$  & const. \\
$m^3$                & $0.0$   & $\mathcal{N}(0.0241,0.01)$  & const. \\
$m^4$                & $0.0$   & $\mathcal{N}(0.0391,0.01)$  & const. \\
$\Delta z^{1}$       & $0.0$ & $\mathcal{N}(0,0.0374)$    & $\mathcal{N}(0,0.012)$    \\
$\Delta z^{2}$       & $0.0$  & $\mathcal{N}(0,0.0124)$    & $\mathcal{N}(0,0.01)$      \\
$\Delta z^{3}$       & $0.0$  & $\mathcal{N}(0,0.0326)$   & $\mathcal{N}(0,0.018)$      \\
$\Delta z^{4}$       & $0.0$  & $\mathcal{N}(0,0.0343)$     & $\mathcal{N}(0,0.021)$   \\\hline
\end{tabular}
\caption{The fiducial parameter values used to generate the mock data vector for the HSC Y3 cosmic shear mock analysis (described in Section~\ref{sec:cosmo:y3_ana}), and priors for both the HSC Y1 re-analysis (described in Section~\ref{sec:cosmo:y1_ana}) and Y3 mock analysis. $U[a,b]$ indicates a uniform distribution from $a$ to $b$, while  $\mathcal{N}(\mu,\sigma)$ indicates a Gaussian distribution with mean $\mu$ and standard deviation $\sigma$.
}
\label{tab:cosmology_parameters}
\end{table}

For the re-analysis of HSC Y1 cosmic shear, we adopted the cosmic shear 2PCF $\ve{\xi}_{\pm}$, its covariance matrix $\mt{\Sigma_{\rm Y1}}$, and the redshift distribution of the four tomographic bins from \cite{2020PASJ...72...16H}\footnote{\url{http://th.nao.ac.jp/MEMBER/hamanatk/HSC16aCSTPCFbugfix/index.html}}. We built the forward model for the data vectors, including cosmological and astrophysical modeling choices, in \textsc{CosmoSIS} \citep{2015A&C....12...45Z}.

The choices of the cosmological model and the priors on the parameters 
are made to be as close as possible to those of \cite{2020PASJ...72...16H}. The only difference is that we marginalize over the multiplicative bias for each tomographic bin, instead of using one nuisance parameter for $m$. Here we briefly review the settings. We used \textsc{CAMB} \citep{2000ApJ...538..473L, Lewis_2002, Howlett_2012} to compute the linear matter power spectrum, and \textsc{halofit} \citep{2012ApJ...761..152T} to compute the non-linear matter power spectrum. The optical depth $\tau$ was set to $0.0561$, and neutrino mass was set to $0.06 {\rm eV}$. The priors on the cosmological parameters are listed in the first section of Table~\ref{tab:cosmology_parameters}.

Regarding the astrophysical and nuisance parameters of the re-analysis, we use the non-linear alignment model \citep[NLA][]{Hirata_2004, Bridle_2007} to model the intrinsic alignments \citep[see][for the specification of the model]{2017MNRAS.470.2100K}. The prior on the NLA parameters $A_{\text{IA}}$, $\eta$ and $z_0$ are listed in the second section of Table~\ref{tab:cosmology_parameters}. The priors on the multiplicative biases $m^1$--$m^4$ and the redshift uncertainty parameters $\Delta z^{1}$--$\Delta z^{4}$ are listed in the third section of Table~\ref{tab:cosmology_parameters}. We use the same redshift distribution, astrophysical and systematics models and priors as \cite{2020PASJ...72...16H}.

We validate our forward modeling inference and model choices by comparing the cosmological parameter results when applying the same PSF systematics model as in \cite{2020PASJ...72...16H}. In \cite{2020PASJ...72...16H}, the fiducial model, which used the second-moment-only PSF systematics, results in the $68\%$ confidence intervals\footnote{Slightly updated from the original version in an erratum, \citet{2022PASJ...74..488H}.} of $0.237< \Omega_m<0.383$ and $0.795<S_8<0.855$. Our second moment model reports the $68\%$ confidence interval of $0.253 <\Omega_m < 0.394$ and $0.795 < S_8 < 0.855$. There is a very small offset ($\sim$0.1$\sigma$) on our $\Omega_m$ confidence interval, and the $S_8$ interval matches perfectly. We therefore conclude that our forward model is validated for the purpose of comparing the PSF systematics model.

In the re-analysis, we compared the original and our fiducial model PSF models for marginalizing the PSF systematics. We tested the model in \cite{2020PASJ...72...16H} by adopting its prior and p-p correlations ($\xi^{pp}$, $\xi^{pq}$, and $\xi^{qq}$). We use our fiducial model to determine another set of priors for $\alpha^{\rm (2)}$, $\beta^{\rm (2)}$, $\alpha^{\rm (4)}$, and  $\beta^{\rm (4)}$, using the HSC Y1 high-SNR star catalog described in \cite{2022arXiv220507892Z}.   Both models lack a constant term. The priors used for both models are listed in Table~\ref{tab:y1_psf_prior}. In addition, for the sake of comparison, we run another analysis with no correction for PSF systematics in shear.

\begin{table}
\centering
\begin{tabular}{lll}
\hline
Param.    & Original  & Fiducial           \\ \hline
$\alpha^{\rm (2)}$ & $\mathcal{N}(0.015,0.05)$  & $\mathcal{N}(0.035,0.05)$\\
$\beta^{\rm (2)}$ & $\mathcal{N}(-0.7,0.6)$  & $\mathcal{N}(-0.67,0.05)$\\
$\alpha^{\rm (4)}$ & $\mathcal{N}(0,0)$ & $\mathcal{N}(0.17,0.02)$\\
$\beta^{\rm (4)}$ & $\mathcal{N}(0,0)$& $\mathcal{N}(-0.32,0.10)$ \\\hline
\end{tabular}
\caption{The prior on the PSF parameters for the HSC Y1 re-analysis. The ``Original'' column presents the priors adopted in the original HSC Y1 cosmic shear analysis \protect\citep{2020PASJ...72...16H}; when fitting with these priors, we also used the p-p correlations from that work. The `Fiducial' column presents the priors on our extended PSF systematics model, which was applied to the p-p and p-q correlations for the HSC Y1 high-SNR star sample.}
\label{tab:y1_psf_prior}
\end{table}

In Fig.~\ref{fig:y1_contour}, we show the 2d contour and 1d errorbars in the $\Omega_m$-$S_8$ plane for the HSC Y1 cosmic shear re-analysis. Our fiducial model reports $\Omega_m = 0.319^{+0.072}_{-0.071}$ and $S_8 = 0.824^{+0.030}_{-0.029}$. The analysis without any correction for PSF systematics shows that $\Omega_m$ would have been biased by $0.2\sigma$ if the PSF systematics is not modelled at all. The analysis using the PSF second moment-based model was able to remove $0.13\sigma$ from the bias, leaving $0.07\sigma$ uncorrected.
We use the effective number of parameters defined in \cite{2019PhRvD..99d3506R}
\begin{equation}
n_{\rm eff} = 2 {\rm ln} \mathcal{L}(\theta_p) - 2\langle{\rm ln}\mathcal{L}\rangle_{\theta},
\end{equation}
where $\mathcal{L}(\theta_p)$ is the posterior of the mean parameter $\theta_p$, and $\langle{\rm ln}\mathcal{L}\rangle_{\theta}$ is the average posterior over the parameter space $\theta$.
The $\chi^2$ values of the ``no correction'', second moment correction, and fiducial model correction are $160.3, 156.9, 143.7$, respectively. The effective degrees of freedom, $170-n_{\rm eff}$, are $159.5,159.4,156.3$, respectively.
The $p$-values are $0.47, 0.54, 0.76$, respectively -- meaning that all models are nominally acceptable, presumably because the PSF systematics in shear are only a small contributor to the data vector that is being fit. Still, fiducial model obtains a substantially better fit while only increasing the number of parameters by $\sim2$.

\subsection{HSC Y3 mock analysis}
\label{sec:cosmo:y3_ana}

\begin{figure}
    \centering
    \includegraphics[width=0.98\columnwidth]{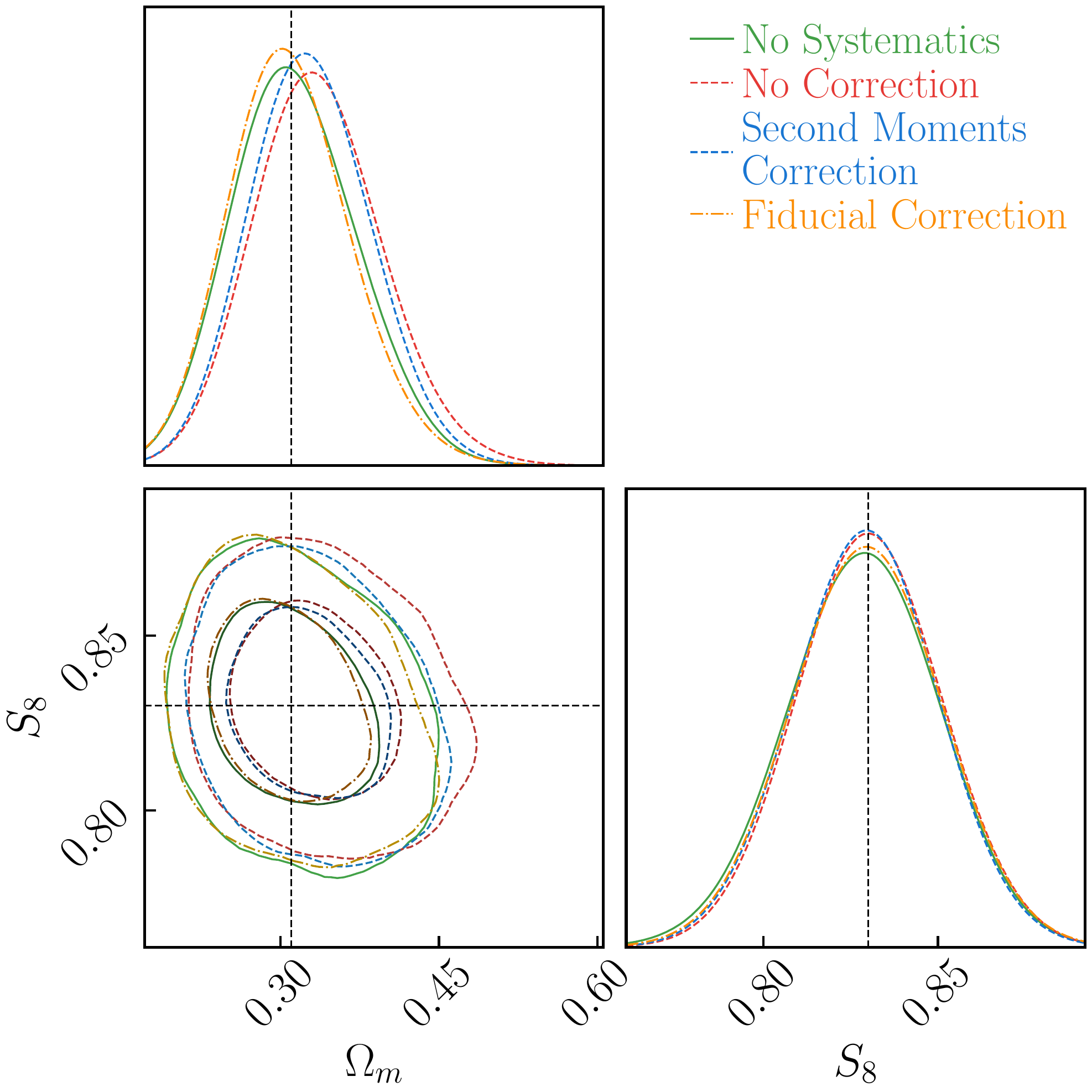}
    \includegraphics[width=0.98\columnwidth]{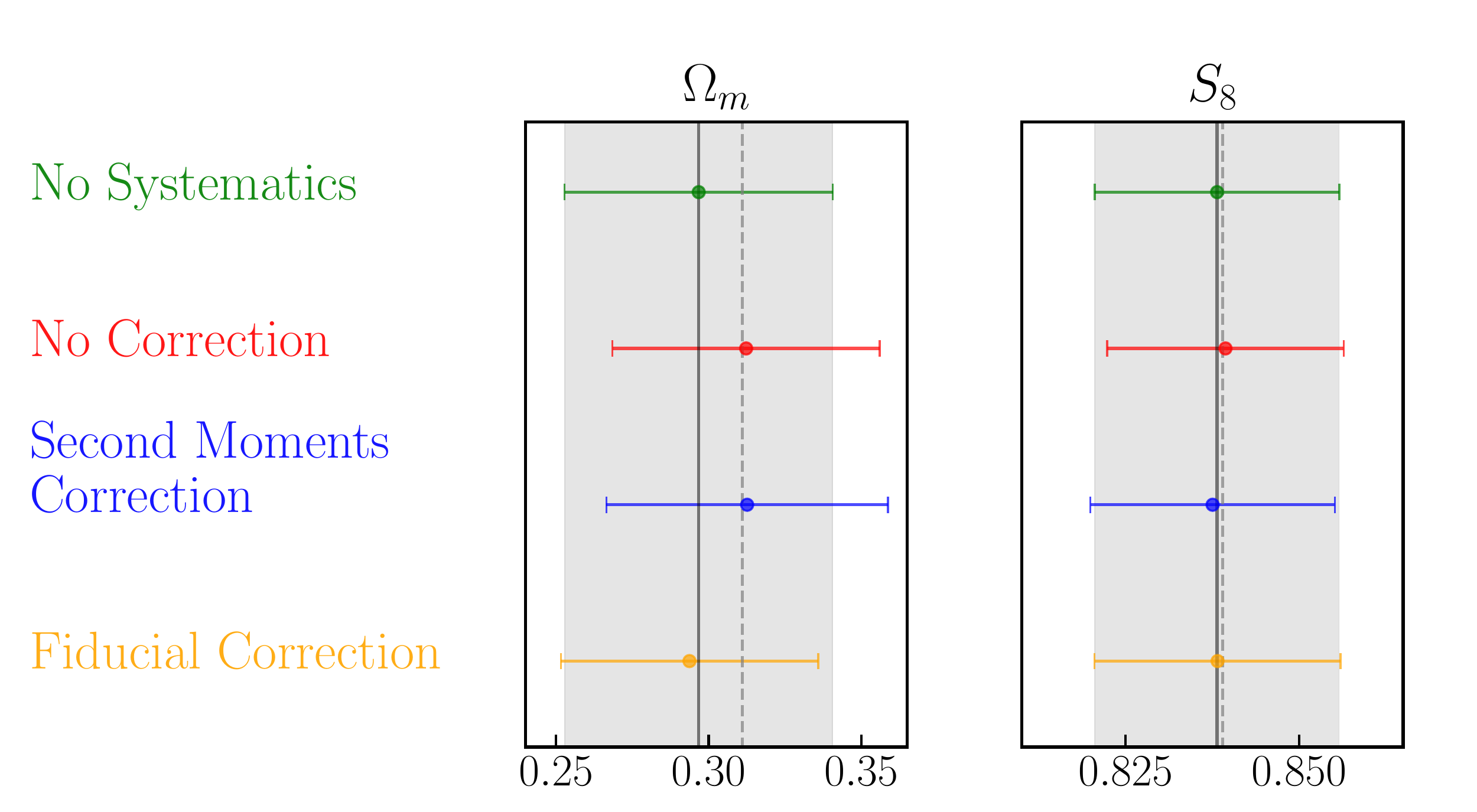}
\caption{ The $\Omega_m$-$S_8$ constraints of the HSC Y3 mock cosmic shear analysis.
The upper panel shows the 2d contours  of the $68\%$ and $95\%$ confidence interval and the 1d marginal posterior distributions, while the lower panel shows the 1d marginalized $1\sigma$ errorbars. The dashed lines show the true input cosmological parameters. The green line shows the results of analyzing the data vector with no PSF systematics added, as a baseline. The vertical lines indicate the mean values of the posterior for the fiducial correction analysis, while the shaded areas indicate the $1\sigma$ errorbar when applying the fiducial correction.
Due the skewness of the $\Omega_m$ posterior distribution, the mean $\Omega_m$ value in the ``No Systematics'' case is biased low. The red shows the worst-case scenario, where a realistic level of redshift-dependent PSF systematics are added but no attempt is made to correct for them. The blue shows the results of analysis using the redshift-independent second moments-only PSF systematics model, and the orange shows the results of analysis using the fiducial redshift-independent PSF systematics model. We see that the second moment-only model provides very similar results to applying no correction at all. The fiducial model is more successful at correcting the PSF systematics.   }
    \label{fig:y3_contour}
\end{figure}

\begin{table}
\centering
\begin{tabular}{llll}
\hline
Param.     & Second moment & Fiducial           \\ \hline
$\alpha^{\rm (2)}$  & $\mathcal{N}(-0.007,0.002)$ & $\mathcal{N}(0.016,0.002)$\\
$\beta^{\rm (2)}$   & $\mathcal{N}(-0.85,0.05)$ & $\mathcal{N}(-0.91,0.05)$\\
$\alpha^{\rm (4)}$  & $\mathcal{N}(0,0)$ & $\mathcal{N}(0.17,0.01)$\\
$\beta^{\rm (4)}$  & $\mathcal{N}(0,0)$ & $\mathcal{N}(-0.6,0.2)$ \\\hline
\end{tabular}
\caption{The priors on the PSF systematics model parameters for the HSC Y3 mock analysis. ``Second moment'' model only fits the second moments galaxy-PSF correlations, setting the fourth moment parameters to zero. The fiducial model fits all the galaxy-PSF correlations with both second and fourth moments leakage and modeling error.  }
\label{tab:y3_psf_prior}
\end{table}

To study the impact of the PSF systematics modeling on the HSC Y3 cosmic shear analysis, we conducted a mock analysis that mimics the analysis scenario. The noise-free cosmic shear data vector is generated using the Planck cosmological parameters from \cite{2020A&A...641A...6P}
and astrophysical values listed in Table~\ref{tab:cosmology_parameters}, without PSF contamination. We refer to this parameter set the ``fiducial cosmology''.  Then, mock PSF systematics are generated using the best-fitting parameters of the redshift-dependent fiducial model, described in Section~\ref{sec:model:zbin},
\begin{equation}
\label{eq:delta_xi_mock}
\Delta \xi^{uv}_+ = \sum_{i=1}^4 \sum_{j=1}^4 \ve{p}^{u}_i \ve{p}^{v}_j \langle S_i S_j\rangle,
\end{equation}
where $\ve{p}^{u}_i$ takes the best-fitting values in Table~\ref{tab:params_redshift}. This PSF contamination term is added to the original noise-free data vector to generate a Y3 mock data vector. The priors on the cosmological, astrophysical, and nuisance parameters are listed in Table~\ref{tab:cosmology_parameters}. The priors on the Y3 cosmological and astrophysical parameters are set to the same ranges as for Y1. The multiplicative biases are set to 0, while the photometric redshift uncertainty parameters take the Gaussian priors given in \cite{2022arXiv220610169Z}.
We use the same scale cuts as the Y1 analysis, i.e., $7$--$56$~arcmin for $\xi_+$ and $28$--$178$~arcmin for $\xi_-$. 
The covariance matrix is estimated using $\mt{\Sigma}_{\rm Y3} = \mt{\Sigma}_{\rm Y1}/3$ to approximately account for the increase in survey area, while neglecting changes due to differences in survey edge effects. We use the redshift distributions and their priors estimated in \cite{2022arXiv221116516R}, for which the marginalization method was validated in \cite{2022arXiv220610169Z}. Although the redshift distributions and their uncertainties are estimated in an earlier version of \cite{2022arXiv221116516R}, and are likely to be slightly different in the actual Y3 analysis, they do not significantly impact our conclusion here. \responsemnras{In the HSC Y3 cosmic shear analyses \citep{2023arXiv230400701D,2023arXiv230400702L}, the scale cuts, covariance and some modeling choices are slightly different from the choice used for this test. However, they carried out the same mock analysis as was done here and found the same conclusion regarding the choice of the PSF systematics model. }

In Fig.~\ref{fig:y3_contour}, we compare the results of using two different PSF systematics models in the Y3 mock analysis. The second moment correction model (in blue) only fits the Eqs.~\eqref{eq:null1} and~\eqref{eq:null2} using two free PSF parameters ($\alpha^{\rm (2)}$ and $\beta^{\rm (2)}$), setting the other parameters in those equations to 0. The fiducial model (in orange) uses the first four PSF parameters in $\ve{p}$, setting $e_c$ to zero, and fits all of Eqs.~\eqref{eq:null1}--\eqref{eq:null4}. The priors on the PSF parameters, which are determined by carrying out our fitting process on the HSC Y3 shear and star catalogs, are listed in Table~\ref{tab:y3_psf_prior}. Both models use the PSF stars for determining the prior and p-p correlations, as it is the better-understood sample of this work, with a larger sample size.
In addition, we include the following two analyses: one with no PSF systematics added to the cosmic shear data vector  and no attempt at PSF systematics correction, as a baseline; and one with PSF systematics added to the cosmic shear data vector, but with no attempt at correction, as the worst-case scenario. The input values of $\Omega_m$ and $S_8$ are shown as the dashed lines. The mean parameters of the analysis with no PSF systematics added to the data vector are shown in the solid vertical lines. The mean value of $\Omega_m$ in the ``No Systematics'' case is biased low compared to the true input value, even though this constraint is meant to be bias-free. We attribute this difference to the ``projection effect'' of the non-Gaussian posterior \citep[e.g., see Section~IV of ][]{2022PhRvD.106d3520P}.

\responsemnras{To fully account for the uncertainty in the PSF systematics parameters, the fiducial model in this test accounts for the correlation between those parameters, by assuming the prior to be a 4D multivariate Gaussian. The details of modeling the correlated prior on PSF parameters are described in Appendix~\ref{sec:ap:correlation}. We find no significant difference between using an uncorrelated versus correlated prior. But for the sake of fully propagating the PSF systematic uncertainties, we recommend that the HSC Y3 analysis should use the correlated prior for the PSF parameters. }

We can see that the second moments-only model barely corrected for the PSF systematics in shear, because it missed the leakage from the PSF fourth moments. The fiducial model comes closer to the baseline (``No Systematics''), although the correction overshoots the truth for $\Omega_m$.
This imperfect correction is likely because the fiducial model does not consider the redshift dependency in the real contamination.
Compared to the ``No Systematics'' run, the PSF contamination causes a $+0.36\sigma$ bias on $\Omega_m$, which the second moments correction does not remove; and the fiducial model over-corrects, resulting in a $-0.06\sigma$ bias. For $S_8$, these effects are smaller: PSF systematics cause a bias of $+0.06\sigma$, while the second moment model overcorrects, resulting in a bias of $-0.03\sigma$, and the fiducial almost perfectly corrects the bias on $S_8$.

Regarding the errorbar size, the choice between the models shown here only affects the errorbars at the few-percent level, so this is not a significant factor in model selection.

We did not use the non-PSF stars to determine the contamination in this mock analysis, since the PSF stars provide better statistics for the p-p correlation functions. In a real analysis, if one uses the non-PSF stars to determine the prior and p-p correlation, the correction made by the second moment-only model will be even smaller than it was here, since $\Delta \xi_+$ is smaller for the non-PSF stars with the second moments model. For the fiducial model, we do not expect the cosmological results to change by much because the predicted $\Delta \xi_+$ for the PSF vs.\ non-PSF stars are similar for the fiducial model, shown in Fig.~\ref{fig:gp_corr}, due at least in part to the dominance of leakage rather than PSF modeling error.

\section{Summary of Methodology}
\label{sec:recipe:0}

In this section, we summarize the process of building and selecting a PSF systematics model for a given cosmic shear survey, while we developed and tested this model with HSC Y3 data. This is a general approach that we recommend for any weak lensing survey, rather than being HSC-specific.

\begin{enumerate}
\item Build a star catalog with measured and residual moment measurements from second to higher moments, as described in Section~\ref{sec:star:moments}. Care should be taken to ensure the purity of this sample, along with adoption of flag cuts and measures to avoid moment contamination due to blending in the images.

\item Derive the true and residual spin-$2$ combinations of those moments, as described in Section~\ref{sec:star:spin2}. (These can either be first order spin-$2$ quantities, or second order spin-$2$ quantities such as spin-$0 \times$ spin-$2$, described in Section~\ref{sec:ap:sot}.)

\item Cross-correlate the  spin-$2$ quantities in the star catalogs with the shear catalog, and conduct a likelihood analysis, including the following steps:
\begin{enumerate}
    \item Estimate covariances through some method that includes relevant sources of uncertainty, including cosmic variance in the shears, and systematic variations in PSF properties across the sky.
    \item Build the systematics model by assigning a parameter to each PSF spin-$2$ quantity, as explained in Section~\ref{sec:model:formalism}.
    \item Define sub-models can be defined by putting very constraining priors on the parent model (Section~\ref{sec:model:define}).
\end{enumerate}
All models should be applied to the same set of galaxy-PSF cross correlations, as described in Section~\ref{sec:model:gp_corr}.

\item Define statistical criteria to distinguish the models. \responsemnras{The preferred model should capture all of the additive systematic contamination to $\Delta \xi_+$ that is significant compared to their statistical uncertainty. This implies that, if a more complex model only changes the inferred $\Delta \xi_+$ insignificantly compared to the error budget, the simpler model should be selected.}  This is described in Section~\ref{sec:model:cosmology}.

\item Test the robustness of the fiducial model by complicating it. These tests include:
\begin{enumerate}
\item testing and understanding the consistency between PSF and non-PSF stars (Section~\ref{sec:model:psf_vs_nonpsf})
\item the redshift dependency of the model (Section~\ref{sec:model:zbin})
\item other spin-$2$ quantities (Section~\ref{sec:ap:six} and Section~\ref{sec:ap:sot})
\item impact on $\xi_-$ (Section~\ref{sec:ap:xim}).
\end{enumerate}

\item Conduct a mock cosmological analysis and confirm that the fiducial PSF systematics model can correct the bias to a level that satisfies the requirement of the given survey (Section~\ref{sec:cosmo:y3_ana}).
\end{enumerate}

\section{Conclusions}
\label{sec:conclu}

The overall goal of this paper was to provide a general framework for describing additive weak lensing shear systematics due to the impact of PSF leakage and modeling error on inferred weak lensing shears.  To do so,
we defined a key concept underlying the PSF contamination in cosmic shear: this contamination is driven by spin-2 combinations of PSF moments  (Section~\ref{sec:star:spin2}). In addition to PSF second moments, all even moments, e.g., fourth moments, contribute to PSF spin-2 quantities.
The overall outline of our method is summarized in Section~\ref{sec:recipe:0}. To apply our method in a real-world scenario, we
generated an HSC Y3 star catalog with higher moment measurement of the PSF and its modeling residuals, applying cuts to avoid contamination by galaxies and provide valid PSF and non-PSF star samples. We compared the moment residuals of the PSF and non-PSF stars, and concluded that the PSF model is overfitted for the HSC Y3 catalog.

Next, we defined a full PSF systematics model that considers PSF spin-2 quantity leakage from second and fourth moments, along with a constant shear systematics term (Section~\ref{sec:model:null}).   Using the HSC Y3 galaxy and mock catalogs (Section~\ref{sec:shape:0}) and the HSC Y3 star catalog with measurements of higher moments measurement (Section~\ref{sec:star:0}), we quantified the level of PSF contamination in cosmic shear data vector in Section~\ref{sec:model:null} using that model. The full model can be considered to have nested models, each of which has a subset of the full model parameters set to zero. Our statistical metrics showed that a constant is not necessary in our particular case, but the second and fourth moments leakage and modeling errors are all impactful for cosmic shear. Therefore, our recommended fiducial model for the PSF systematics for HSC Y3 cosmic shear is a four parameters formula (Eq.~\eqref{eq:psf-sys-full} with $e_c = 0$).

In addition to the direct leakage and modeling error of the PSF fourth moments, we also investigated other possible contamination terms to $\xi_\pm$ from the PSF. These additional tests include the redshift dependency of the PSF contamination (Section~\ref{sec:model:zbin}), contamination to $\xi_-$ (Appendix~\ref{sec:ap:xim}), contamination caused by moments higher than the fourth order (Appendix~\ref{sec:ap:six}), and contamination caused by the second-order systematics (Appendix~\ref{sec:ap:sot}). These effects and additional contamination from the PSF are demonstrated to be subdominant in HSC Y3. Therefore, we do not recommend directly modeling them in the HSC Y3 cosmic shear analyses. However, we suggest that future surveys with different shear estimation methods and PSF modeling algorithms check for the importance of these effects, in case they become a significant contribution in a different setting.

Last but not least, we conducted a cosmological analysis to assess the impact of PSF systematics model selection on the cosmological results. We conducted a re-analysis on the HSC Y1 cosmic shear using our fiducial PSF systematics model, and obtain an cosmological results of $\Omega_m = 0.319^{+0.072}_{-0.071}$ and $S_8 = 0.824^{+0.030}_{-0.029}$. Both parameter are shifted from the original mean posterior by $<0.1\sigma$.
We produced a Y3-like mock data vector with redshift-dependent PSF systematics. This introduce a $+0.36\sigma$ bias on $\Omega_m$, and $+0.06\sigma$ bias on $S_8$. After the correction by the fiducial PSF model, the bias is $-0.06\sigma$ bias on $\Omega_m$, and no bias on $S_8$, which means fiducial model is sufficient for HSC Y3. The second moment model is insufficient because the bias on $\Omega_m$ is $+0.36\sigma$ after correction.

There are several caveats in this work that are worth mentioning:  (a) The cosmological mock analysis of the HSC Y3, which drives some of our conclusions, includes simple assumptions about the model for cosmological parameters and astrophysical systematics, redshift distributions, covariance, and scale cuts in relation to those from HSC Y1. These assumptions may not hold in the real Y3 analysis, though we do not think it will impact the overall conclusion. (b) In our prior for PSF systematics parameters for the mock and re-analysis, we do not consider correlations between the parameters. We leave such development to future work. (c) We do not consider a redshift-dependent PSF systematics model in the cosmological analysis, which might explain the imperfect correction made by the fiducial model in the mock analysis. We leave this implementation and its testing to future work.  Note that these features do not affect the framework for modeling additive shear systematics that we have developed, and are simply limitations of how we applied it to HSC Y3.

This work motivates a few future studies: (a) This motivates other ongoing (DES, KiDS) and future (LSST, Roman, Euclid) weak lensing surveys to investigate the potential contamination by the PSF higher moments. \responsemnras{As the survey area grows larger with the next generation of photometric surveys, the statistical uncertainties of both the shear-shear 2PCF (the error budget for the cosmological measurement) and the PSF-PSF/galaxy-PSF 2PCF (the detectability of systematics) will go down with the area, so the method will remain powerful for detecting the spin-2 leakage in shear signal. The depth increase of the Stage-IV surveys over the current surveys will increase the galaxy number density faster than the star number density, which can benefit this framework, as the uncertainty in the galaxy-PSF correlation functions for HSC Y3 is limited by the shape noise and cosmic variance.} (b) Although we found that the PSF model of HSC Y3 is overfitting the PSF, we did not account for it by using non-PSF stars only, because there were too few of them to enable a study with reasonable uncertainties. A self-consistent selection of PSF stars in all exposures in the future data release will slightly increase the fraction of non-PSF stars in the catalog. (c) An extensive study of the PSF leakage from different shear estimation methods will be of interest for future weak lensing surveys, e.g., LSST and Euclid. Furthermore, a list of typical values of $\alpha^{\rm (2)}$, $\beta^{\rm (2)}$, $\alpha^{\rm (4)}$, and $\beta^{\rm (4)}$ will help translating the requirements on additive shear biases to the requirement on the image processing pipeline and PSF models, which is normally developed at an earlier stage of the survey, to increase the chances of meeting the ever more stringent requirement on shear systematics.

A final lesson learned from this work is that a systematic approach to null testing, including reliable uncertainty estimates, is a really important part of the validation for weak lensing analysis. The leading contributor to the PSF systematics in our case -- the fourth moment leakage -- was not previously considered as a potentially significant factor until the results of this work. With that said, any factor characterized as ``minor'' in this study, whether it is $\Delta \xi_-$, sixth moments, redshift dependency of the PSF contamination, or second-order spin-2 terms, could become a leading factor in a specific setting and silently bias the cosmological results. Therefore, the main future work that this work motivates is a comprehensive set of null testings that is used to make principled decisions about the model for PSF systematics in cosmological weak lensing analyses in any surveys.

\subsection*{Acknowledgments}

\responsemnras{We thank the referee for their helpful feedback on this paper. We thank Mike Jarvis for the helpful comments and discussion. } 

TZ, XL and RM are supported in part by the Department of Energy grant DE-SC0010118 and in part by a grant from the Simons Foundation (Simons Investigator in Astrophysics, Award ID 620789). 
RD acknowledges support from the NSF Graduate Research Fellowship Program under Grant No.\ DGE-2039656. Any opinions, findings, and conclusions or recommendations expressed in this material are those of the authors and do not necessarily reflect the views of the National Science Foundation. This work was supported in part by JSPS KAKENHI Grant Numbers 21J10314. 
SS is supported by International Graduate Program for Excellence in Earth-Space Science (IGPEES), World-leading Innovative Graduate Study (WINGS) Program, the University of Tokyo.
HM is supported by JSPS KAKENHI Grant Number JP20H01932.

The Hyper Suprime-Cam (HSC) collaboration includes the astronomical communities of Japan and Taiwan, and Princeton University. The HSC instrumentation and software were developed by the National Astronomical Observatory of Japan (NAOJ), the Kavli Institute for the Physics and Mathematics of the Universe (Kavli IPMU), the University of Tokyo, the High Energy Accelerator Research Organization (KEK), the Academia Sinica Institute for Astronomy and Astrophysics in Taiwan (ASIAA), and Princeton University. Funding was contributed by the FIRST program from the Japanese Cabinet Office, the Ministry of Education, Culture, Sports, Science and Technology (MEXT), the Japan Society for the Promotion of Science (JSPS), Japan Science and Technology Agency (JST), the Toray Science Foundation, NAOJ, Kavli IPMU, KEK, ASIAA, and Princeton University.

This paper makes use of software developed for Vera C. Rubin Observatory. We thank the Rubin Observatory for making their code available as free software at http://pipelines.lsst.io/.

This paper is based on data collected at the Subaru Telescope and retrieved from the HSC data archive system, which is operated by the Subaru Telescope and Astronomy Data Center (ADC) at NAOJ. Data analysis was in part carried out with the cooperation of Center for Computational Astrophysics (CfCA), NAOJ. We are honored and grateful for the opportunity of observing the Universe from Maunakea, which has the cultural, historical and natural significance in Hawaii.

The Pan-STARRS1 Surveys (PS1) and the PS1 public science archive have been made possible through contributions by the Institute for Astronomy, the University of Hawaii, the Pan-STARRS Project Office, the Max Planck Society and its participating institutes, the Max Planck Institute for Astronomy, Heidelberg, and the Max Planck Institute for Extraterrestrial Physics, Garching, The Johns Hopkins University, Durham University, the University of Edinburgh, the Queen’s University Belfast, the Harvard-Smithsonian Center for Astrophysics, the Las Cumbres Observatory Global Telescope Network Incorporated, the National Central University of Taiwan, the Space Telescope Science Institute, the National Aeronautics and Space Administration under grant No. NNX08AR22G issued through the Planetary Science Division of the NASA Science Mission Directorate, the National Science Foundation grant No. AST-1238877, the University of Maryland, Eotvos Lorand University (ELTE), the Los Alamos National Laboratory, and the Gordon and Betty Moore Foundation.

\section*{Data Availability}

This work is part of the HSC Year 3 cosmological analysis. The data and analysis products, as well as the software, will be made publicly available via the HSC-SSP website \url{https://hsc.mtk.nao.ac.jp/ssp/data-release/} upon journal acceptance.
The correlation function data vectors, model fitting software, and cosmological analysis software will be shared upon reasonable request to the authors.

\bibliographystyle{mnras}

\bibliography{main,xcite}

\appendix

\section{Moments that contribute to spin-2 quantities}
\label{sec:ap:spin2}

In this section, we prove that only the even moments with $n = p+q \geq 2$ has
the spin-2 property, which supports our choice to only consider those moments
in Section~\ref{sec:star:spin2}.
\xlrv{
A spin-2 moment negates under image rotation by $\pi/2$. As a result, it is
invariant under rotation of $n\pi$ (for integer values of $n$) and negates under the rotation of
$(2n+1)\pi/2$.
}
The moments $M_{pq}$ defined in Eq.~\eqref{eq:moment_define} is the projection
of the image onto the basis polynomial function of $x^p y^q$ (or $u^p v^q$
depending on whether it is defined in standardized coordinate).
\xlrv{Note that the moment has the same spin property as the basis polynomial
function $x^p y^q$. To be more specific, if the basis polynomial function
negates under $\pi/2$ image rotation\footnote{
    Note, we rotate the image but do not rotate the basis polynomial function
    (see Appendix~A of \citealt{FPFS_Li2022b}). This is consistent with real
    observations, where we fix the basis polynomial function in the moment
    measurement and galaxy images are randomly orientated.
},
the corresponding moment negates under $\pi/2$ rotation \citep{FPFS_Li2022b}.
Therefore, we focus on the spin-2 component of basis function $x^p y^q$ by
projecting it onto the $m=2$ spinor --- $e^{2i\phi}$:}
\begin{align}\nonumber
    &\int_{-\infty}^{\infty} \, \int_{-\infty}^{\infty} \mathrm{d}x \,\mathrm{d}y\, x^p y^q e^{i2\phi} \\\nonumber
    =& \int_{0}^{\infty} r^{p+q+1} \mathrm{d}r  \int_0^{2\pi} \mathrm{d}\phi \cos^p(\phi) \sin^q(\phi) e^{2i\phi}\\\nonumber
    =& \int_{0}^{\infty} r^{p+q+1} \mathrm{d}r  \int_0^{2\pi} \mathrm{d}\phi\, [2^{-p} (e^{i\phi} + e^{-i\phi})^p][ (2i)^{-q}(e^{i\phi}-e^{-i\phi})^q] e^{2i\phi}\\\nonumber
    =& 2^{-p-q} i^{-q} \sum_{k=0}^p \sum_{j=0}^q  \int_{0}^{\infty} r^{p+q+1} \mathrm{d}r  \\
     &  \int_0^{2\pi} \mathrm{d}\phi (-1)^{q-j} {p \choose k} {q \choose j} e^{i(2k+2j-p-q+2)\phi}\,.
    \label{eq:spin-derive}
\end{align}
The last step uses the binomial theorem, and $p$, $k$, $q$, and $j$ are all
integers. Since $\int_{0}^{2\pi} \mathrm{d}\phi\, e^{im\phi} = 0$ if the
integer $m\neq0$. Therefore, Eq.~\eqref{eq:spin-derive} can only be nonzero if
$2k+2j-p-q+2=0$. This means the order $n = p+q$ must obey
\begin{equation}
    n = 2k+2j+2.
\end{equation}
Since $k$ ($j$) takes any natural number between 0 to $p$ ($q$), $n$ must be an
even number that is greater than or equal to 2.

We further notice that $M_{pq}$ contributes to the real part of the spin-2
quantity if $q$ is even, and contributes to the imaginary part if $q$ is odd, due to the $i^{-q}$ factor in
Eq.~\eqref{eq:spin-derive} (and that the rest of the integral is real).

\responsemnras{Alternatively, one could derive the moment combinations with a specific spin number by expanding $(x+iy)^k (x-iy)^l$. Under this definition, the order $N=k+l$, and spin number $s=k-l$. For the fourth moment spin-2 combination, one can derive Eq.~\eqref{eq:spin-2-polynomial} with $k=3$ and $l=1$. One can also derive the sixth moment spin-2 with $k=4$ and $l=2$, and show that there is no spin-2 combination for odd number moments.}

\section{Problematic Region in \code{GAMA09H}}
\label{sec:ap:cutaway}

\begin{figure*}
    \centering
    \includegraphics[width=1.0\columnwidth]{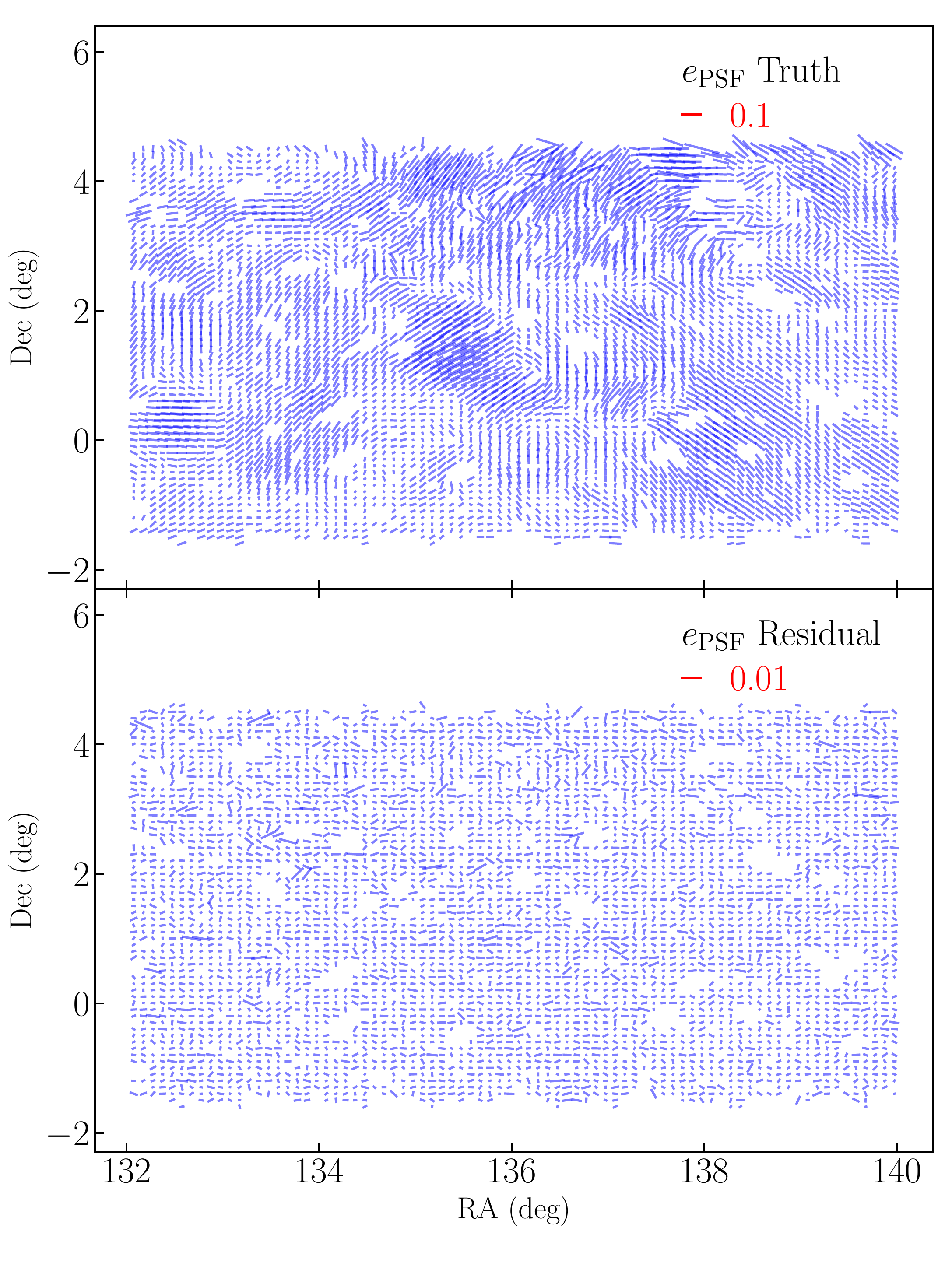}
    \includegraphics[width=1.0\columnwidth]{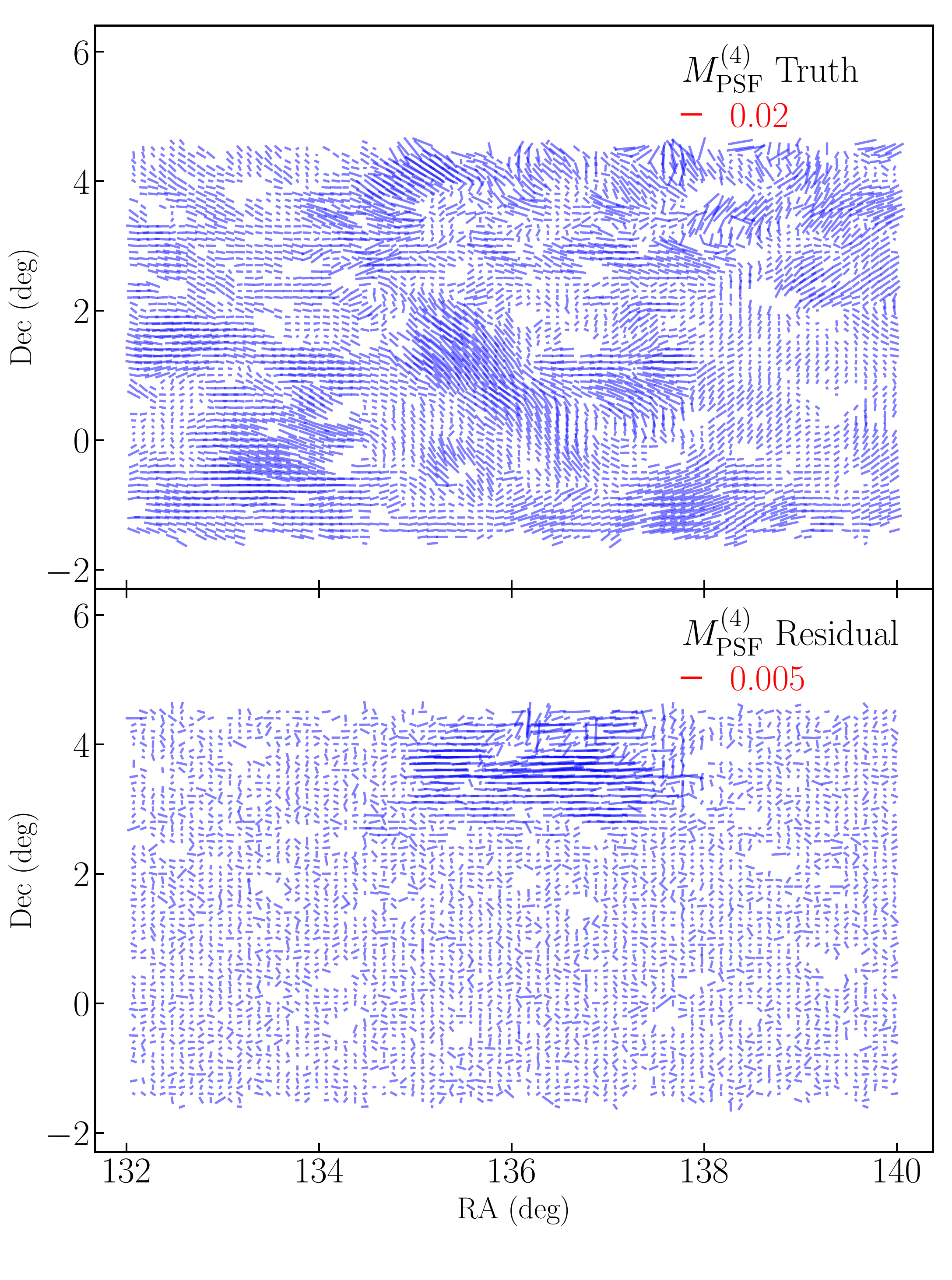}
    \caption{The truth and residual whisker plots of the spin-2 components of the PSF second (top) and fourth moments (bottom) in the field \code{GAMA09H} between ${\rm RA} \in  [132, 140]~({\rm deg})$. The region between ${\rm RA} \in [135,138]~({\rm deg})$ and ${\rm Dec} \in [3,5]~({\rm deg})$ has a particularly large fourth moment modeling error, which do not manifest in the second moment model residual. We have eliminated this problematic region in this work and the upcoming Y3 cosmic shear analysis.}
    \label{fig:gama09h_whisker}
\end{figure*}

In Fig.~\ref{fig:gama09h_whisker}, we show a region within the \code{GAMA09H} field that has a particularly large PSF fourth moment residual. This region is also found to be responsible for a strong B-mode cosmic shear signal in Li et al.~{\it in prep.} The region has a good seeing, and significant proportion of visits are lost due to the overflowing the warning flag \code{maxScaledSizeScatter}\responsemnras{, which sets a maximum scatter in the PSF size residual  allowed for a visit.} As a results, this region has an lower visits, higher galaxy number density (due to good seeing, thus better resolution), and a significant B-mode signal on cosmic shear. In the HSC Y3 cosmic shear analysis and this work, we remove this region from the star and shear catalog.

It is worth noticing that the PSF modeling residual in this region only manifested itself in the fourth moment, rather than the second moment residual. We search through all six fields in the HSC Y3 star catalogs, and found a few other spots with a similar pattern, but the condition in Fig.~\ref{fig:gama09h_whisker} is the most severe. Understanding any potential causal connection between these fourth moment residual hot-spots and the B-mode in the cosmic shear signal is left for future work.

\section{Alternative Definition for Higher Moments}
\label{sec:ap:alt_def:0}

\responsemnras{In this work, the higher moments are defined in a transformed coordinate system where the second moments are standardized, hereafter referred to as the standardized moments. There is an alternative way to define the higher moments, i.e., measuring the higher moments in the image coordinate, hereafter referred to as the raw moments. The raw moments are what functionally affect the raw second moments used for shear inference, but in practice we find it useful to measure standardized moments to separate out the contributions of moments at different orders.  In Section~\ref{sec:ap:alt_def:raw_moments}, we define the raw higher moments, and discuss how to separate their second moments and higher moments parts. In Section~\ref{sec:ap:alt_def:connection}, we establish the analytical connection between the raw higher moments and standardized higher moments, which are used in the main text of this work. In Section~\ref{sec:ap:alt_def:cosmology}, we use the raw higher moments to capture the PSF systematics using the same framework introduced in Section~\ref{sec:model:0}, and compare the impact on the cosmological probes between the two definitions of higher moments.} 

\subsection{Raw Moments}
\label{sec:ap:alt_def:raw_moments}

\responsemnras{
The raw moments are measured in the image coordinates. In our case, we use coadded images, which are aligned with the equatorial coordinate system. 
 In this work, we define the raw moments to be
\begin{equation}
\label{eq:moment_define}
    \mathcal{M}_{pq} = \frac{\int \mathrm{d}x \, \mathrm{d}y \, x^p \, y^q \, \omega(x,y)
    \, I(x,y)}{\int \mathrm{d}x \, \mathrm{d}y \, \omega(x,y) \, I(x,y) }.
\end{equation}
Again, $I(x,y)$ is the image profile, and $\omega(x,y)$ is the adaptive Gaussian weight defined in Eq.~\eqref{eq:weight_function}. The raw higher moments defined here are measured by \textsc{PSFHOME}. We cross-checked our code with the functionality that measures raw higher moments in \textsc{Piff}\footnote{\url{https://github.com/rmjarvis/Piff}}, and find consistent results. }

\responsemnras{
Similar to the standardized moments, there is a combination of the raw moments that forms a spin-$2$ quantity. We call that $\mathcal{M}^{\rm (4)}$.
\begin{equation}
\label{eq:raw_spin_2}
\mathcal{M}^{\rm (4)} = \mathcal{M}_{40} - \mathcal{M}_{04} + 2i(\mathcal{M}_{13}+\mathcal{M}_{31})
\end{equation}}

\responsemnras{
Because of how raw moments are defined, $\mathcal{M}^{\rm (4)}$ not only carries higher order information but also the second order information (Gaussian part). In order to use the raw moments for capturing the spin-2 components of the PSF systematics, we need to find the Gaussian part of the $\mathcal{M}^{\rm (4)}$. It turns out that $\mathcal{M}^{\rm (4)}$ of an elliptical Gaussian PSF profile is just $3e_{\rm PSF} T_{\rm PSF}^2$, where $e_{\rm PSF}$ is the ellipticity of the PSF, and $T_{\rm PSF}$ is the trace.}


\responsemnras{
This relationship can be proved by analytically finding the fourth moments of the Gaussian distribution. We start by defining the Moment Generating Function (MGF) of a two-dimensional Gaussian distribution
\begin{equation}
\label{eq:gaussian_mgf}
M_{X}(\ve{t}) = D^2 e^{\frac{1}{2}\ve{t}^T\mt{M}^{-1}\ve{t}}.
\end{equation}
Here, $\ve{t}^T = \begin{bmatrix}t_1,t_2\end{bmatrix}$ is the two-dimensional dummy variable of the MGF. $\mt{M}^{-1}$ is the inverse of second moment matrix
\begin{equation}
\label{eq:inv_M}
\mt{M}^{-1} = \begin{bmatrix} M_{20} & M_{11} \\  M_{11} & M_{02}\end{bmatrix}^{-1} = \frac{1}{D} \begin{bmatrix}
M_{02} & - M_{11} \\ -M_{11} & M_{02}.
\end{bmatrix}
\end{equation}
And $D$ is the determinant of $\mt{M}$.}

\responsemnras{
The fourth moments are the fourth derivative of the MGT evaluated at $\ve{t}=\ve{0}$.  
One can show that 
\begin{align}
    \label{eq:mgt_driv_1}
    \mathcal{M}_{40} - \mathcal{M}_{04} &= \frac{\mathrm{d}^4 M_{X}(t)}{\mathrm{d}t_1^4} \biggr\rvert_{t_1=t_2=0} - \frac{\mathrm{d}^4 M_{X}(t)}{\mathrm{d}t_2^4} \biggr\rvert_{t_1=t_2=0}\\
     &= 3 (M_{20}^2 - M_{02}^2) = 3e_1T^2
\end{align}
Similarly, for the imaginary part,
\begin{align}
    \label{eq:mgt_driv_1}
    2\left(\mathcal{M}_{13} +\mathcal{M}_{31}\right) &= \frac{\mathrm{d}^4 M_{X}(t)}{\mathrm{d}t_1 \mathrm{d}t_2^3} \biggr\rvert_{t_1=t_2=0} - \frac{\mathrm{d}^4 M_{X}(t)}{\mathrm{d}t_1^3 \mathrm{d}t_2} \biggr\rvert_{t_1=t_2=0}\\
     &= 6M_{11}(M_{20}+M_{02})= 3e_2T^2.
\end{align}}

\begin{figure}
    \centering
    \includegraphics[width=0.49\columnwidth]{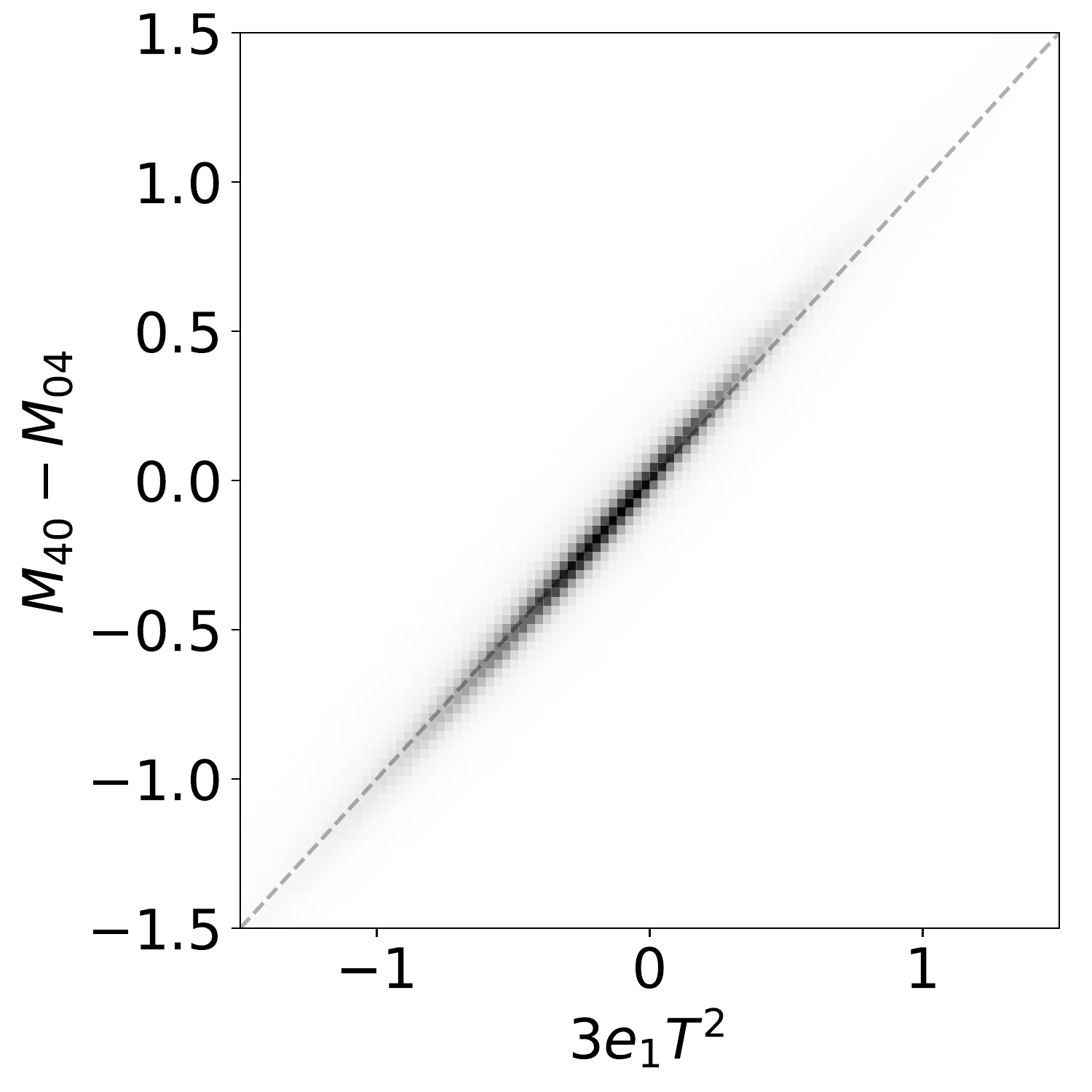}
    \includegraphics[width=0.49\columnwidth]{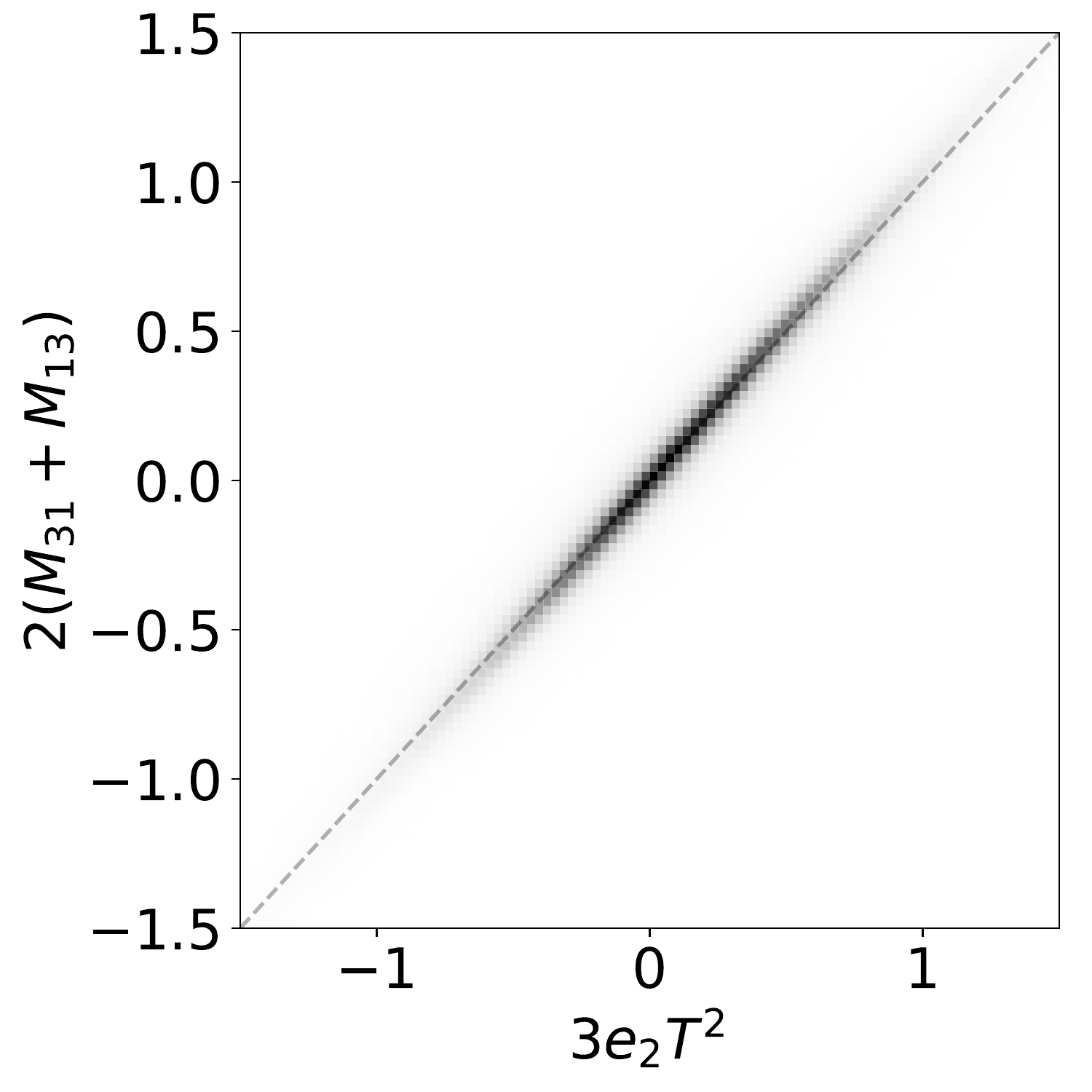}
    \caption{2-d histograms of the raw higher moments $\mathcal{M}^{\rm (4)}$ versus \responsemnrassecond{information from the} Gaussian part -- $3eT^2$. Since the Gaussian part dominates over the information on the departure of the PSF from an elliptical Gaussian, the distribution is sharply peaked along the grey dashed $y=x$ lines. 
    }
    \label{fig:gaussian_part}
\end{figure}

{
We confirmed using our PSF star catalog that the raw fourth moments mostly consist of the Gaussian part. In Fig.~\ref{fig:gaussian_part}, we show the 2-d histogram of the real and imaginary parts of $\mathcal{M}^{\rm (4)}$ and $3eT^2$. The two quantities match closely, which justifies our choice to use standardized fourth moments for our analysis, as the raw fourth moments are so highly correlated with the second moments. However, an alternative approach using raw moments is to construct a dimensionless quantity that only carries \responsemnrassecond{higher moments} spin-2 information
\begin{equation}
\label{eq:reduce_raw}
m^{\rm (4)} = \frac{\mathcal{M}^{\rm (4)}}{T^2} - 3e.
\end{equation}
We will call $m^{\rm (4)}$ the reduced raw fourth moment spin-2 quantity. In Section~\ref{sec:ap:alt_def:cosmology}, we demonstrate that $m^{\rm (4)}$ can be effectively used to track down PSF systematics in cosmic shear. }

\subsection{Connection between the Raw and Standardized Moments}
\label{sec:ap:alt_def:connection}

\responsemnras{
In this section, we analytically derive the connection between the raw and standardized moments. This is a useful formalism in the circumstance that one wants to calculate one definition from the other. }

\responsemnras{
We start by deriving the standardized moments from the raw moments. The standardized coordinates $(u,v)$ in Eq.~\eqref{eq:moment_define} can be expressed in terms of $(x,y)$ using the second moments of the image,
\begin{equation}
    \begin{bmatrix} u \\ v\end{bmatrix} =
    \mt{M}^{-\frac{1}{2}}
    \begin{bmatrix} x \\ y\end{bmatrix}.
\end{equation}
where
\begin{equation}
    \mt{M}^{-\frac{1}{2}} = \frac{1}{\sqrt{\zeta}}
    \begin{bmatrix} M_{02}+\sqrt{D} & -M_{11}\\ -M_{11} & M_{20}+\sqrt{D}\end{bmatrix}.
\end{equation}
Here $D$ is the determinant of $\mt{M}$ and $\zeta=D(M_{20}+M_{02}+2\sqrt{D})$.}

\responsemnras{
We can express $u$ and $v$ as linear functions of $x$ and $y$, 
\begin{align}
    u &= \frac{M_{02}+\sqrt{D}}{\sqrt{\zeta}}x - \frac{M_{11}}{\sqrt{\zeta}}y \\
    v &= - \frac{M_{11}}{\sqrt{\zeta}}x + \frac{M_{20}+\sqrt{D}}{\sqrt{\zeta}}y .
\end{align}
Let's denote 
\begin{align}
    A &\equiv \frac{M_{02}+\sqrt{D}}{\sqrt{\zeta}}\\
    B &\equiv - \frac{M_{11}}{\sqrt{\zeta}}\\
    C &\equiv \frac{M_{20}+\sqrt{D}}{\sqrt{\zeta}}
\end{align}
The standardized fourth moments are then
\begin{align}
    \nonumber M_{40} =& A^4 \mathcal{M}_{40} + 4A^3B\mathcal{M}_{31}+6A^2B^2\mathcal{M}_{22}\\
    & +4AB^3\mathcal{M}_{13}+B^4\mathcal{M}_{04}\\
    \nonumber M_{31} =& A^3B \mathcal{M}_{40} + (A^3C+3A^2B^2)\mathcal{M}_{31}+ (3A^2BC+3AB^3)\mathcal{M}_{22} \\&+(3AB^2C+B^4)\mathcal{M}_{13}+B^3C\mathcal{M}_{04}.\\
    \nonumber M_{22} =& A^2B^2 \mathcal{M}_{40} + (2A^2BC+2AB^3)\mathcal{M}_{31} \\\nonumber &+(A^2C^2+4AB^2C+B^4)\mathcal{M}_{22} \\&+(3AB^2C+B^4)\mathcal{M}_{13}+B^3C\mathcal{M}_{04}.\\
    \nonumber M_{13} = & AB^3 \mathcal{M}_{40} + (3AB^2C+B^4)\mathcal{M}_{31} \\\nonumber &+ (3ABC^2+3B^3C)\mathcal{M}_{22} \\ &+ (AC^3+3B^2C^2)\mathcal{M}_{13} + BC^3\mathcal{M}_{04}\\
    \nonumber M_{04} =& B^4 \mathcal{M}_{40} + 4B^3C\mathcal{M}_{31}+\\&6B^2C^2\mathcal{M}_{22}+4BC^3\mathcal{M}_{13}+C^4\mathcal{M}_{04}.
\end{align}
Here $\mathcal{M}_{pq}$ are the raw higher moments.  
\textsc{PSFHOME} has the functionality to carry out this transformation. We compared the standardized higher moments measured on the image and predicted by this formalism, and found the fractional difference to be on the order of $10^{-10}$, which is an exquisite consistency. This formalism shows that given the second moments, the 5 raw fourth moments can be remapped to standardized fourth moments. Using a similar formalism, one can remap in the other direction, but we will not derive those equations. }

\responsemnras{
We confirmed that changing the higher order spin-2 quantity in the raw moments will not only change the standardized spin-2 quantity, but also the standardized spin-0 quantity, and vice versa.  This was implied by the above equations and can be demonstrated easily with image simulations as well. }

\subsection{Raw Moments for Capturing PSF Systematics}
\label{sec:ap:alt_def:cosmology}

\responsemnras{
In this section, we demonstrate that one can use raw moments to measure the PSF systematics contamination in the cosmic shear 2PCF using our HSC catalog. Further, we empirically show that despite the complex mapping between the standardized and raw moments shown in Section~\ref{sec:ap:alt_def:connection}, using raw moments to trace PSF systematics gives results for the cosmological contamination that are no different from using the standardized moments.}

\responsemnras{
To remove the contribution from the second moments, we use the reduced raw higher moments spin-2 $m^{\rm (4)}_{\rm PSF}$ defined in Eq.~\eqref{eq:reduce_raw} to model the higher moments leakage and modeling error. Namely, Eq.~\eqref{eq:psf-sys-full} is modified to be
\begin{equation}
\label{eq:psf-sys-full_raw}
g_{\rm sys} = \alpha^{\rm (2)} e_{\rm PSF} + \beta^{\rm (2)} \Delta e_{\rm PSF}
+ \alpha^{\rm (4)} m^{\rm (4)}_{\rm PSF} + \beta^{\rm (4)} \Delta m^{\rm (4)}_{\rm PSF}.
\end{equation}
With the raw moments, we only conducted cross-correlations using the PSF star catalog, and implemented the 4-parameter fiducial model. By cross-correlating with galaxy shapes (described in Section~\ref{sec:model:gp_corr}) and maximizing the likelihood function defined in Eq.~\eqref{eq:likelihood}, we get $\alpha^{{\rm (2)}} = -0.024\pm 0.003$, $\beta^{\rm (2)}=-0.72\pm0.06$, $\alpha^{\rm (4)}=-0.15\pm 0.01$, and $\beta^{\rm (4)}=-0.6\pm 0.2$. The reduced higher moments are still correlated with the second-moment shape. As a result, the correlation coefficient between $\alpha^{2}$ and $\alpha^{4}$ is 0.85, higher than the value of 0.62 found with standardized moments.}

\begin{figure}
    \centering
    \includegraphics[width=1.0\columnwidth]{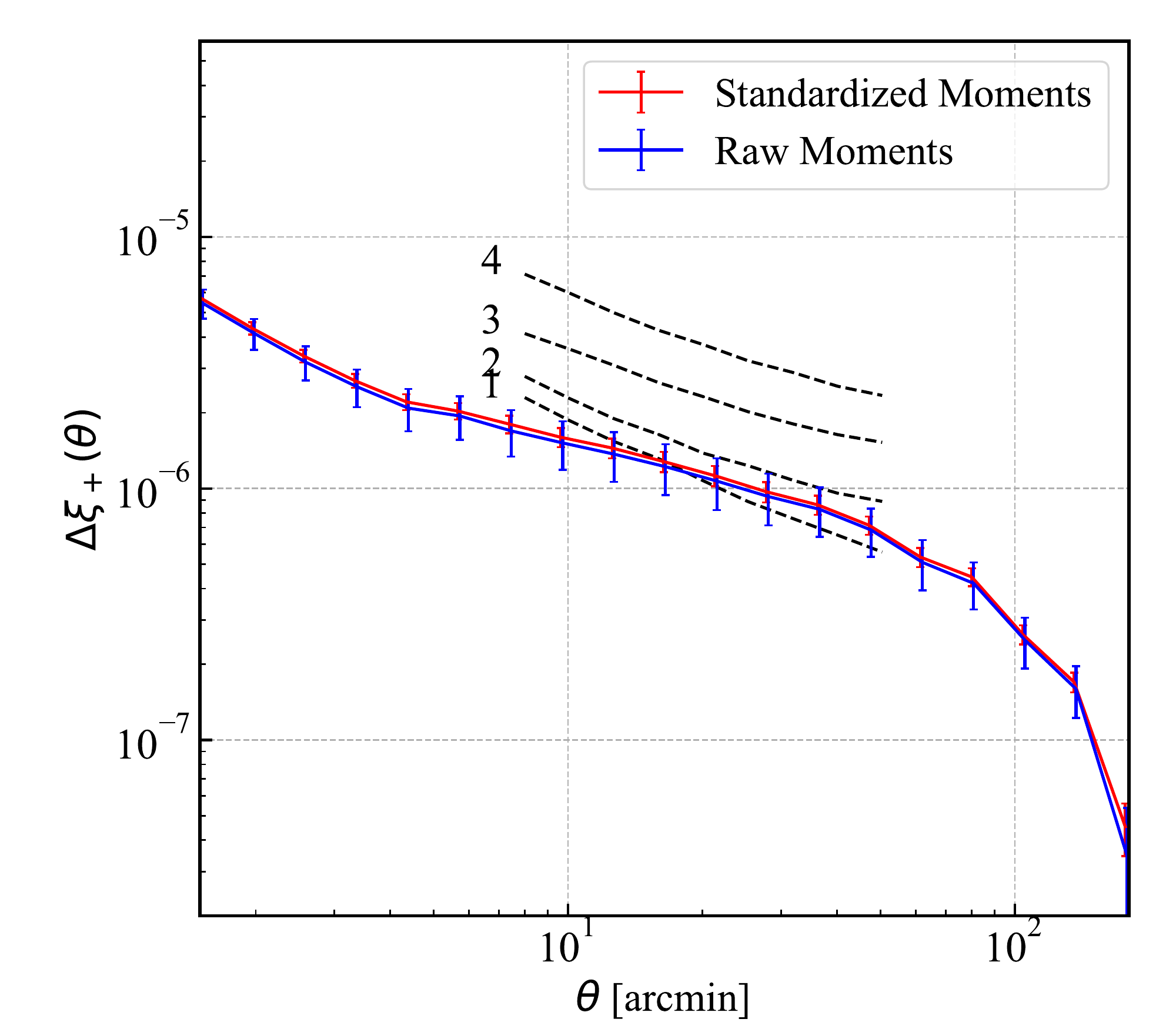}
    \caption{Total impact on shear-shear 2PCF caused by the PSF additive bias. The statistical uncertainty of the shear-shear auto correlations are plotted as dashed lines. We can see that the $\Delta \xi_+$ calculated using raw and standardized higher moments matches very well across all angular scales shown in this plot. }
    \label{fig:raw_vs_stan}
\end{figure}

\responsemnras{
The most important quantity that we want to compare between standardized moments and raw moments is the impact on the shear-shear 2PCF. In Fig.~\ref{fig:raw_vs_stan}, we show that the $\Delta \xi_+$ calculated using the reduced raw moments is highly consistent with the one calculated using the standardized moments. This means that both choices can effectively capture the additive bias due to second and higher PSF moments, as long as the data-driven procedure is followed. }

\responsemnras{Although we successfully demonstrated that the two approaches toward defining the PSF higher moments yield the same cosmological impact for HSC Y3, this is still an empirical demonstration that may be contingent on the moment distribution of the PSF in HSC Y3. We leave the study of the potential mathematical origin of this equivalence to future work. Before that, we suggest that future surveys conduct higher moments null tests using both definitions. }

\section{Mock Catalog Test}
\label{sec:ap:mock_test}

We conducted a mock catalog test to validate the inference of the PSF systematics model parameters (as defined in Section~\ref{sec:model:formalism}). The crucial element of this test is to generate mock star and galaxy catalogs with  systematics that we know follow our model on all scales.  The steps for generating the mock star and galaxy catalogs are as follows:
\begin{enumerate}
\item Populate a healpix map \citep{2005ApJ...622..759G} with \code{nside=512} with stars from the HSC Y3 star catalog (Section~\ref{sec:star:0}). Compute the average values of $e_{\rm PSF}$, $\Delta e_{\rm PSF}$, $M^{\rm (4)}_{\rm PSF}$ and $\Delta M^{\rm (4)}_{\rm PSF}$ using all stars within each pixel. Assign the average PSF moments in a pixel to the stars in that pixel to produce the mock star catalog.
\item Compute shear bias from Eq.~\eqref{eq:psf-sys-full} using the average PSF moments in the healpix pixel, and a set of input PSF parameters $\alpha^{\rm (4)} = 0.04$, $\beta^{\rm (2)} = -1$, $\alpha^{\rm (4)} = 0.19$, and $\beta^{\rm (4)} = -0.5$. Assign the shear bias to the mock galaxy catalogs (see Section~\ref{sec:shape:mock}) based on their corresponding pixels in the map to produce the mock shear catalog.
\end{enumerate}

We use these mock star and shear catalogs to infer the PSF parameters using the pipeline developed for inference from the real data, to ensure that the pipeline is able to recover the input parameters. In doing so, we use the covariance matrix measured using the real data (as described in Section~\ref{sec:model:gp_corr}).
We produced 10 mock catalogs with shear biases, and individually inferred their PSF systematics model parameters. Over the 10 mocks, we retrieve the averaged PSF parameters $\alpha^{\rm (2)} = 0.040\pm0.001$, $\beta^{\rm (2)} = -1.10\pm0.02$, $\alpha^{\rm (4)} = 0.185\pm0.01$, and $\beta^{\rm (4)} = -0.53\pm 0.01$. Although there appears to be a statistically significant bias on the $\beta^{\rm (2)}$ and $\beta^{\rm (4)}$ parameter, the differences are within $\pm 10\%$ of the true PSF parameters. Further investigation is needed for understanding the discrepancy between the inferred and true  modeling error parameters in the mock catalog tests. We inspected the $\Delta \xi_+$s predicted by the true PSF parameters and by the inferred PSF parameters, and seeing no significant difference between the two.   

\section{Subdominant effects}
\label{sec:ap:consistency}



In this section, we discuss different aspects of the PSF systematics that could complicate the model. We implemented these extra terms on top of the fiducial model from Section~\ref{sec:model:null}, which describes the PSF systematics as an additive bias on $\xi_+$, including the leakage and modeling error caused separately by the PSF second moments and fourth moments. Most of these complications to the model do not significantly contribute to the HSC Y3 PSF systematics. However, they might be significant in other cosmic shear surveys with different shear estimation methods and PSF modeling algorithms. Therefore, we elaborate on these phenomena below.

In Appendix~\ref{sec:ap:xim}, we generalize the formalism in Section~\ref{sec:model:formalism} and \ref{sec:model:null} from $\xi_+$ alone to include $\xi_-$ as well. In Appendix~\ref{sec:ap:six} and~\ref{sec:ap:sot}, we considered other spin-$2$ quantities--the PSF sixth moments and second order spin-$2$ quantities (product of spin-$2$ and spin-$0$, etc.), and proved it is unnecessary to model these quantities for HSC Y3. 

\subsection{$\Delta \xi_-$}
\label{sec:ap:xim}

\begin{figure}
    \centering
    \includegraphics[width=0.98\columnwidth]{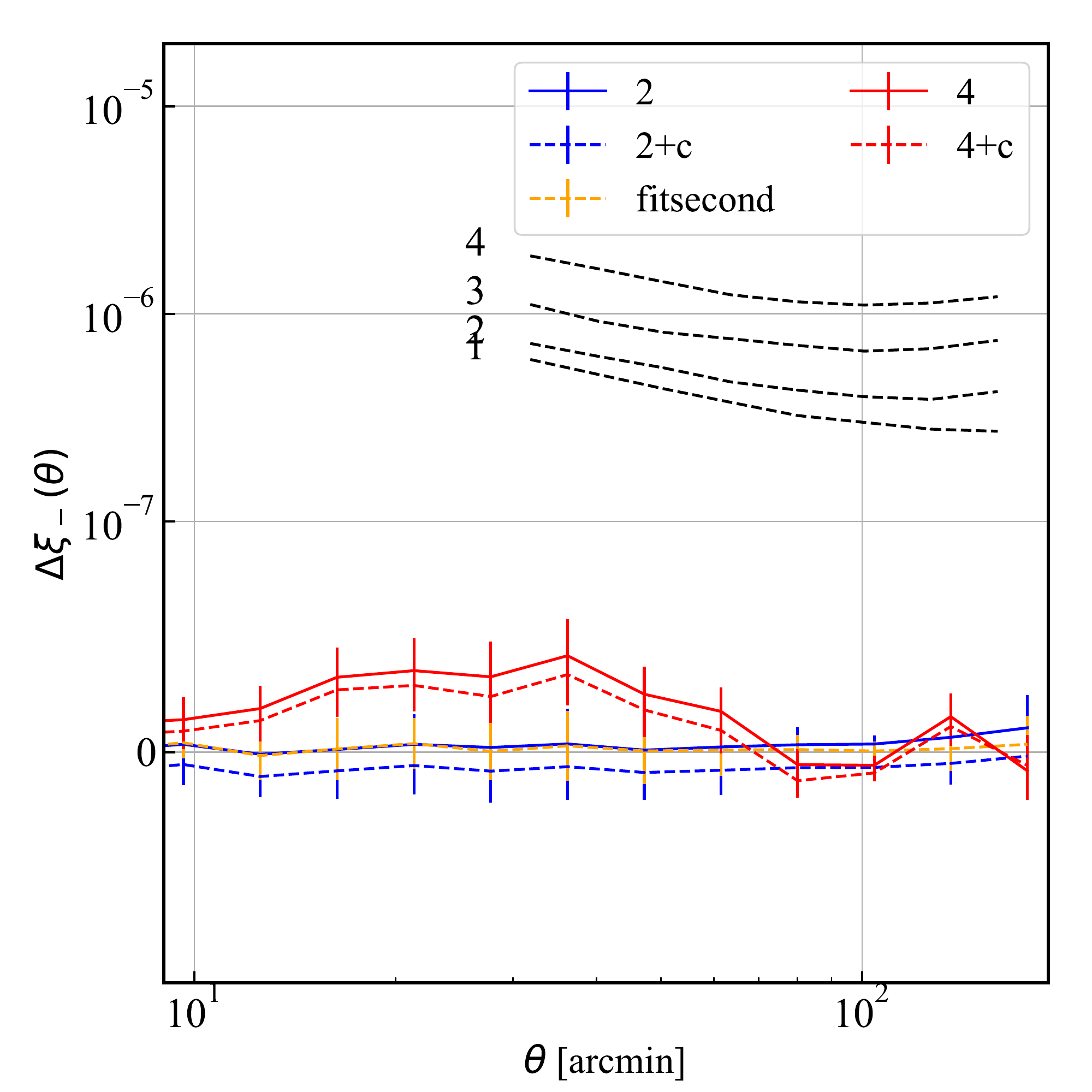}
\caption{ The additive bias on the cosmic shear 2PCF $\xi_-$. We find the $\Delta \xi_-$ to be below $10\%$ of the predicted statistical uncertainty of $\xi_-$ for all the tomographic bins, and therefore it can be ignored.}
    \label{fig:delta_xim}
\end{figure}

In this section, we discuss the additive PSF systematics in $\xi_-$. Previous studies have shown that the impact on $\xi_-$ from PSF second moment contamination is sufficiently small that it can be ignored in the cosmic shear analysis \citep[e.g.,][]{2020PASJ...72...16H,2021MNRAS.501.1282J}. \cite{2022arXiv220507892Z} also found the additive bias on $\xi_-$ due to PSF fourth moment contamination to be consistent with zero. Here, we simply repeat the formalism in Section~\ref{sec:model:formalism} and \ref{sec:model:null}, and take $\xi_-$ for all the correlation functions. In Fig.~\ref{fig:delta_xim}, we present the $\Delta \xi_-$ in comparison to the cosmic shear signal predicted by the fiducial cosmology. We found the $\Delta\xi_-$ to be below 1 per cent of the predicted shear signal in all of the tomographic bins, \responsemnras{with a total statistical significance equal to 0.22$\sigma$} and it therefore can be safely ignored.

\subsection{Sixth Moment Terms}
\label{sec:ap:six}

\begin{figure}
    \centering
    \includegraphics[width=0.98\columnwidth]{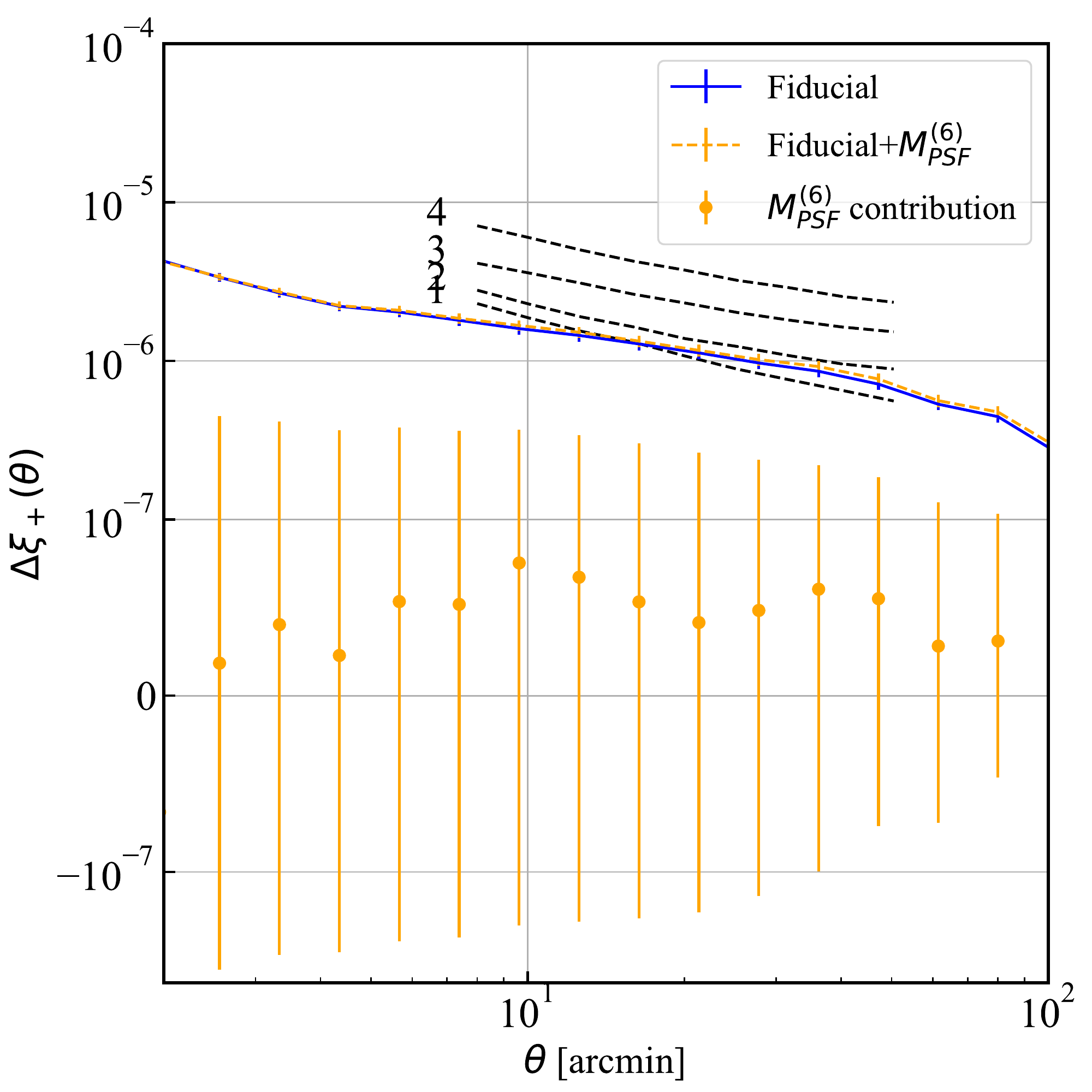}
\caption{ The additive bias on cosmic shear $\xi_+$ if the PSF sixth moments leakage and modeling error are considered. We include the PSF six moments as an extension to the fiducial model, which has the second and fourth moments. The PSF six moment contributes $<10\%$ to the overall $\Delta \xi_+$, \responsemnras{ as well as to the statistical uncertainty}, therefore is subdominant. 
}
    \label{fig:delta_xip_six}
\end{figure}

In Section~\ref{sec:star:spin2}, we pointed out that not just the second and fourth moments can combine to form a spin-2 quantity, but rather all even moments can do so (proof in Appendix~\ref{sec:ap:spin2}). So a natural question is whether even higher order PSF moments need to be considered. In this section, we expand our model to accommodate the spin-2 combination of PSF sixth moments, which can be expressed as
\begin{equation}
\label{eq:sixth_moments}
M^{\rm (6)}_{\rm PSF} = (M_{60} + M_{42} - M_{24} - M_{06}) + i (2M_{51}+4M_{33}+2M_{15}).
\end{equation}
Similarly, we included $\langle \hat{g}_{\rm gal} M^{\rm (6)}_{\rm PSF} \rangle $ and $\langle \hat{g}_{\rm gal} \Delta M^{\rm (6)}_{\rm PSF} \rangle$ in the data vector and added sixth moments leakage and modeling error terms to the model ($\alpha^{\rm (6)}$ and $\beta^{\rm (6)}$, respectively). In this expanded framework, the data vector has a length of 122 and the parameter space grows to 6 from the fiducial model's 4.

In Fig.~\ref{fig:delta_xip_six}, we show the additive bias $\Delta \xi_+$ with and without the sixth moment leakage and modeling error, and the difference, which is the contribution of $M^{\rm (6)}$.
We see that the additional additive bias induced by the PSF sixth moments is $\lesssim$ $10\%$ of that from the fiducial model. \responsemnras{The increase in statistical significance is only 0.1$\sigma$.} We therefore neglect the spin-2 combination of PSF sixth moments, $M^{\rm (6)}$, due to its subdominant impact.

We speculate that the reason that PSF sixth moments do not add much more additive bias to the overall $\Delta \xi_+$ is that (a) they are more susceptible to noise, which increases their statistical error; (b) they are shown to be highly correlated with the fourth moments 
\citep{2022arXiv220507892Z}.  Most likely this correlation would be reduced if the sixth moments are measured on images with standardized second and fourth moments, instead of only standardized second moments.


\subsection{Second Order Terms}
\label{sec:ap:sot}

\begin{figure}
    \centering
    \includegraphics[width=0.98\columnwidth]{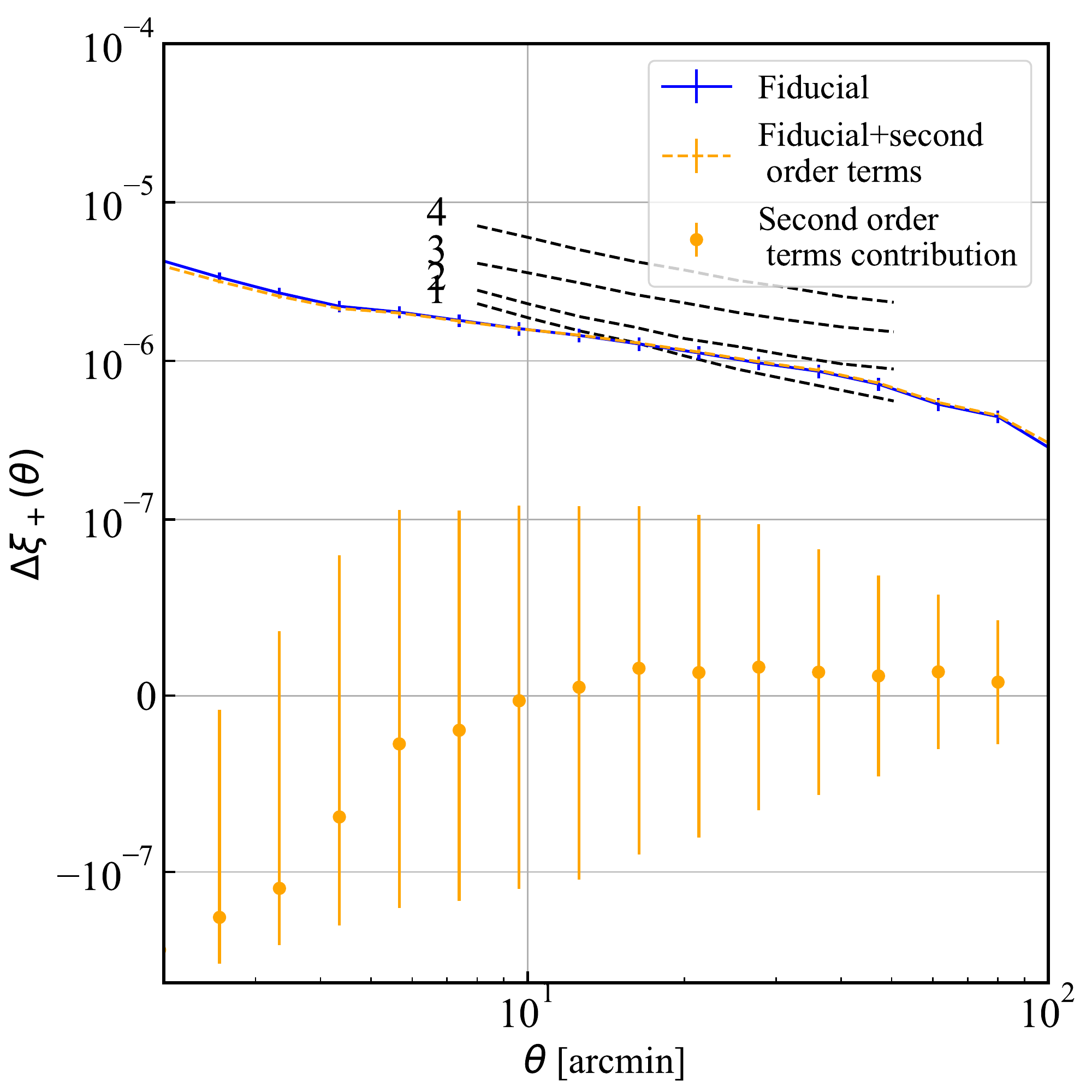}
\caption{ The additive bias on cosmic shear $\xi_+$ considering all second-order spin-2$ \times$ spin-0 systematics.  We included four spin-2 $\times$ spin-0 systematics described in Section~\ref{sec:ap:sot} as an extension to the fiducial model, which has first-order contributions from the spin-2 combinations of PSF second and fourth moments. The  second-order systematic biases induced by  spin-2 $\times$ spin-0 terms are subdominant compared to those from the first order terms. 
}
    \label{fig:delta_xip_sot}
\end{figure}

So far, we limited our discussion to the first order terms of the PSF moments, which means they are either a single moment like $e_{\rm PSF}$, or a moment residual like $\Delta e_{\rm PSF}$. In this section, we discuss the second-order spin-2 quantities, which can take the form of a spin-2 quantity multiplied by a spin-0 quantity, e.g., $e_{\rm PSF} \Delta T_{\rm PSF}/T_{\rm PSF}$, which gives rise to the higher-order $\rho$ statistics  \citep{2016MNRAS.460.2245J}. Another possibility is a spin-4 quantity multiplied by a spin-2 quantity, \responsemnras{or a spin-1 multiplied by a spin-1 quantity, which could arise from the product of two $N=3$ moments}; we will leave that for future work.

Since the first order spin-$2$ quantities $e_{\rm PSF}$, $\Delta e_{\rm PSF}$, $M^{\rm (4)}_{\rm PSF}$, $\Delta M$, $\Delta T$ are defined such that they are $\ll$1, their second-order terms should be negligible (given that their pre-factors are  of order 1). Therefore, we focused on first-order terms in the model. In this section, we discuss the potential impact of the second-order terms in PSF systematics.



\begin{figure}
    \centering
    \includegraphics[width=1.0\columnwidth]{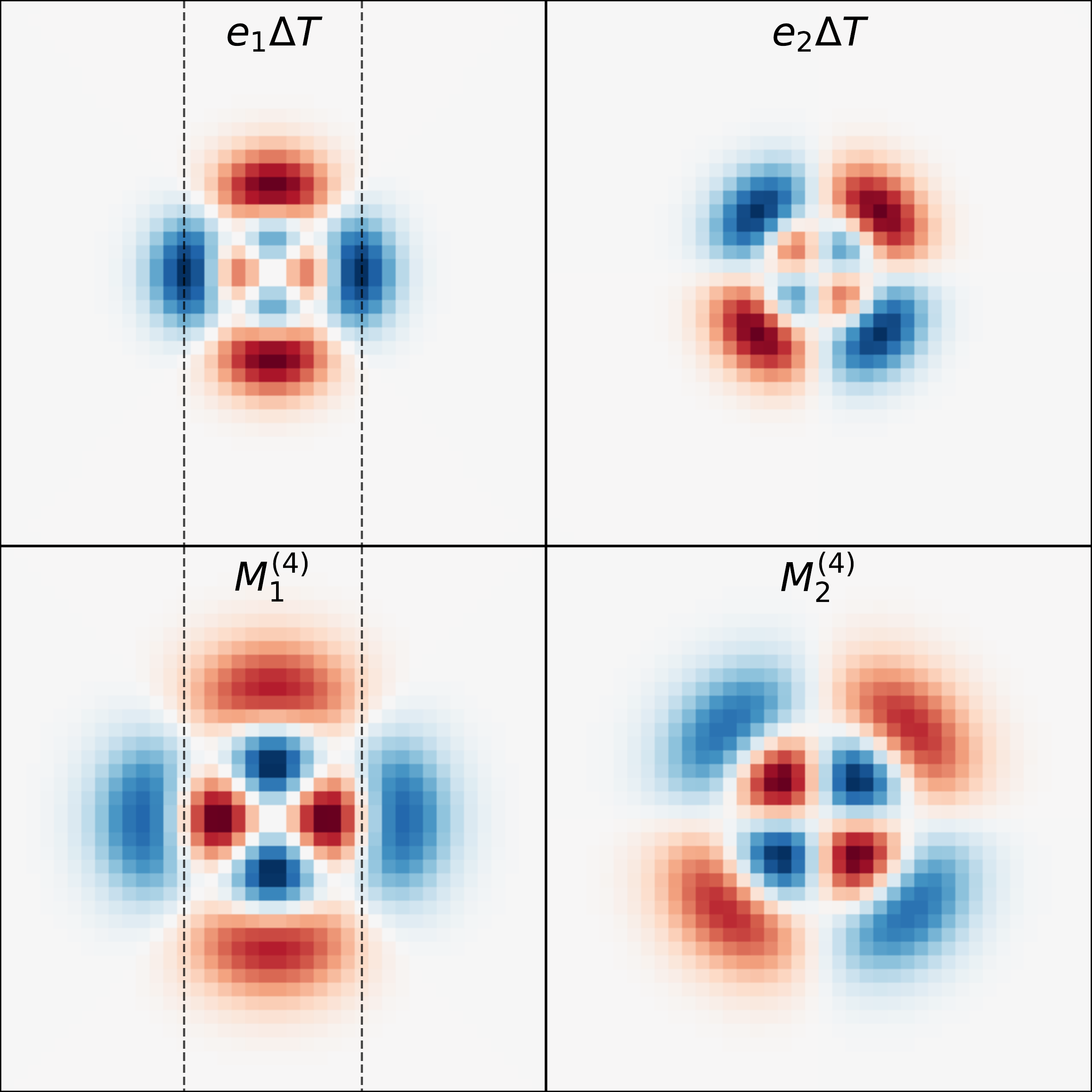}
    \caption{
    A comparison between the image response to $e_\text{PSF}\Delta T_\text{PSF}/T_\text{PSF}$ and that to $M^{\rm (4)}_{PSF}$. These two terms have very similar patterns, just with sensitivities to different scales, as the reference lines show. 
    }
    \label{fig:image_response1}
\end{figure}

The spin-2 combination of PSF fourth moments that serves as a counterpart to $e_{\rm PSF}$ is $M^{\rm (4)}_{\rm PSF}$, defined in Eq.~\eqref{eq:spin-2-fourth}. The spin-0 combination of PSF fourth moments that serves as a counterpart to $T_{\rm PSF}$ (trace, defined in Eq.~\ref{eq:define_trace}) is called the radial kurtosis, defined in Eq.~\eqref{eq:define_kurtosis}.  Errors in modeling either of these spin-0 quantities in the PSF can be a source of multiplicative bias in shear. As a demonstration, we show the image response to one of the second-order terms, $e_\text{PSF}\Delta T_\text{PSF}/T_\text{PSF}$, and compare that to the image response to $M^{\rm (4)}_{PSF}$ in Fig.~\ref{fig:image_response1}. Because of the multiplication by $T_{\rm PSF}$, $e_\text{PSF}\Delta T_\text{PSF}/T_\text{PSF}$ now has a very similar pattern to $M^{\rm (4)}_{PSF}$, but is sensitive to pixels with different radii compared to $M^{\rm (4)}_{PSF}$.

If we were to include $M^{\rm (4)}_{\rm PSF}$ and $\rho^{\rm (4)}_{\rm PSF}$ to form  second order spin-2 terms, this would give rise to 3 more terms beyond the second-order term that is already in the $\rho$ statistics ($e_\text{PSF}\Delta T_\text{PSF}/T_\text{PSF}$): $e_{\rm PSF} \Delta \rho^{\rm (4)}_{\rm PSF}/\rho^{\rm (4)}_{\rm PSF}$, $M^{\rm (4)}_{\rm PSF} \Delta \rho^{\rm (4)}_{\rm PSF}/\rho^{\rm (4)}_{\rm PSF}$, and $M^{\rm (4)}_{\rm PSF}  \Delta T_{\rm PSF}/T_{\rm PSF}$. We define these four spin-$2$ quantities as $\Psi_j$ for $j=1\dots4$, and define their corresponding pre-factors in $\gamma^{\rm sys}$ to be $\eta_j$.

Including all four of these expands the total number of PSF spin-2 quantities from 4 to 8, which in principle generates $8\times 9/2 = 36$ $\rho$ statistics for which we want to know their impact on the overall $\Delta \xi_+$. We quantify the impact of the second order terms by adding four $\langle \hat{g}_{\rm gal} \Psi_j \rangle$ to the g-p correlation data vector, and adding four $\eta_j \Psi_j$ terms, for $j = 1\dots4$, to $g_{\rm sys}$ in addition to the fiducial model terms. The $\Psi_j$ and $\eta_j$ are defined in the previous paragraph.  
We conducted a joint fitting process that considers all the second-order terms, their g-p correlations, and their auto- and cross-correlation with other PSF first-order terms. The additional additive bias in $\xi_+$ is shown in Fig.~\ref{fig:delta_xip_sot}. We observe that the impact of the second-order spin-2$\times$spin-0 terms is subdominant, \responsemnras{only increasing the statistical significance of $\Delta \xi_+$ by 0.02$\sigma$}. We include them in this work for completeness, even though they do not need to be modelled in the HSC Y3 analysis, and we recommend future cosmic shear surveys consider these possible sources of contamination to the cosmic shear.

\section{Correlation in PSF parameters}
\label{sec:ap:correlation}

\responsemnras{
It is visually evident from Figure~\ref{fig:whisker_plots} that the fourth moment spin-2 $M^{\rm (4)}_{\rm PSF}$ and the second moments $e_{\rm PSF}$ are anti-correlated. This correlation is also manifested in the correlation matrix in Figure~\ref{fig:correlation}, and the posterior of the PSF parameters in Figure~\ref{fig:prior_nobin}.}

\responsemnras{
To account for the correlation in the PSF parameters $\ve{p}$, we sample a standard multivariate Gaussian distribution $\tilde{\ve{p}} = [\tilde{\alpha}^{\rm (2)},
\tilde{\beta}^{\rm (2)},\tilde{\alpha}^{\rm (4)}, \tilde{\beta}^{\rm (4)}]$, drawn from $\mathcal{N}(\ve{0}, \mt{I})$, where the null vector $\ve{0} \in R^4$ and $\mt{I}$ is a $4\times4$ identity matrix. We then transform $\tilde{\ve{p}}$ to get $\ve{p}$ in the fiducial model, by
\begin{equation}
\label{eq:params_transformation}
\ve{p} = \mt{\Lambda} \mt{U}^{1/2} \ve{\tilde{p}} + \bar{\ve{p}}.
\end{equation}
Here $\mt{\Lambda}$ is the eigenvalue vector of $\ve{p} - \bar{\ve{p}}$, $\mt{U}^{1/2}$ is the eigenvector matrix of $\ve{p} - \bar{\ve{p}}$, both inferred from the prior distribution of PSF parameters. 
}

\begin{figure}
    \centering
    \includegraphics[width=1.0\columnwidth]{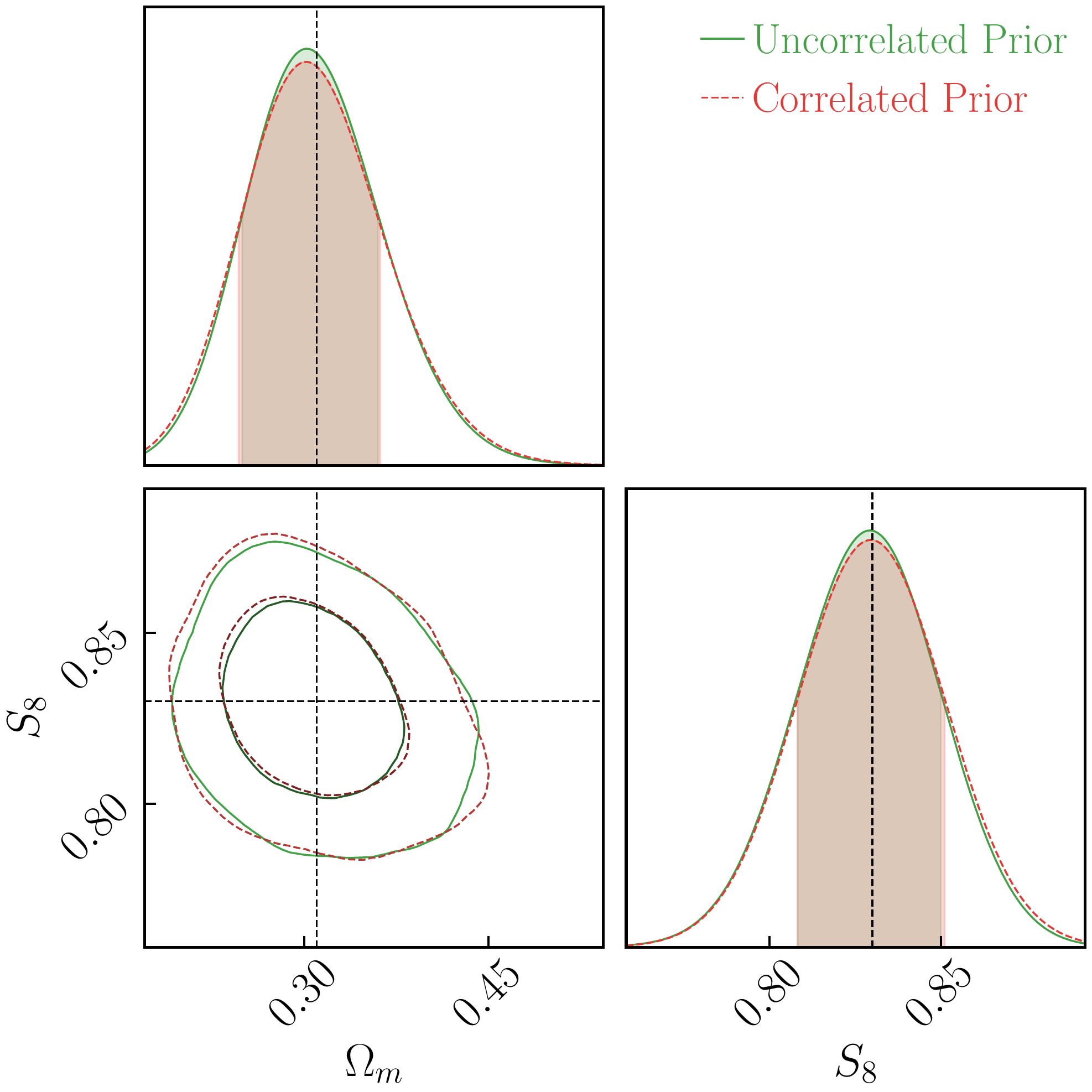}
    \caption{
    Comparison between the HSC Y3 mock cosmological analysis using an uncorrelated prior (green) versus correlated prior (red) for the two $\alpha$ parameters. We find no significant change in the cosmological constraints due to this difference in model choices. 
    }
    \label{fig:correlated_contour}
\end{figure}

\responsemnras{
In Figure~\ref{fig:correlated_contour}, we show that including the correlation of the PSF parameters in the cosmological parameter analysis does not cause a significant difference to the HSC Y3 mock analysis. Nonetheless, we recommend including the correlation for the completeness of the error propagation. 
}

\section{Fourier Space Cosmic Shear Analysis}
\label{sec:ap:fourier}


Cosmic shear are probed and analyzed in the configuration space by two-point correlation function, and also probed and analyzed in Fourier space using pseudo-$C_\ell$ \citep[e.g.,][]{2019PASJ...71...43H, 2021JCAP...03..067N}. In this section, we discuss the PSF systematics formalism in Fourier space (Section~\ref{sec:ap:fourier_formalism}), and the real-Fourier space consistency for the PSF additive bias modeling (Section~\ref{sec:ap:rf_consistency}).

\subsection{PSF systematics formalism in Fourier Space}
\label{sec:ap:fourier_formalism}

We also investigated the impact of PSF systematics on cosmic shear power spectra, $C_{\ell}$s, in addition to the above analysis using two-point correlation functions. In doing so, we use the model for $g_{\rm sys}$ given by Eq.~\eqref{eq:psf-sys-full}, without the mean ellipticity parameter, $e_c$, as the analysis with 2PCFs has shown that this parameter has negligible impact for HSC Y3:
\begin{equation}
\label{eq:psf_sys_full_fourier}
g_{\rm sys} = \alpha^{\rm (2)} e_{\rm PSF} + \beta^{\rm (2)} \Delta e_{\rm PSF}
+ \alpha^{\rm (4)} M^{\rm (4)}_{\rm PSF} + \beta^{\rm (4)} \Delta M^{\rm (4)}_{\rm PSF}.
\end{equation}

Upon adding $g_{\rm sys}$ to the observed galaxy ellipticity, the measured cosmic shear power spectrum becomes:

\begin{equation}
\label{eq:delta_cl_fourier}
C_{\ell} \rightarrow C_{\ell} +
\sum_{i=1}^4 \sum_{j=1}^4 \ve{p}_i \ve{p}_j C_{\ell}^{\ve{S}_i \ve{S}_j}.
\end{equation}
where, as before, we define the parameter vector $\ve{p} = [\alpha^{\rm (2)},
\beta^{\rm (2)},\alpha^{\rm (4)}, \beta^{\rm (4)}]$, and the PSF moments vectors $\ve{S} = [e_{\rm PSF}, \Delta e_{\rm PSF},
M^{\rm (4)}_{\rm PSF}, \Delta M^{\rm (4)}_{\rm PSF}]$. We refer to the additive term in Eq.~\eqref{eq:delta_cl_fourier} as $\Delta C_{\ell}$.

To get the best-fitting values of the parameters  $\ve{p}$, we repeat the process carried out  with 2PCFs, measuring the p-p power spectra and the g-p power spectra ($\vect{D}_{gp}$), in 6 $\ell$ bins, from $300\leq \ell \leq 1800$ (the provisional scale cuts for the Fourier space cosmology analysis). We use the pseudo-$C_{\ell}$ code \texttt{NaMaster} \citep{2019MNRAS.484.4127A} to measure the power spectra. Although the pseudo-$C_{\ell}$ method requires subtracting a noise spectrum from auto-correlations \citep{2021JCAP...03..067N}, this term is negligible for the PSF moments (unlike the case of galaxy shape auto-correlations). We then predict the theory data vector ($\vect{T}_{gp}$) for the g-p power spectra, given the p-p power spectra, which is equivalent to the real space fiducial model in Eq.~\eqref{eq:null1}--\eqref{eq:null4} with $e_c = 0+0j$:

\begin{strip}
\noindent\rule{8.8cm}{0.4pt}
\begin{align}
    \label{eq:null1_fourier}C_{\ell}^{\hat{g}_{\rm gal} e_{\text{PSF}}} &= \alpha^{\rm (2)} C_{\ell}^{e_{\text{PSF}} e_{\text{PSF}}}  + \beta^{\rm (2)}C_{\ell}^{\Delta e_{\text{PSF}}  e_{\text{PSF}}}  + \alpha^{\rm (4)} C_{\ell}^{ M^{\rm (4)}_{\text{PSF}}  e_{\text{PSF}}} + \beta^{\rm (4)} C_{\ell}^{\Delta M^{\rm (4)}_{\text{PSF}}  e_{\text{PSF}}} \\
    \label{eq:null2_fourier} C_{\ell}^{\hat{g}_{\rm gal} \Delta e_{\text{PSF}}} &= \alpha^{\rm (2)} C_{\ell}^{e_{\text{PSF}} \Delta e_{\text{PSF}}}  + \beta^{\rm (2)} C_{\ell}^{\Delta e_{\text{PSF}} \Delta e_{\text{PSF}}} + \alpha^{\rm (4)} C_{\ell}^{M^{\rm (4)}_{\text{PSF}} \Delta e_{\text{PSF}}}  + \beta^{\rm (4)} C_{\ell}^{\Delta M^{\rm (4)}_{\text{PSF}} \Delta e_{\text{PSF}}}\\
    \label{eq:null3_fourier} C_{\ell}^{\hat{g}_{\rm gal}  M^{\rm (4)}_{\text{PSF}}} &= \alpha^{\rm (2)} C_{\ell}^{e_{\text{PSF}}  M^{\rm (4)}_{\text{PSF}}}  + \beta^{\rm (2)} C_{\ell}^{\Delta e_{\text{PSF}} M^{\rm (4)}_{\text{PSF}}} + \alpha^{\rm (4)} C_{\ell}^{M^{\rm (4)}_{\text{PSF}}  M^{\rm (4)}_{\text{PSF}}}  + \beta^{\rm (4)} C_{\ell}^{\Delta M^{\rm (4)}_{\text{PSF}}  M^{\rm (4)}_{\text{PSF}}}\\
    \label{eq:null4_fourier}C_{\ell}^{\hat{g}_{\rm gal} \Delta M^{\rm (4)}_{\text{PSF}}} &= \alpha^{\rm (2)} C_{\ell}^{e_{\text{PSF}} \Delta M^{\rm (4)}_{\text{PSF}}}  + \beta^{\rm (2)} C_{\ell}^{\Delta e_{\text{PSF}} \Delta M^{\rm (4)}_{\text{PSF}}}  + \alpha^{\rm (4)} C_{\ell}^{M^{\rm (4)}_{\text{PSF}} \Delta M^{\rm (4)}_{\text{PSF}}}  + \beta^{\rm (4)} C_{\ell}^{\Delta M^{\rm (4)}_{\text{PSF}} \Delta M^{\rm (4)}_{\text{PSF}}}.
\end{align}
\begin{flushright}\rule{8.8cm}{0.4pt}\end{flushright}
\end{strip}

\begin{figure}
    \centering
    \includegraphics[width=0.98\columnwidth]{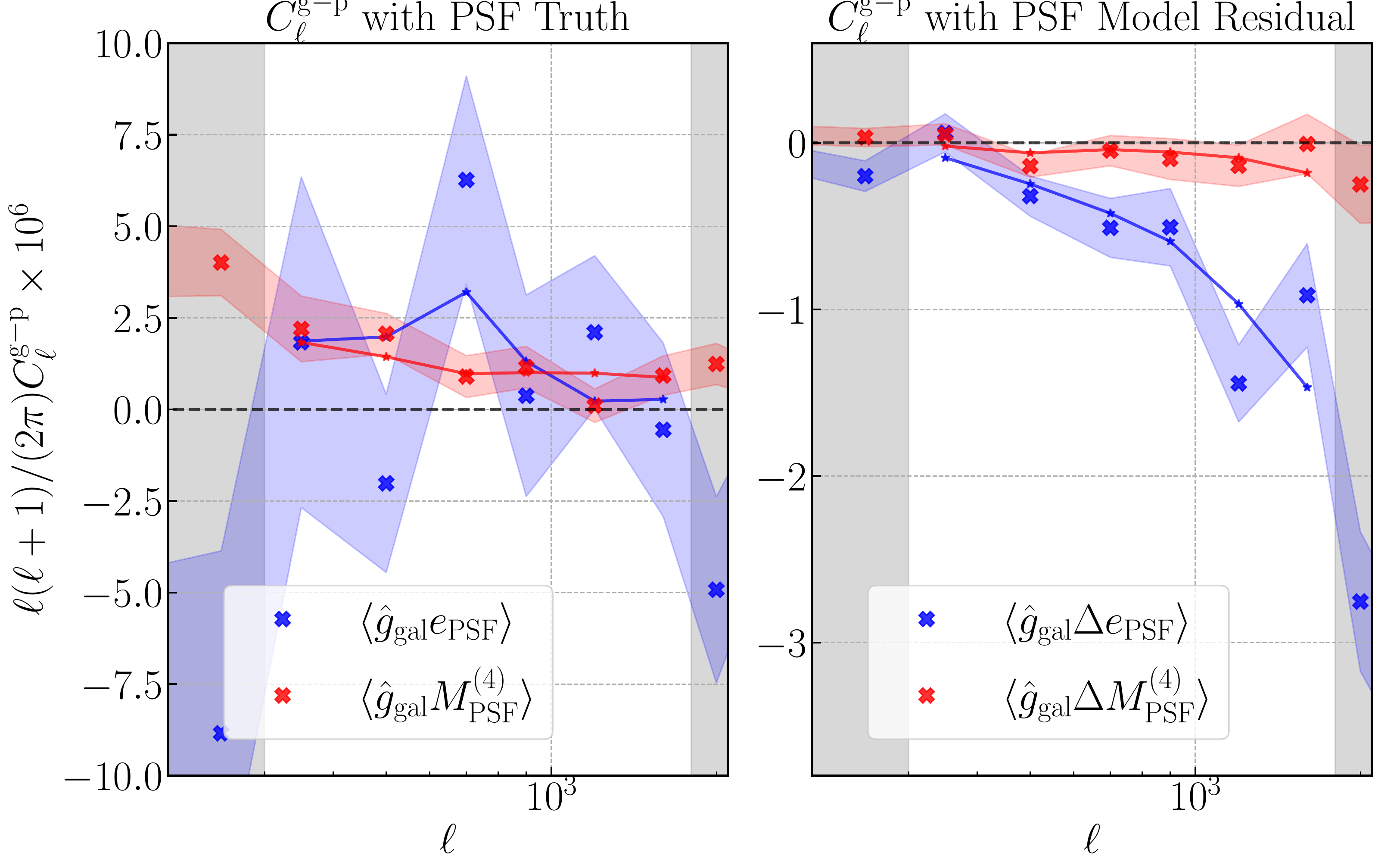}
\caption{The measured Fourier space g-p angular power spectrum $\ve{D}_{gp}$ and the bestfit $\ve{T}_{gp}$ for the fiducial PSF systematics model. The left panel shows the g-p power spectra with the PSF truth terms expressed by Eqs.~\eqref{eq:null1_fourier} and~\eqref{eq:null3_fourier}, and the right panel shows the power spectra with the PSF model residual expressed by Eqs.~\eqref{eq:null2_fourier} and~\eqref{eq:null4_fourier}. We only use scales between $300\le \ell\le 1800$ in our fit (unshaded region).
}
\label{fig:cell_gp}
\end{figure}

\begin{figure}
    \centering
    \includegraphics[width=1.0\columnwidth]{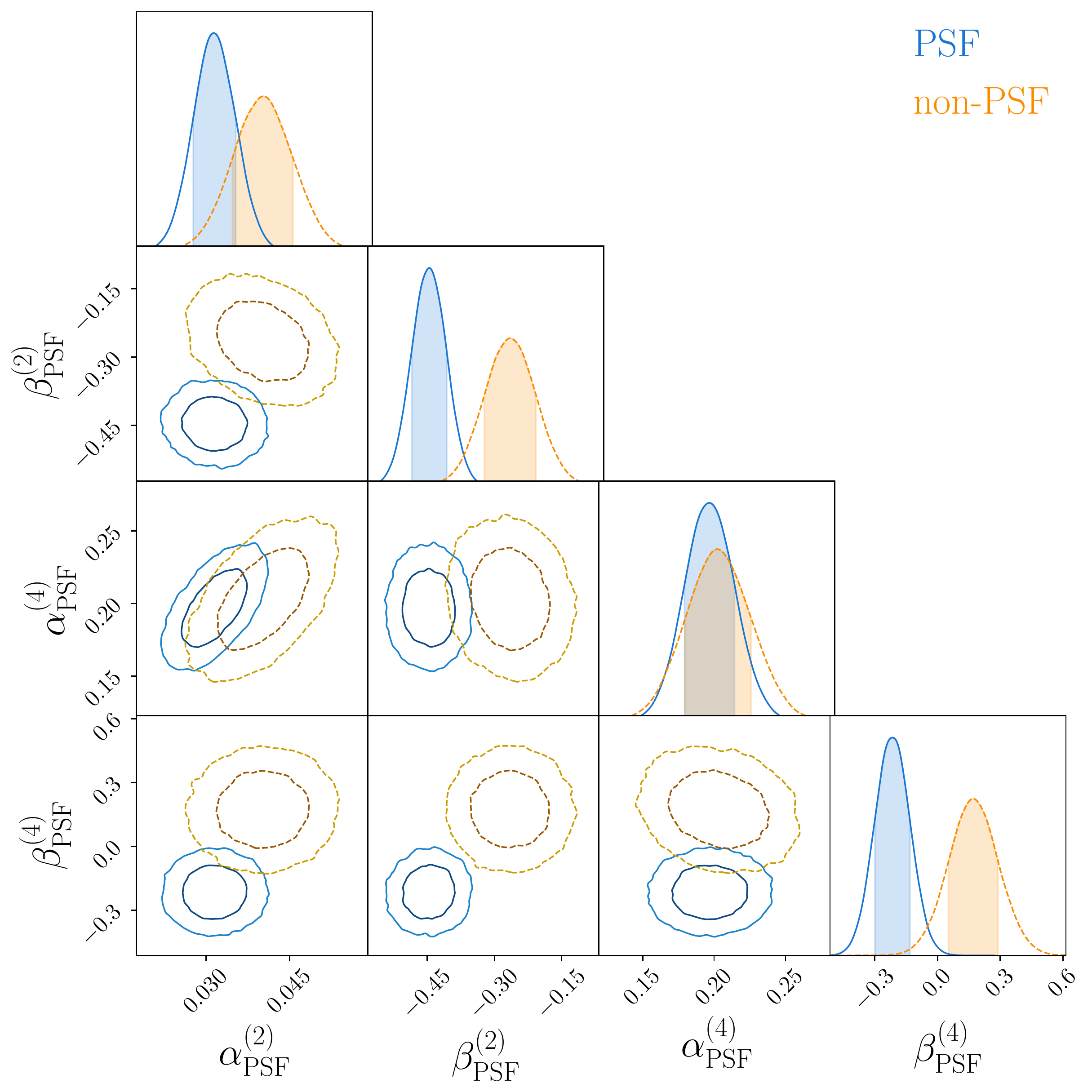}
\caption{The posterior probability distribution of the fiducial PSF systematics model parameters applied to the angular power spectra, using the PSF and non-PSF stars.
}
    \label{fig:fourier_alpha_beta_posterior}
\end{figure}

We find the values of the parameters $\alpha^{\rm (2)}$,
$\beta^{\rm (2)}$, $\alpha^{\rm (4)}$, and $\beta^{\rm (4)}$ which maximize the log-likelihood function given by Eq.~\eqref{eq:likelihood_define}. The covariance matrix of $\vect{D}_{gp}$ for the Fourier space analysis is computed from the HSC Y3 mock catalog, described in Section~\ref{sec:shape:mock}. Note that the covariance for the Fourier space cross power spectra does not include the uncertainty of the p-p power spectra, which is different from the real space analysis.
The best-fitting $\ve{T}_{gp}$, as well as the measured g-p correlations, $\ve{D}_{gp}$, are shown in Fig.~\ref{fig:cell_gp}. As in the case of the 2PCF analysis, we also ran a Markov Chain Monte Carlo (MCMC) to measure the posterior of the PSF parameters $P(\vect{p} | \ve{D}_{gp})$, using a flat prior for the PSF parameters from $-\infty$ to $+\infty$. These posteriors are shown in Figure~\ref{fig:fourier_alpha_beta_posterior}. We validate the parameter inference using the mock catalog test, described in Appendix~\ref{sec:ap:mock_test}.

\begin{figure}
    \centering
    \includegraphics[width=0.98\columnwidth]{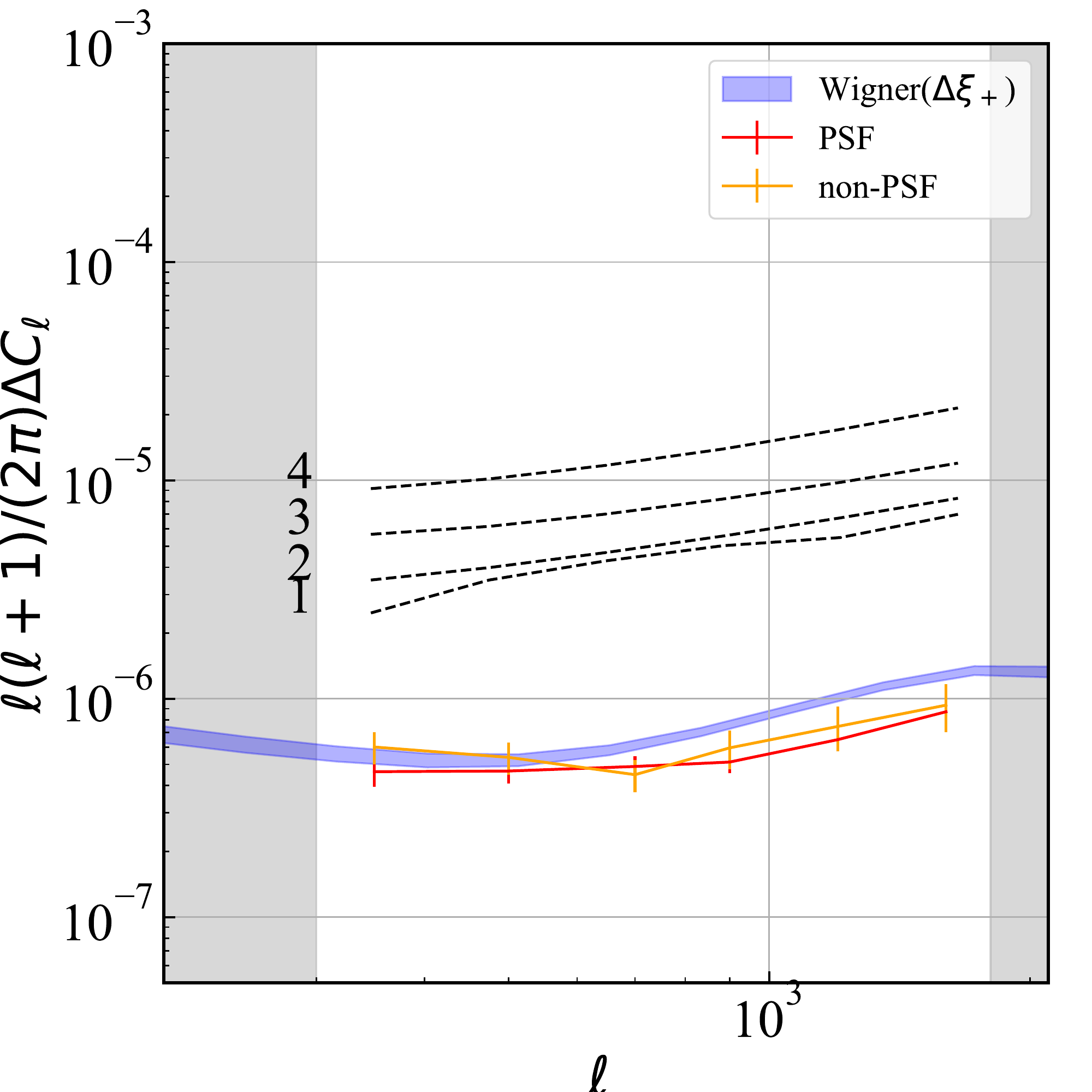}
\caption{
The additive bias in cosmic shear power spectra from PSF systematics (see Eq.~\eqref{eq:delta_cl_fourier}), based on the best-fitting values of $\vect{p}$ (red for PSF stars, yellow for non-PSF), compared to the expected bias based on an inverse-Wigner transform of the bias in the cosmic shear 2PCF predicted by the fiducial model in real space $\Delta\xi_+$ (blue). 
The theory cosmic shear power spectra in each tomographic bin, based on the fiducial cosmology (see Table~\ref{tab:cosmology_parameters}), are shown in black. This figure is the Fourier space equivalence to Fig.~\ref{fig:delta_xip_nobin}.}
    \label{fig:delta_cl}
\end{figure}

Finally, we use the best-fitting values of the PSF parameters to compute the bias in the cosmic shear power spectra, $\Delta C_{\ell}$, for the parameter values estimated from both PSF and non-PSF stars. As shown in Figure~\ref{fig:delta_cl}, the additive biases inferred from the two star catalogs are consistent with one another.

\subsection{Consistency between Real and Fourier Space}
\label{sec:ap:rf_consistency}

In this section, we discuss the internal consistency between the PSF systematics manifested in the real and Fourier space analyses. By checking that the real space and Fourier space analysis provide consistent results across different stages of the analysis, we further validate our PSF systematics treatment\footnote{If the model is not sufficient to describe the data, we expect results to differ in real space and Fourier space, because they implicitly weight scales differently, which can affect how the model mismatch manifests in the fits.  If the model is sufficient, however, they should agree within the uncertainties.}.

\begin{table}
\centering
\begin{tabular}{lllll}
\hline
Sample & Parameter  & Real Space   & Fourier Space     \\ \hline
        & $\alpha^{\rm (2)}$ & $0.016\pm0.002$ & $0.032\pm 0.004$ \\
PSF     & $\beta^{\rm (2)}$  & $-0.84\pm0.03$ & $-0.45\pm 0.04$ \\
        & $\alpha^{\rm (4)}$ & $0.17\pm0.01$ & $0.20\pm 0.02$ \\
        & $\beta^{\rm (4)}$ & $-0.6\pm0.10$ & $-0.21\pm 0.08$ \\\hline
        & $\alpha^{\rm (2)}$ & $0.020\pm0.004$ & $0.040\pm 0.005$ \\
non-PSF & $\beta^{\rm (2)}$  & $-0.57\pm0.07$ & $-0.26\pm 0.06$ \\
        & $\alpha^{\rm (4)}$ & $0.17\pm0.01$ & $0.20\pm 0.02$ \\
        & $\beta^{\rm (4)}$ & $0.11\pm0.12$ & $0.18\pm 0.12$
\\\hline
\end{tabular}
\caption{
The best-fitting parameters of the fiducial model in real space and Fourier space analysis, for both PSF and non-PSF stars. The dominant contributor to the additive bias in the power spectra/2PCFs, the fourth moment leakage parameter $\alpha^{\rm (4)}$ matches well between real space and Fourier space, while $\alpha^{\rm (2)}$, $\beta^{\rm (2)}$, and $\beta^{\rm (4)}$ are inconsistent between the two analyses.}
\label{tab:real_fourier_parameters}
\end{table}

In Table~\ref{tab:real_fourier_parameters}, we compare the best-fitting parameters of the fiducial PSF systematics model in real space and Fourier space, for both PSF and non-PSF samples. $\alpha^{\rm (2)}$, $\beta^{\rm (2)}$, and $\beta^{\rm (4)}$ appear to be inconsistent, although the dominant contributor to the additive bias, $\alpha^{\rm (4)}$, is consistent between the two analyses, for both the PSF and non-PSF samples. As a result of the consistency in $\alpha^{\rm (4)}$, we expect the additive bias on the data vectors inferred from both methods to be roughly consistent. 
We compute the predicted $\Delta \Tilde{C}_\ell$ by inverse-Wigner transforming the shear-shear contamination $\Delta \xi_+(\theta)$  \citep{2021MNRAS.508.1632S}
\begin{equation}
    \Delta \Tilde{C}_\ell = 2 \pi \int_0^{2\pi} \mathrm{d}\theta \sin(\theta) d^\ell_{2,2}(\theta) \Delta \xi_+(\theta).
\end{equation}
Here $d^\ell_{2,2}$ is the Wigner matrix for two spin-2 fields at the given  $\ell$ \citep{1999IJMPD...8...61N}. We fit the $\Delta \xi_+(\theta)$ predicted by the real space fiducial model in the range $[1,200]$ arcmin using a double exponential model (determined empirically), while setting the value outside the angular range to zero:
\begin{equation}
\Delta \xi_+ = a_1 e^{-s_1 \theta} + a_2 e^{-s_2 \theta}.
\end{equation}
The best-fitting parameters of the double exponential model are $a_1 = 1.33\times 10^{-5}$, $a_2 = 2.19\times 10^{-6}$, $s_1 = 54.3~\mathrm{deg}^{-1}$, $s_2 = 1.38~\mathrm{deg}^{-1}$. We show the predicted $\Delta \Tilde{C}_\ell$ using the $1\sigma$ uncertainty on the PSF systematics model parameters with the blue region in Fig.~\ref{fig:delta_cl}. Despite having different $\alpha^{\rm (2)}$, $\beta^{\rm (2)}$ and $\beta^{\rm (4)}$, the impact on the cosmological observable still marginally matches, due to the fact that the fourth moment leakage is the largest contributor to the additive bias.
The $\Delta \Tilde{C}_\ell$ predicted from the real space $\Delta \xi_+$ matches the $\Delta C_\ell$s predicted by the PSF and non-PSF stars of the Fourier space fiducial model, expressed in Eq.~\eqref{eq:delta_cl_fourier}, within $2\sigma$.

\begin{figure}
\centering
\includegraphics[width=0.98\columnwidth]{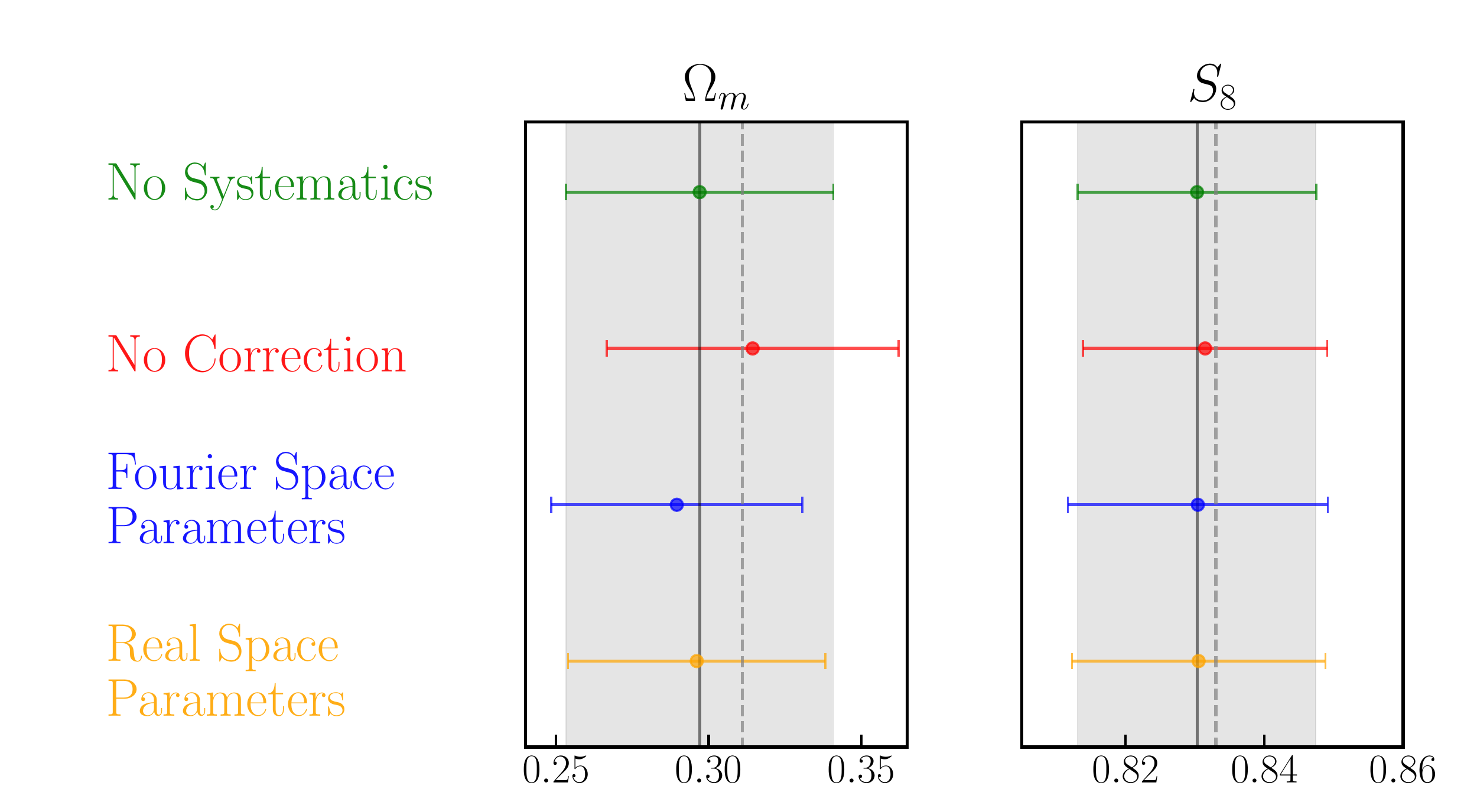}
\caption{The 1d constraints on $\Omega_m$ and $S_8$ in the HSC Y3 mock cosmic shear analysis. The green and red lines are the same as in Fig.~\ref{fig:y3_contour}. The orange lines are the parameter constraints using the fiducial model with PSF parameters inferred in real space, while the blue lines are the parameter constraints using the PSF systematics parameters obtained in the Fourier space. The difference between the correction using real and Fourier space parameters with the fiducial model causes a $\Omega_m$ bias around $0.15\sigma$, which is subdominant.}
\label{fig:1d-fourier}
\end{figure}

To demonstrate that the difference in Fourier and real space for the PSF systematics parameters will not significantly impact the cosmological results, we run an additional mock cosmological analysis on the Y3-like data vector and covariance. In Fig.~\ref{fig:1d-fourier}, we show the 1-d $\Omega_m$-$S_8$ constraints of the Y3 mock cosmological analysis. In addition to the green, red and orange lines that were shown in Fig.~\ref{fig:y3_contour}, we include another fiducial correction with the PSF parameters obtained in the Fourier space analysis. The difference results in a bias on $\Omega_m$ of about $0.15\sigma$, and a $0.01 \sigma$ bias on $S_8$. We conclude that these remaining systematics are subdominant for the Y3 cosmological results. 

These results suggest that our PSF systematics model may not be fully sufficient to describe the data, but the real versus Fourier space comparison suggests this is not a problem for an analysis at our current level of precision.  We therefore leave this issue for future work; with a larger area catalog it will be more important to understand this issue, if it persists.  Since most image systematics are tied to particular physical scales (such as the size of the image focal plane, the typical correlation length of the atmospheric PSF anisotropies, etc.) we suspect that the issue could arise because the adopted range of $\ell$ values include information from values of $\theta$ on which our model does not include all relevant physics.

\label{LastPage}

\end{document}